\documentclass[11pt,a4paper,notitlepage]{report}
\usepackage[utf8]{inputenc}
\usepackage[english]{babel}
\usepackage{amsmath}
\usepackage{amsfonts}
\usepackage{amssymb}
\usepackage{lmodern}
\usepackage{geometry}
\usepackage{physics}
\usepackage{dsfont}
\usepackage{graphicx}
\usepackage[numbers,compress]{natbib}
\usepackage{wrapfig}
\usepackage{pgfplots}
\usepackage{tikz}
\usepackage{nomencl}
\usepackage{relsize}
\usepackage{pdfpages}
\usetikzlibrary{decorations.pathmorphing}
\usepackage{color}   %May be necessary if you want to color links
\usepackage{hyperref}
\usepackage{titling}
\hypersetup{colorlinks=true,linktoc=all,linkcolor=blue,}
\makenomenclature
\makeindex
\title{A Review of the Holographic Relation between Linearized Gravity and the First Law of Entanglement Entropy}
\author{Rasmus Jaksland}
\begin{document}
\begin{titlingpage}
    \maketitle
    \begin{abstract}
        This thesis reviews the conjectured holographic relation between entanglement and gravity due to Mark van Raamsdonk and collaborators. It is accounted how the linearized Einstein equations both with and without matter in a $d+1$-dimensional AdS background can be derived from the first law of entanglement entropy in a $d$-dimensional CFT. This derivation builds on the Ryu-Takayanagi formula that relates entanglement entropy for CFT subsystems to extremal surfaces in the AdS bulk. The relation between gravity and entanglement is also corroborated by a qualitative investigation of the duality between the thermofield double state and the maximally extended AdS/Schwarzschild black hole using the Bekenstein-Hawking formula. Furthermore, this qualitative argument is generalized to generic CFT states with a classical spacetime dual using the Ryu-Takayanagi.

The thesis also reviews the most relevant prerequisites  for this holographic relation between gravity and entanglement: Anti-de Sitter spacetime, entanglement and entanglement entropy, gauge/gravity duality, the Ryu-Takayanagi formula, and linearized gravity.
\normalsize
    \end{abstract}
\end{titlingpage}

\tableofcontents
%\printnomenclature
\chapter{Introduction}
Ever since its discovery by Maldacena in 1998, the AdS/CFT correspondence has intrigued researchers in quantum gravity. The correspondence entails a duality between theories with and without gravity, and recently, Mark van Raamsdonk and collaborators \citep{van_raamsdonk_building_2010,van_raamsdonk_patchwork_2011,lashkari_gravitational_2014,faulkner_gravitation_2014}
have proposed that entanglement here plays a crucial role. The suggestion relies on a single entry from the AdS/CFT dictionary; the Ryu-Takayanagi formula \citep{ryu_holographic_2006} and its covariant generalization \citep{hubeny_covariant_2007}. This formula relates entanglement entropy on the CFT side to areas of extremal co-dimension two surfaces on the AdS side, thereby relating an intrinsically quantum mechanical phenomena on the CFT side to the geometry of the bulk in the spacetime dual. This result, van Raamsdonk conjectures, proves an intimate relation between entanglement on the CFT side with spacetime itself on the AdS side. He argues “that the intrinsically quantum phenomenon of entanglement appears to be crucial for the emergence of classical spacetime geometry” \citep[pp. 4–5]{van_raamsdonk_building_2010}.

A qualitative justification for this claim employs the example of a thermofield double state and its dual description as a maximally extended AdS-Schwarzschild black hole (the eternal black hole). In this case, the spacetime connectivity between the two exterior regions of the eternal black hole changes when the amount of entanglement between the two double states is changed. While this result may be established employing a holographic interpretation of the well known Bekenstein-Hawking formula, it may be generalized using the Ryu-Takayanagi formula to all quantum states with a classical spacetime dual. Thus, for such pairs this qualitative argument indicates that there is an intimate relation between entanglement in a CFT state and spacetime in the dual spacetime. The details of this qualitative argument will be the topic of chapter \ref{Entanglement and Spacetime} of this thesis.

Chapter \ref{Entropy and Field Equations} will be dedicated to more rigorous and quantitative support for a relation between entanglement on the CFT side and gravity on the AdS side. More precisely, it will be demonstrated how the first law of entanglement entropy on the CFT side is equivalent to imposing the linearized Einstein equations without matter on the AdS side. 

The linearized Einstein equations without matter are the $G_N \rightarrow 0$ limit of the Einstein equations, where $G_N$ is Newton's constant. Going to the first subleading order in the $G_N$ expansion, bulk matter fields will have to be taken into account. Since these, when promoted to quantum fields, give rise to entanglement in the bulk, a correction to the Ryu-Takayanagi formula is required \citep{faulkner_quantum_2013}. Chapter \ref{Beyond linearized gravity} will explore this correction and argue that the same entanglement constraint on the CFT side with this correction is equivalent to linearized Einstein equations including a source term.

Thus in summary, it is reviewed in this thesis how the Ryu-Takayanagi formula alone entails that the linearized Einstein equations both with and without matter are satisfied in a spacetime if the first law of entanglement entropy is satisfied in a dual CFT state. Before proceeding to this, however, chapters two through six will introduce relevant aspects of anti-de Sitter spacetime (chapter \ref{Anti-de Sitter Spacetime}), entanglement and entanglement entropy (chapter \ref{Entanglement and Entropy}), the AdS/CFT correspondence (chapter \ref{A brief introduction to gauge/gravity duality}), the Ryu-Takayanagi formula (chapter \ref{The Ryu-Takayanagi formula}), and linearized gravity (chapter \ref{Linearized gravity in AdS background}).

\section{Notation and conventions}
Throughout this thesis, we will work in units where Planck's constant, $\hbar$, Boltzmann's constant, $k_B$, and the speed of light, $c$ to unity -- $\hbar = c = k_B = 1$ -- we will, however, keep the dependence on Newton's constant, $G_N$, explicit since this will be instructive for later purposes. Adopting Einstein notation, we will assume an implicit sum over repeated indices. When ambiguities may occur, we will indicate the number of dimensions in which constants are defined by raised numbers in parenthesis. For instance, the notation $G_N^{(26)}$ signifies that this is Newton's constant in 26-dimensional spacetime. Greek lower case indices starting with $\mu$ -- $\mu, \nu, \rho, \sigma$ -- will be used for Minkowski spacetime coordinates, latin lower case indices starting from $i$ -- $i,j,k,l$ -- will be used for Minkowski space coordinates.  Latin lower case indices starting from $a$ -- $a,b,c,d,e,f$ -- will be used for Poincaré coordinates of AdS spacetime and occasionally for coordinate independent relations. To avoid confusing in the numerical ordering of the Poincaré coordinates, we will use $t$ for the time component of tensors (for instance $A^t$) and $z$ for the AdS scaling coordinates component of tensors (for instance $A^z$). Thus, raised and lowered $t$ and $z$ are not indices but tensor components. Most often, we will consider $d+1$-dimensional AdS spacetime. Here $0 \leq a \leq d$, $0 \leq \mu \leq d-1$, and $1 \leq i \leq d-1$. We will use $D$ to denote the dimensionality of a spacetime. Thus, we will often find that $D = d+1$. Finally, we will use capital indices starting with $A$ -- $A, B, C, D$ -- for target space coordinates and greek lower case indices starting with $\alpha$ -- $\alpha, \beta$ -- for world sheet coordinates.

Throughout, $g$ will denote the background metric, and a metric induced by $g$ on a surface (including the world sheet metric) will be denoted by $\gamma$. Unless stated otherwise, metrics will have Lorentzian signature and we adopt the sign convention $\text{diag}(-,+,\cdots,+)$.

\chapter{Anti-de Sitter Spacetime}\label{Anti-de Sitter Spacetime}
Throughout this thesis, we will be interested in so-called anti-de Sitter (AdS) spacetime: It is one of the cornerstones in gauge/gravity duality, it will serve as the background in which we will consider linearized gravity, and one of the central examples will deal with asymptotically AdS black holes. Consequently, we will subsequently spend some time introducing AdS spacetime and briefly asymptotically AdS black holes. To connect with the subsequent chapters, we introduce AdS in $d+1$ dimensions. The material in this section is based on chapter 2 of \citep{ammon_gauge/gravity_2015} and chapter 5 of \citep{carroll_spacetime_2014}.
\section{Global AdS}
According to the theory of general relativity, spacetimes must satisfy the Einstein equations
\begin{equation}
G^E_{ab} = 8 \pi G_N T_{ab}
\end{equation}
where $0 \leq a \leq d$, $T_{ab}$ is the energy momentum tensor and $G^E_{ab}$ is the Einstein tensor
\nomenclature{$d$}{The number of spatial dimensions.}
\nomenclature{$G_N$}{Newton's constant.}
\nomenclature{$T_{ab}$}{The energy-momentum tensor.}
\nomenclature{$G_{ab}$}{The Einstein tensor.}
\nomenclature{$g_{ab}$}{The metric tensor of the spacetime.}
\begin{equation}
G_{ab} \equiv R_{ab} - \frac{1}{2} R g_{ab}  + \Lambda g_{ab}.
\end{equation}
Here $\Lambda$ is the cosmological constant, $R_{ab}$ is the Ricci tensor and $R$ is the Ricci scalar defined in terms of the Riemann tensor as
\begin{equation}
R_{ab} \equiv R_{acb}^{\; \; \; \; c}, \; \; \; \; \; \; \; \; \; \; \; \; R \equiv g^{ab} R_{ab}.
\end{equation}
%\nomenclature{$R_{abcd}$}{The Riemann tensor.}
%\nomenclature{$R_{ab}$}{The Ricci tensor. Defined as $R_{ab} \equiv R_{acb}^{\; \; \; \; \; c}$}
%\nomenclature{$R$}{The Ricci scalar. Defined as $R \equiv g^{ab} R_{ab}$.}
%\nomenclature{$\mathtt{R}$}{Radius of ball shaped regions.}

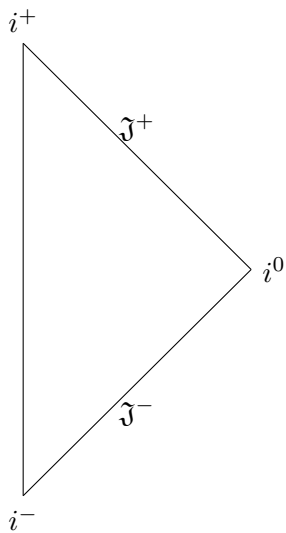
\begin{figure}[h]
\begin{center}
\begin{tikzpicture}
  % Four corners of left diamond
     \path  (-1.5,3)  coordinate  (top);
     \path  (-1.5,-3) coordinate (bot);
     \path  (1.5,0) coordinate (right);
     \path  (-1.5,0)   coordinate (left);
    
         \node at (-1.5,3.3) {$i^+$};
         \node at (1.8,0) {$i^0$};
     \node at (-1.5,-3.3) {$i^-$};
     
\draw (top) -- (right) node[midway, above, inner sep=2mm] {$\mathfrak{J}^+$};
\draw (right)  -- (bot) node[midway, below, inner sep=2mm] {$\mathfrak{J}^-$};
\draw (top) -- (bot);

\end{tikzpicture}
\caption{\label{PMST} Penrose diagram of $d$-dimensional Minkowski spacetime. A sphere $S^{d-2}$ in each point is implicit. $i^+$ denotes future timelike infinity, $i^-$ denotes past timelike infinity, $i^0$ denotes spacelike infinity, $\mathfrak{J}^+$ denotes future null infinity, and $\mathfrak{J}^-$ denotes past null infinity.}
\end{center}
\end{figure}

Minkowski spacetime is arguably the most well known solution to the vacuum Einstein equations without cosmological constant, i.e. Einstein equations for which $T_{ab} = 0$ and $\Lambda = 0$. The causal structure of this and other spacetimes can be depicted in a Penrose diagram, where the possibly infinite spacetime is brought into finite size nice by choosing an appropriate conformal factor. Null curves are always $\pm 45^{\circ}$ with vertical such that timelike directions are those directions bounded by the null curves. The Penrose diagram of Minkowski spacetime is depicted in figure \ref{PMST}. 

A characteristic feature of Minkowski spacetime is that it is a maximally symmetric spacetime in the sense that is has the maximal number of spacetime symmetries as given by Killing vector fields, $K_{a}$, that satisfy the equation
\begin{equation}
\nabla_a K_b + \nabla_b K_a = 0.
\end{equation}
Here $\nabla_a$ is the covariant derivative, which is given in terms of the partial derivative, $\partial_a$, and the Christoffel connection, $\Gamma^c_{ab}$,
\begin{equation}
\nabla_a V_b = \partial_a V_b - \Gamma^c_{ab} V_c
\end{equation}
where $\Gamma^c_{ab}$ may be expressed in terms of the metric as
\begin{equation}
\Gamma^c_{ab} = \frac{1}{2} g^cd (\partial_a g_{db} + \partial_b g_{da} - \partial_d g_{ab} ).
\end{equation}
\nomenclature{$\nabla_a$}{The covariant derivative.}
One finds that the maximal number of symmetries and therefore Killing vector fields is $\frac{1}{2}D(D+1)$ where $D$ is the number of spacetime dimensions of the manifold ($D = d+1$).
\nomenclature{$D$}{The number of spacetime dimension of a manifold.} Thus, any manifold with this many symmetries/Killing vector fields is maximally symmetric. 
In a maximally symmetric spacetime, the curvature must be the same everywhere; something that is trivially obeyed by Minkowski spacetime. 

That curvature is the same everywhere entails that the Riemann tensor may be expressed in terms of the Ricci scalar, $R$,
\begin{equation}
R_{abcd} = \frac{R}{D(D-1)} \left( g_{ab} g_{cd} - g_{ad} g_{bc} \right)
\end{equation}
For Minkowski spacetime, where $R=0$ everywhere, this entails that the Riemann tensor vanishes everywhere. However, the relation indicates that we may consider two other maximally symmetric solutions to the Einstein equations in vacuum: One where $R>0$ and one where $R<0$. The latter type of spacetimes -- the maximally symmetric spacetimes with $R<0$ -- will be our primary interest in the following. Such spacetimes are known as anti-de Sitter spacetimes and solve the Einstein equations with negative cosmological constant as may be seen from
\begin{equation}
\begin{split}
0 & = G^E_{ab}
\\& = \left( R_{ab} - \frac{1}{2} R g_{ab} + \Lambda g_{ab} \right) 
\\& = 2 g^{ab} \left( R_{ab} - \frac{1}{2} R g_{ab} + \Lambda g_{ab} \right) 
\\& = 2 R - D R + 2 \Lambda D
\end{split}
\end{equation}
from which it follows $R = 2 \Lambda D /(D-2)$. Thus, $R \propto \Lambda$.

For convenience, we will consider anti-de Sitter spacetimes in (d+1) dimensions. This may be embedded in a (d+2) dimensional Minkowski spacetime, $R^{d,2}$, with the metric
\begin{equation}
ds^2 = - (dX^0)^2 + (dX^1)^2 + \cdots + (dX^d)^2 - (dX^{d+1})^2
\end{equation}
and given by the hypersurface subject to the restriction
\begin{equation}\label{HS}
-(X^0)^2 + \sum_{i=1}^d (X^i)^2 - (X^{d+1})^2 = -L^2.
\end{equation}
This hypersurface may also be parametrized by
\begin{equation}
\begin{split}
X^0 & = L \, \cosh(\rho) \, \cos(\tau)
\\ X^i & = L \, \Omega_i \, \sinh(\rho),  \; \; \; \; \text{for }i \in \lbrace 1, \dots , d \rbrace
\\ X^{d+1} & = L \, \cosh(\rho) \, \sin(\tau)
\end{split}
\end{equation}
where $\rho \in \mathbb{R}_+$, $\tau \in [0,2\pi[$, and $\Omega_i$ parametrizes a $(d-1)$-dimensional sphere, $S^{d-1}$, such that it satisfies $\sum_{i=1}^d \Omega_i^2 = 1$. These coordinates are known as global coordinates of $AdS_{d+1}$. In these global coordinates, the metric takes the form 
\begin{equation}
ds^2 = L^2 \left( - \cosh^2(\rho) d\tau^2 + d\rho^2 + \sinh^2(\rho) d\Omega_{d-1}^2 \right).
\end{equation}
where $d\Omega_{d-1}^2$ is the standard metric on $S^{d-1}$. The vector field $\partial_{\tau}$ is a Killing vector field on the entire manifold and $\tau$ may therefore serve as a global time coordinate. It is worth noting that $\tau$ and therefore time is periodic in $2 \pi$. Usually, we will mean by global anti-de Sitter spacetime, the universal covering of anti-de Sitter spacetime in global coordinates, where $\tau$ is unwrapped such that $\tau \in \mathbb{R}$.

For purposes below, we will be interested in the conformal boundary of anti-de Sitter spacetime. This is more easily studied if we replace $\rho$ by $\theta$ such that $\tan(\theta) = \sinh(\rho)$ which entails $\theta = [0,\frac{\pi}{2}]$. In these coordinates, the metric takes the form
\begin{equation}\label{GAdS}
ds^2 = \frac{L^2}{\cos^2(\theta)} (- d\tau^2 + d\theta^2 + \sin^2(\theta) d \Omega_{d-1}^2 ).
\end{equation}
Since we are interested in a conformal boundary, the overall scaling factor $\frac{L^2}{\cos^2(\theta)}$ may be ignored and we see that for $\theta = \pm \frac{\pi}{2}$, we obtain $\partial AdS_{d+1}^{Global} = \mathbb{R} \times S^{d-1}$, where $\partial AdS_{d+1}^{Global}$ denotes the conformal boundary of (the universal covering of) global anti-de Sitter spacetime.  

\begin{figure}[h]
\begin{center}
\begin{tikzpicture}
  % Four corners of left diamond
     \path  (-1.5,3)  coordinate  (topl);
     \path  (-1.5,-3) coordinate (botl);
     \path  (1.5,3) coordinate (topr);
     \path  (1.5,-3)   coordinate (botr);
     \path  (-5,0) coordinate (mid);
     
     \node at (-1.5,-3.3) {$\theta = 0$};
     \node at (1.5,-3.3) {$\theta = \frac{\pi}{2}$};

\draw (topl) -- (topr) node[midway, above, inner sep=2mm] {$i^+$};
\draw (botl) -- (botr) node[midway, below, inner sep=2mm] {$i^-$};
\draw (topl) -- (botl) node[midway, left, inner sep=2mm] {$\rho= 0$};
\draw (topr) -- (botr) node[midway, right, inner sep=2mm] {$\rho= \infty$};

%\draw (4,-3) -- (4,3) node[midway, right, inner sep=2mm] {$r=\infty$};
%\draw (4,3) -- (1,0) node[midway, below, sloped, inner sep=2mm] {$r=0$};
%\draw (4,-3) -- (1,0) node[midway, above, sloped, inner sep=2mm] {$r=0$};

\end{tikzpicture}
\caption{\label{PADS} Penrose diagram of $d+1$-dimensional universal covering of global AdS spacetime. A contribution $\sin^2(\theta)  d \Omega_{d-1}^2$ in each point is implicit. $i^+$ detones future timelike infinity, $i^-$ denotes past timelike infinity. Due to the causal structure of the spacetime, there is no spacelike, future null, or past null infinity.}
\end{center}
\end{figure}
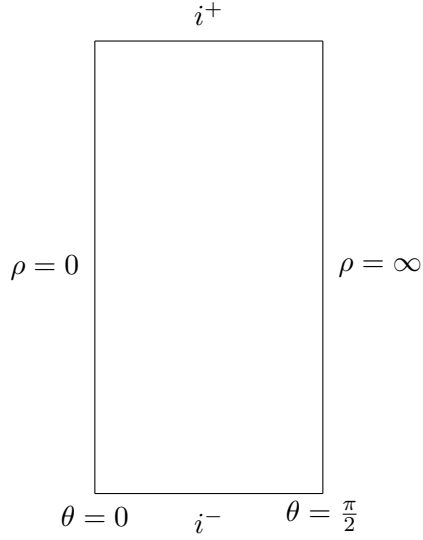

Figure \ref{PADS} depicts the Penrose diagram for the universal covering of global anti-de Sitter spacetime; still suppressing the $\sin^2(\theta) d \Omega_{d-1}^2$ part of the metric. To include this, regard the Penrose diagram in figure \ref{PADS} as a radial slice of a hypercylinder whose angular part is given by $\sin^2(\theta) d \Omega_{d-1}^2$. Thus, $\theta=0$ is the center of the cylinder and $\theta = \frac{\pi}{2}$ is the boundary. Contrary to Minkowski spacetime, there is no spatial infinity since a light signal may travel to boundary, $\rho = \infty$, and back again in finite time. To see this, consider the metric (\ref{GAdS}) for a light ray (null curve) for which $ds^2 = 0$. Disregarding the factor $L^2/\cos(\theta)$ and the scaled sphere $\sin^2(\theta)  d \Omega_{d-1}^2$, we have
\begin{equation}
\begin{split}
0 & = -d \tau^2 + d \theta^2
\\ \Rightarrow d \tau & = d \theta
\\ \Rightarrow \tau & = \int_0^{\pi/2} d \theta = \pi/2
\end{split}
\end{equation}
from which we see that light may travel to the boundary in finite time.

\section{Poincaré patch of AdS}
Another parametrization of (\ref{HS}) that will be used repeatedly below is given in terms of the coordinates $t \in \mathbb{R}$, $\vec{x} = (x^1, \dots , x^{d-1} ) \in \mathbb{R}^{d-1}$ and $r \in \mathbb{R}_+$. The parametrization then reads
\begin{equation}
\begin{split}
X^0 & = \frac{L^2}{2r} \left( 1 + \frac{r^2}{L^4} \left[ \vec{x}^2-t^2+L^2 \right] \right),
\\ X^i & = \frac{r x^i}{L}, \; \; \; \; \text{for }i \in \lbrace 1, \dots , d-1 \rbrace,
\\ X^{d} & = \frac{L^2}{2r} \left( 1 + \frac{r^2}{L^4} \left[ \vec{x}^2 - t^2 - L^2 \right] \right),
\\ X^{d+1} & = \frac{rt}{L}.
\end{split}
\end{equation}
These are only local coordinates since the restriction $r > 0$ entails that we only cover half of $AdS_{d+1}$. This is the so-called Poincaré patch and the coordinates therefore are known as Poincaré coordinates. The metric of this Poincaré patch of $AdS_{d+1}$ is
\begin{equation}\label{Poincaremetric}
ds^2 = \frac{L^2}{r^2} dr^2 + \frac{r^2}{L^2} \left( -dt^2 + d\vec{x}^2 \right).
\end{equation}
In subsequent chapters, we will most often take it to be implicit that $AdS_{d+1}$ is the Poincaré patch of $AdS_{d+1}$. 

We will also be interested in the conformal boundary of the Poincaré patch of $AdS_{d+1}$. This can be shown to be d-dimensional Minkowski spacetime i.e. $\mathbb{R}^{1,d-1}$, which obtains for $r \rightarrow \infty$. This is more readily seen, if we replace $r$ by defining $z = \frac{L^2}{r}$ such that the metric becomes 
\begin{equation}\label{dsAdS}
ds^2 = \frac{L^2}{z^2} \left( - dt^2 + dz^2 + d\vec{x}^2 \right).
\end{equation}
Here the conformal boundary is found at $z \rightarrow 0$ and ignoring the pre-factor we exactly get d-dimensional Minkowski spacetime.\footnote{More precisely stated, the metric $d\tilde{s}^2 = ds^2/z^2$ has the boundary $\mathbb{R}^{1,d-1}$ at $z \rightarrow 0$ and $d\tilde{s}^2$ is conformally equivalent to $ds^2$.} Thus, for any slice of the Poincaré patch with $z=\text{constant}$, the metric is that of Minkowski spacetime. The coordinate $z$ takes the form of a warpfactor. 

For this reason, the Penrose diagram for the Poincaré path of $AdS_{d+1}$ is the same as that for $(d)$-dimensional Minkowski spacetime (figure \ref{PMST}). This may seem odd. Minkowski spacetime is after all very different from anti-de Sitter spacetime. The reason that the Poincaré patch can have the same Penrose diagram as Minkowski spacetime is that the Poincaré patch -- as the name indicates -- only covers a patch of AdS. This patch, it turns out, has the same causal structure as Minkowski spacetime in agreement with the claim the the Poincaré patch is conformally equivalent to Minkowski spacetime.

%Since they will be used repeatedly below, the Christoffel connections for the Poincaré patch are calculated below. The Christoffel connections are given as (\textsc{Some are missing???})
%\begin{equation}
%\Gamma_{ab}^c = \frac{1}{2} g^{cd} \left( \partial_a g_bd + \partial_b g_{da} - \partial_d g_{ab} \right)
%\end{equation}
%In the Poincaré patch of anti-de Sitter spacetime given by the metric (\ref{dsAdS}), we find:
%\begin{equation}
%\begin{split}
%\Gamma_{zz}^c v_c& = \frac{1}{2} g^{cd} \left( \partial_z g_zd + \partial_z g_{dz} - \partial_d g_{zz} \right) v_c
%\\& = \frac{1}{2} g^{cz} \left( \partial_z g_zz + \partial_z g_{zz} - \partial_z g_{zz} \right) v_c
%\\& = - \frac{1}{z} \delta^{c}_z v_c
%\end{split}
%\end{equation}
%
%\begin{equation}
%\begin{split}
%\Gamma_{zi}^c v_c& = \frac{1}{2} g^{cd} \left( \partial_z g_{id} + \partial_i g_{dz} - \partial_d g_{zi} \right) v_c
%\\& = \frac{1}{2} g^{cd} \left( \partial_z g_{id} \right) v_c
%\\& = - \frac{1}{z} \delta^{c}_i v_c
%\\& = - \frac{1}{z} \delta^c_i v_c
%\end{split}
%\end{equation}

%Finally, a remark about perturbation of the Poincaré patch of AdS spacetime

\section{Schwarzschild black holes}\label{Schwarzschild black holes}
Another interesting class of solutions to the Einstein equations are black hole solutions that feature event horizons. The so-called maximally extended asymptotically AdS-Schwarzschild black hole in global coordinates will be of particular interest since it will play a crucial role in developing the qualitative argument in favour of a relation between entanglement and spacetime. 

%\textsc{What is the dimensionality in the following?}
However, we will begin with the well known Schwarzschild black hole. This is a spherically symmetric solution to the Einstein equations for $\Lambda = 0$. The metric may be expressed as
\begin{equation}\label{dsflat}
ds^2 = - f(r) dt^2 + \frac{dr^2}{f(r)} + r^2 d\Omega^2_{d-2}
\end{equation}
where $d\Omega^2_{d-2}$ is the usual metric for $S^{d-2}$, $r$ is a radial coordinate and $f(r) = 1 - \frac{2 \mu}{r^{d-3}}$. The constant, $\mu$ is given by
\begin{equation}
\mu = \frac{8 \pi G_N M}{(d-1) Vol(S^{d-1})}
\end{equation}
such that in 4 dimensions $\mu_{d=4} = G_N M$ from which we obtain the well known expression for the Schwarzschild black hole in 4 dimensions
\begin{equation}
ds^2 = - \left( 1 - \frac{2 G_N M}{r} \right) dt^2 + \left( 1 - \frac{2 G_N M}{r} \right)^{-1} dr^2 + r^2 d\Omega^2_{d-2}.
\end{equation}
Since our primary interest is in the causal structure of black holes and this is the same in any number of dimensions, we will for convenience work in 4 dimensions in the following. Also, we will suppress the last part of the metric $r^2 d\Omega^2_{d-2}$, since this part of the metric merely contributes with a $(d-2)$-dimensional sphere in all points in the conformal diagram of the spacetime and may be reinserted thus.

The metric of interest -- the $(t,r)$ subspace -- therefore takes the form
\begin{equation}
ds^2 = - \left( 1 - \frac{2 G_N M}{r} \right) dt^2 + \left( 1 - \frac{2 G_N M}{r} \right)^{-1} dr^2.
\end{equation}
Two points stand out. One is $r=0$, which is a curvature singularity and the other is $r=2G_N M$, which turns out to be a coordinate singularity. In general dimensions, the curvature singularity is still situated at $r=0$ and the coordinate singularity (the black hole horizon) is situated at $r = r_h \equiv (2 \mu)^{\frac{1}{d-3}}$.

To investigate the causal structure, we want to consider the behaviour of light-cones and one finds that the slope of these -- defined as $\frac{dr}{dt}$ -- are given by
\begin{equation}
\frac{dr}{dt} = \pm \left(1 - \frac{2 G_NM}{r} \right)^{-1}
\end{equation}
since the light-cone is spanned by the radial null curves $ds^2 = 0$. For $r \rightarrow 2 G_NM$, we see that $\frac{dr}{dt}=0$; the light-cone closes up as we approach the event horizon. This, however, is merely a feature of these particular coordinates. A better choice of coordinates will resolve this and allow us to study the causal structure as we pass the event horizon. The problem that the light-cone closes up may be resolved, if we can find a coordinate, $r^*(r)$, such that
\begin{equation}
\frac{dr}{dt} = \pm r^*(r).
\end{equation}
This coordinate, known as the tortoise coordinate, must satisfy 
\begin{equation}
dr^* = \frac{dr}{1 - \frac{2G_NM}{r}}
\end{equation}
and integrating on both sides, one obtains
\begin{equation}
r^* = r - 2G_NM + 2G_NM \ln\left( \frac{r}{2G_NM} - 1 \right).
\end{equation}
We see that the event horizon at $r = 2 G_NM$ is located, in the tortoise coordinate, at $r^* \rightarrow - \infty$. This also entails that $r^*$ is only defined for $r \geq 2 G_NM$. In the tortoise coordinates, the metric takes the form
\begin{equation}
ds^2 = \left( 1 - \frac{2 G_NM}{r(r^*)} \right) \left( -dt^2 + {dr^*}^2 \right).
\end{equation}
In these coordinates, the metric is no longer singular at $r = 2 G_N M$, however, it is at the price that the event horizon is pushed to $r^* \rightarrow - \infty$. This may be resolved by replacing the time coordinate with either one of the light-cone tortoise coordinates:
\begin{equation}
v \equiv t + r^*, \; \; \; \; \; \; \; \; \; \; \; \; u \equiv t - r^*
\end{equation}
where the infalling radial null geodesics have $v = \text{constant}$ and outgoing radial null geodesics have $u = \text{constant}$. The causal structure of infalling objects may then be investigated with the coordinates $(v,r)$, where the metric takes the form
\begin{equation}
ds^2 = - \left( 1 - \frac{2 G_NM}{r} \right) dv^2 + dv \, dr + dr \, dv.
\end{equation}
The radial null curves, $ds^2 = 0$, satisfies
\begin{equation}
\frac{dv}{dr} = \begin{cases} 0 \\ 2 \left( 1- \frac{2 G_NM}{r} \right)^{-1} \end{cases}
\end{equation}
from which we see that the outgoing radial null curve simply tilts over as the object approaches and passes $r = 2 G_NM$ such that all future directed paths for $r < 2G_NM$ goes in the direction of decreasing $r$; there is no way out of the black hole once in. 

However, we could also choose to consider the radial null curves in the coordinates $(u,t)$, where the metric takes the form
\begin{equation}
ds^2 = - \left( 1 - \frac{2 G_NM}{r} \right) du^2 - du \, dr - dr \, du.
\end{equation}
In this case, the radial null curves satisfies
\begin{equation}
\frac{du}{dr} = \begin{cases} - 2 \left( 1- \frac{2 G_NM}{r} \right)^{-1} \\ 0 \end{cases}
\end{equation}
where we again see that the light-cone tilts when $r = 2 G_NM$ is approached and passed. In these coordinates, however, the situation is reversed. The horizon may be passed by following past directed curves such that for $r < 2G_NM$ all such paths go in the direction of decreasing $r$. This is a white hole; once you are out, there is no way back in. 

In order to find coordinates that includes the inside of both the black and the white hole, one could propose to use $(u,v)$. This gives the metric 
\begin{equation}
ds^2 = - \frac{1}{2} \left( 1 - \frac{2 G_NM}{r} \right) \left( dv \, du + du \, dv \right)
\end{equation}
where $r$ is related to $u$ and $v$ by
\begin{equation}
\frac{1}{2} (v-u) = r + 2 G_NM \ln \left( \frac{r}{2 G_NM} - 1 \right).
\end{equation}
We see that the horizon at $r = 2 G_NM$ has once again been pushed to infinity and is found at either $u= \infty$ or $v = - \infty$. To bring the horizon to a finite value in $v$ and $u$, we define
\begin{equation}
\begin{split}
v' & = e^{\frac{v}{4G_NM}}
\\ u' & = - e^{- \frac{u}{4G_NM}}
\end{split}
\end{equation}
and see that the horizon is now found at either $v' = 0$ or $u' = 0$ and that $0 < v' < \infty$, $-\infty < u < 0$. Since $v$ and $u$ are null coordinates so are $v'$ and $u'$. To express the metric in terms of one timelike coordinate and the rest spacetime, we introduce the coordinates known as Kruskal-Szekeres coordinates,
\begin{equation}
\begin{split}
T & = \frac{1}{2} (v' + u')
\\ R & = \frac{1}{2} (v' - u').
\end{split}
\end{equation}
such that $-\infty < T < \infty$ and $0 < R < \infty$. The metric becomes 
\begin{equation}\label{KSM}
ds^2 = \frac{32 G_N^3 M^3}{r} e^{-\frac{r}{2G_NM}} \left( -dT^2 + dR^2 \right) 
\end{equation}
and the horizon in these coordinates is found at $T = R$ for $T > 0$ and at $T = -R$ for $T < 0$. We can approach and pass the horizon following null curves given by $T = \pm R + \text{constant}$. Thus, these coordinates include both the inside and the outside of the white and black holes. The coordinates, however, may be analytically extended even further. In these coordinates, we can extend the range of $R$ from $0 < R < \infty$ to $-\infty < R < \infty$. The region where $R < 0$ mirrors the region $R>0$ and therefore includes a second region exterior to the black and white holes. As it turns out, no choice of coordinates can extend this spacetime further and we therefore denote the metric (\ref{KSM}) with coordinate range $-\infty < T < \infty$ and $-\infty < R < \infty$ as the maximally extended Schwarzscild black hole.

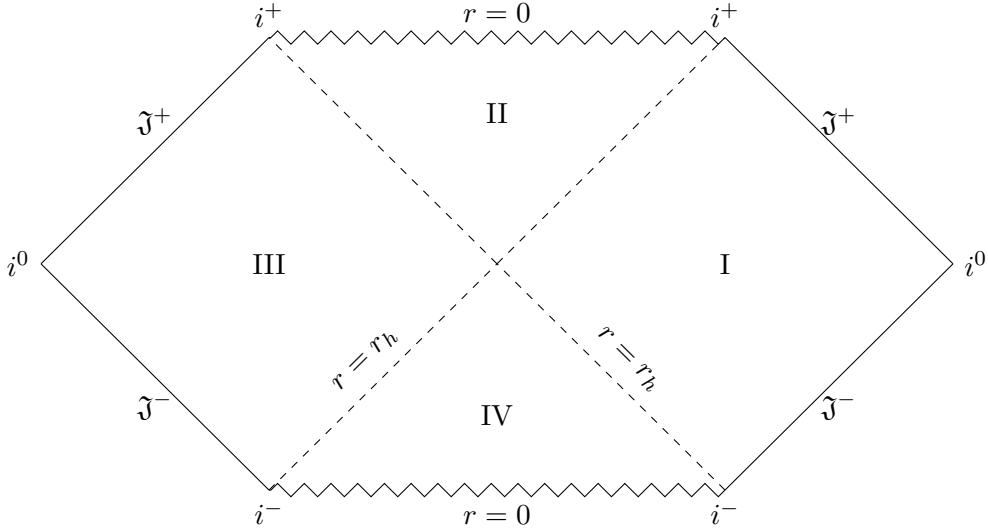
\begin{figure}[h]
\begin{center}
\begin{tikzpicture}
\node (I)    at ( 3,0)   {I};
\node (II)   at (-3,0)   {III};
\node (III)  at (0, 2) {II};
\node (IV)   at (0,-2) {IV};

  % Four corners of left diamond
     \path  (3,3)  coordinate  (IItop);
     \path  (3,-3) coordinate (IIbot);
     \path  (6,0) coordinate (IImid);
     \path  (0,0)   coordinate (mid);
    
         \node at (3,3.3) {$i^+$};
     \node at (-3,3.3) {$i^+$};
     \node at (-6.3,0) {$i^0$};
     \node at (6.3,0) {$i^0$};
     \node at (3,-3.3) {$i^-$};
     \node at (-3,-3.3) {$i^-$};

\draw[dashed] (IIbot) -- (mid) node[midway, above, sloped, inner sep=2mm] {$r=r_h$};
\draw[dashed] (mid) -- (IItop);
\draw (IItop) -- (IImid) node[midway, above, inner sep=2mm] {$\mathfrak{J}^+$};
\draw (IImid)  -- (IIbot) node[midway, below, inner sep=2mm] {$\mathfrak{J}^-$};

% Four conners of the right diamond (no labels this time)
 \path      (-3,3)  coordinate (Itop);
 \path      (-3,-3) coordinate (Ibot);
 \path      (-6,0) coordinate (Imid);
       
% No text this time in the next diagram
\draw  (Itop) -- (Imid) node[midway, above, inner sep=2mm] {$\mathfrak{J}^+$};
\draw (Imid) -- (Ibot) node[midway, below, inner sep=2mm] {$\mathfrak{J}^-$};
\draw[dashed] (Ibot) -- (mid) node[midway, above, sloped, inner sep=2mm] {$r=r_h$};
\draw[dashed] (mid) -- (Itop);

% Squiggly lines
\draw[decorate,decoration=zigzag] (IItop) -- (Itop)
      node[midway, above, inner sep=2mm] {$r=0$};

\draw[decorate,decoration=zigzag] (IIbot) -- (Ibot)
      node[midway, below, inner sep=2mm] {$r=0$};

\end{tikzpicture}
\caption{\label{PMESBH} Penrose diagram of the $d$-dimensional maximally extended Schwarzschild black hole geometry. A sphere $S^{d-2}$ scaling as $r^2$ must be added to each point. Regions I and III cover regions that lie outside the horizon (dashed, $r=r_h \equiv (2 \mu)^{\frac{1}{d-3}}$) of the black hole interior covered by region II. Region IV is the interior of a white hole.}
\end{center}
\end{figure}

The causal structure of the maximally extended Schwarzschild black hole may be depicted in a Penrose diagram (see figure \ref{PMESBH}) where we use $\tilde{u} = \arctan \left( \frac{u'}{\sqrt{2G_NM}} \right)$ and $\tilde{v} = \arctan \left( \frac{v'}{\sqrt{2G_NM}} \right)$ to bring the coordinates into finite range. Still, we have suppressed the $(d-2)$-dimensional sphere, i.e. we consider the $(t,r)$ subspace. The full spacetime may be obtained merely by including a $(d-2)$-dimensional sphere scaling as $r^2$ in each point in the diagram. As seen on figure \ref{PMESBH}, the Penrose diagram for the maximally extended Schwarzschild black hole shares with Minkowski space the same future null infinity, $\mathfrak{J}^+$, past null infinity, $\mathfrak{J}^-$, and spacelike infinity, $i^0$, which verifies that the maximally extended Schwarzschild black hole is indeed asymptotically Minkowski. One subtlety about the diagram is that future timelike infinity, $i^+$, and past timelike infinity, $i^-$, are distinct from the singularity at $r=0$ despite the appearance to the contrary. There are many timelike paths that do not end on the singularity, but go to other values of $r$. From the diagram we see that no future directed timelike path goes from one of the exterior region to the other; these two regions are causally disconnected.
\section{Asymptotically AdS black holes} \label{Asymptotically AdS black holes}
As mentioned in the beginning of this chapter, what we are really interested in is the maximally extended AdS-Schwarzscild black hole. This is similar to the maximally extended Schwarzscild black hole developed above but differs in that it is asymptotically AdS rather than asymptotically flat. This entails that the function $f(r)$ in (\ref{dsflat}) takes the form
\begin{equation}
f(r) = 1 - \frac{2 \mu}{r^{d-3}} + \frac{r^2}{L^2}.
\end{equation}

\begin{figure}[h]
\begin{center}
\begin{tikzpicture}

\node (I)    at (-3,-0.5)   {I};
\node (II)   at (-7,-0.5)   {III};
\node (III)  at (-5, 2) {II};
\node (IV)   at (-5,-2) {IV};

  % Four corners of left diamond
     \path  (-2,3)  coordinate  (IItop);
     \path  (-2,-3) coordinate (IIbot);
     \path  (-5,0)   coordinate (mid);
     
     \node at (-8,3.3) {$i^+$};
     \node at (-2,3.3) {$i^+$};
     \node at (-8,-3.3) {$i^-$};
     \node at (-2,-3.3) {$i^-$};

\draw (IItop) -- (mid) node[scale=0.7,midway, above, sloped, inner sep=2mm] {$r=r_h$};
\draw (mid) -- (IIbot);
\draw (IItop) -- (IIbot) node[midway, right, inner sep=2mm] {$A$};

% Four conners of the right diamond (no labels this time)
 \path      (-8,3)  coordinate (Itop);
 \path      (-8,-3) coordinate (Ibot);
       
% No text this time in the next diagram
\draw  (Itop) -- (Ibot) node[midway, left, inner sep=2mm] {$B$};
\draw (Ibot) -- (mid);
\draw (mid) -- (Itop) node[scale=0.7,midway, above, sloped, inner sep=2mm] {$r=r_h$};

% Squiggly lines
\draw[decorate,decoration=zigzag] (IItop) -- (Itop)
      node[scale=0.7,midway, above, inner sep=2mm] {$r=0$};

\draw[decorate,decoration=zigzag] (IIbot) -- (Ibot) node[scale=0.7,midway, above, inner sep=2mm] {$r=0$};

\draw[dashed] (-8,0) -- (-2,0) node[scale=0.7,near start, below right, inner sep=2mm] {$T=0$};
\draw[dashed] (-8,1) -- (-2,1) node[scale=0.7,midway, below, inner sep=2mm] {$T=\mathsmaller{\mathrm{const}^+}$};

%Upper right hand figure

  \draw (0,3) to[out=250,in=110] (0,1);
     \draw (0,3) to[out=290,in=70] (0,1);
     \draw (0,3) to[out=0,in=180] (2.5,2.5);
     \draw (0,1) to[out=0,in=180] (2.5,1.5);
     \draw (2.5,2.5) to[out=250,in=110] (2.5,1.5);
     \draw (2.5,2.5) to[out=290,in=70] (2.5,1.5);
     \draw (2.5,2.5) to[out=0,in=180] (5,3);
     \draw (2.5,1.5) to[out=0,in=180] (5,1);
     \draw (5,3) to[out=250,in=110] (5,1);
     \draw (5,3) to[out=290,in=70] (5,1);
     \node  at (2.5,1) {$T=0$};
     \node[scale=0.5,rotate=-90] at (2.75,2) {$r=r_h$}; 
     
%Lower right hand figure
     
     \draw (0,0) to[out=250,in=110] (0,-2);
     \draw (0,0) to[out=290,in=70] (0,-2);
     \draw (0,0) to[out=0,in=180] (2.5,-0.6);
     \draw (0,-2) to[out=0,in=180] (2.5,-1.4);
     \draw (2.5,-0.6) to[out=250,in=110] (2.5,-1.4);
     \draw (2.5,-0.6) to[out=290,in=70] (2.5,-1.4);
     \draw (3.2,-0.5) to[out=250,in=110] (3.2,-1.5);
     \draw (3.2,-0.5) to[out=290,in=70] (3.2,-1.5);
     \draw (1.8,-0.5) to[out=250,in=110] (1.8,-1.5);
     \draw (1.8,-0.5) to[out=290,in=70] (1.8,-1.5);
     \draw (2.5,-0.6) to[out=0,in=180] (5,0);
     \draw (2.5,-1.4) to[out=0,in=180] (5,-2);
     \draw (5,0) to[out=250,in=110] (5,-2);
     \draw (5,0) to[out=290,in=70] (5,-2);
     \node  at (2.5,-2) {$T=\mathrm{const^+}$};
     \draw[decorate,decoration=brace] (1.8,-0.4) -- (3.2,-0.4) node[scale=0.7,midway, above, inner sep=2mm] {\footnotesize black hole interior};
     \node[scale=0.5,rotate=-90] at (3.45,-1) {$r=r_h$};
     \node[scale=0.5,rotate=-90] at (1.55,-1) {$r=r_h$};
     \node[scale=0.45,rotate=-90] at (2.75,-1) {$r<r_h$};

     \node  at (-5,-3.5) {a};
     \node at (2.5,3.5) {b};

\end{tikzpicture}
\caption{\label{PMSASBH} a) Penrose diagram of the maximally extended AdS-Schwarzschild black hole with a $d-1$ dimensional sphere over each point that scales as $r^2$. b) Depiction of two spacelike slices of the the eternal black hole ($T=0$ and $T$ equals a positive constant) with one angular coordinate restored.}
\end{center}
\end{figure}
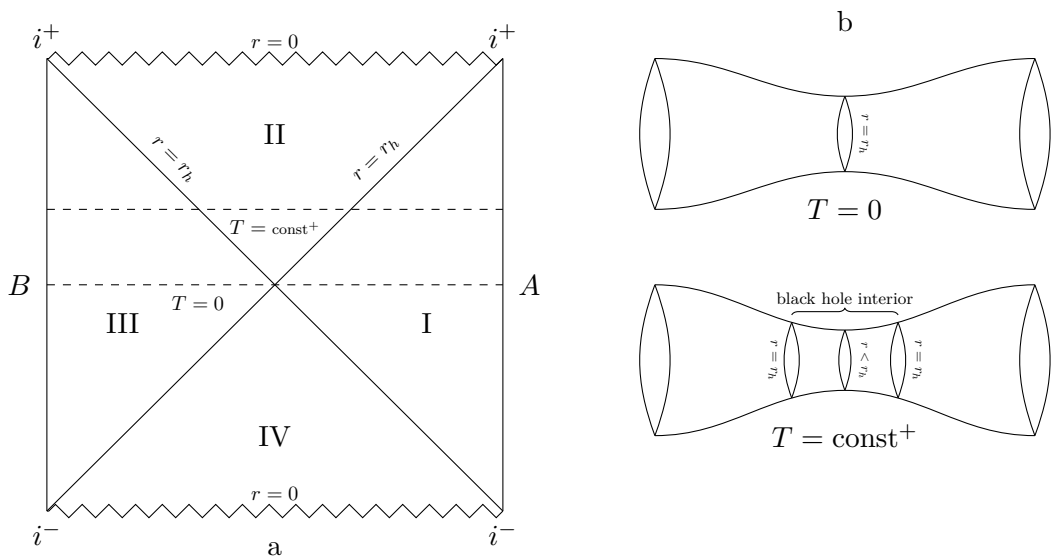

Compared to the maximally extended Schwarzschild black hole, the causal structure therefore remains the same with the exception that the spacetime is asymptotically AdS. The Penrose diagram for the maximally extended AdS-Schwarzscild black hole is therefore similar to figure \ref{PMESBH} with the exception that the future null infinity, the past null infinity and the spacelike infinity must be identical to that of the universal covering of global AdS spacetime. The Penrose diagram is seen in figure \ref{PMSASBH}. Compared to figure \ref{PMESBH}, the asymptotic boundaries of regions I and III are now the universal covering of global AdS spacetime. Due to the nature of this asymptotic boundary, light signals may reach and return from the boundary in finite time as shown above. This has peculiar consequences for the maximally extended AdS-Schwarzscild black hole where $r_h \gg L$. If $r_h \ll L$, we may neglect the term $\frac{r^2}{L^2}$, and the maximally extended AdS-Schwarzscild black hole will have properties similar to that of the asymptotically flat maximally extended Schwarzscild black hole. Particularly, both will evaporate due to the emission of Hawking radiation. For $r_h \gg L$, however, the maximally extended AdS-Schwarzscild black hole may be in an equilibrium with its Hawking radiation since the radiation can reach the asymptotic AdS boundary and return in finite time. Thus, if emission and reabsorption rates are equal then the maximally extended AdS-Schwarzscild black hole will not evaporate. For this reason, we denote such black holes as eternal black holes.
%\begin{figure}[h]
%\begin{center}
%\begin{tikzpicture}
%\node (I)    at ( 2,0)   {I};
%\node (II)   at (-2,0)   {III};
%\node (III)  at (0, 2) {II};
%\node (IV)   at (0,-2) {IV};

  % Four corners of left diamond
%     \path  (3,3)  coordinate  (IItop);
 %    \path  (3,-3) coordinate (IIbot);
  %   \path  (6,0) coordinate (IImid);
   %  \path  (0,0)   coordinate (mid);
    
    %     \node at (3,3.3) {$i^+$};
%     \node at (-3,3.3) {$i^+$};
 %    \node at (3,-3.3) {$i^-$};
  %   \node at (-3,-3.3) {$i^-$};

%\draw[dashed] (IIbot) -- (mid) node[scale=0.7,near end, above, sloped, inner sep=2mm] {$r=r_h$};
%\draw[dashed] (mid) -- (IItop);
%\draw (IItop) -- (IIbot);

% Four conners of the right diamond (no labels this time)
% \path      (-3,3)  coordinate (Itop);
% \path      (-3,-3) coordinate (Ibot);
% \path      (-6,0) coordinate (Imid);
       
% No text this time in the next diagram
%\draw  (Itop) -- (Ibot);
%\draw[dashed] (Ibot) -- (mid) node[scale=0.7,near end, above, sloped, inner sep=2mm] {$r=r_h$};
%\draw[dashed] (mid) -- (Itop);

% Squiggly lines
%\draw[decorate,decoration=zigzag] (IItop) -- (Itop)
%      node[midway, above, inner sep=2mm] {$r=0$};

%\draw[decorate,decoration=zigzag] (IIbot) -- (Ibot)
%      node[midway, below, inner sep=2mm] {$r=0$};

%\end{tikzpicture}
%\caption{\label{PMSASBH} Penrose diagram of the $d$-dimensional maximally extended AdS-Schwarzschild black hole geometry. A sphere $S^{d-2}$ must be added in each point.}
%\end{center}
%\end{figure}
For the eternal black hole we find -- as we did for the maximally extended Schwarzschild black hole -- two causally disconnected regions, I and III. No signal can travel from one to the other. Still, we will describe the eternal black hole as a connected spacetime since a signal from the region I can intersect a signal from region III. The signals, however, can only intersect inside the black hole thereby precluding any causal connection between the two exterior regions I and III. The picture emerging is that of a wormhole so long as it is remembered that the wormhole is simply a two sided black hole, i.e. a black hole with two distinct exterior regions. That we may conceive of the eternal black hole as a wormhole is further signified by the depiction in figure \ref{PMSASBH}b of $T=0$ and a constant positive $T$ slice of the spacetime. 

%These depictions of timelike slices of the eternal black more offers a perhaps more intuitive image of the spacetime. 
These depiction more explicitly demonstrates features of the eternal black hole; most explicitly the scaling of the sphere $S^{d-2}$ with $r^2$. Also, these depictions will prove useful in a qualitative assessment of the behavior of the spacetime when the area of the black hole horizon is decreased. 

\chapter{Entanglement and Entropy}\label{Entanglement and Entropy}
Entanglement is an inherently quantum mechanical phenomena. To see the peculiarity of this phenomena let us consider two discernible particles, 1 and 2, and two positions, A and B. Both particles behave in such a way that when measured it is found at site A half of the times and at site B the other half of the times. However, the two particles are never found at the same site simultaneously. Thus, when we measure and find particle 1 at site A, then a similar measurement at site B will find particle 2 there. Similarly, if particle 2 is found at site A, we can be sure to find particle 1 at site B. 

We may initially attempt to ascribe a probability wave to each particle to account for the behaviour. Each particle is described by a wave with two sharp peaks at A and B and with an amplitude of $1/2$ at each site, i.e. they are ascribed a probability of $1/2$ for being at site A and $1/2$ for being at site B. However, describing the particles with an individual probability wave cannot account for their behaviour never to be found at the same site. Generally, there is no way to describe the behaviour with a function only ranging over the sites A and B (usually this will be a three dimensional position space). This will inevitably leave out some information about the relation between the two particles. This is the information about their entanglement.

To account for their entanglement, i.e. for the effect that the particles are never found at the same site, we must instead describe the particles as living in configuration space. In configuration space there are four possible configurations; not sites since the states are no longer in position space. These may be denoted $AA$, $AB$, $BA$ and $BB$ where the first letter denotes the position of the first particle and the second letter denotes the position of the second particle. Associating probability to each of these configurations, it follows from the example that only $AB$ and $BA$ have a non-zero probability to be occupied. More precise, the probability is $1/2$ for each. The lesson to learn is that it is necessary to describe the observed correlation of these particles in configuration space. It is not possible to describe it in position space as we are used to from classical physics.

For convenience we will subsequently use correlation between the spin state of two particles rather than a correlation between positions. We will assume that the particles are fermions and therefore have spin-$1/2$. Thus, each particle is either found to have spin up, $\ket{\uparrow}$, or spin down, $\ket{\downarrow}$. Calling the system composed of first particle, $A$, and the system composed of the second particle, $B$, their state-spaces is spanned by the states $\lbrace \ket{\uparrow}_A, \ket{\downarrow}_A \rbrace$ and $\lbrace \ket{\uparrow}_B, \ket{\downarrow}_B \rbrace$, respectively. The full system, therefore, is spanned by
\begin{equation}
\lbrace \ket{\uparrow}_N, \ket{\downarrow}_N \rbrace \otimes \lbrace \ket{\uparrow}_B, \ket{\downarrow}_B \rbrace
\end{equation}
Now, the state of the full system, $\ket{\Psi}$, is such that if one particle is found to have spin up, then the other particle will be found to have spin down and vice versa. Again, we cannot account for this correlation if we describe each particle individually. While we can get the statistics right by assigning probability $1/2$ to each state for each particle, this does not account for the correlation. We cannot describe the full state as a product of $A$ and $B$ states. 

The two systems (the two particles) are correlated, and 
$\ket{\Psi}$ therefore becomes:
\begin{equation}
\ket{\Psi} = \frac{1}{ \sqrt{2} } \left( \ket{\uparrow}_A \ket{\downarrow}_B + \ket{\downarrow}_A \ket{\uparrow}_B \right)
\end{equation}
this is what is known as a Bell pair. This is the maximally entangled state consisting of two qubits.

\section{The density matrix}
The example demonstrates that two entangled particles cannot be described as two separable subsystems, but must be regarded as one whole. To introduce this notion of entanglement more formally, we will consider a quantum state in some Hilbert space, $\ket{\Psi} \in \mathcal{H}$ that can be decomposed into at least two spatially separated subsystems.\footnote{$\ket{\Psi}$ is a pure state of $\mathcal{H}$} This implies that $\mathcal{H}$ can be expressed as the tensor product of the Hilbert spaces of each subsystem. Decomposing $\mathcal{H}$ into two subsystems, we have
\begin{equation}
\mathcal{H} = \mathcal{H}_A \otimes \mathcal{H}_B
\end{equation}
where $\mathcal{H}_A$ and $\mathcal{H}_B$ are the Hilbert spaces of the two subsystems (denoted $A$ and $B$ respectively). 

Given this decomposition of the Hilbert space, we may write a quantum state -- a vector $\ket{\Psi} \in \mathcal{H}$ -- in terms of basis states in the subsystem Hilbert spaces, $\ket{\psi^A_i} \in \mathcal{H}_A$ and $ \ket{\psi^B_j} \in \mathcal{H}_B$:
\begin{equation}
\ket{\Psi} = \sum_{i,j} c_{i,j} \ket{\psi^A_i} \otimes \ket{\psi^B_j}
\end{equation}
where $\sum_{i,j} \vert c_{i,j} \vert^2 = 1$.

Generally this is not a product state, i.e. a product of two pure states. Decomposing a system into two or more subsystems, it is generally \textit{not} the case that we may express the full quantum state as a direct product of the states of the subsystems
\begin{equation}\label{product}
\ket{\Psi} = \left( \sum_i c_i \ket{\psi^A_i} \right) \otimes \left( \sum_j c_j \ket{\psi^B_j} \right)
\end{equation}
even though the Hamiltonian of either subsystem -- due to locality -- is independent of each other. When this equality does not hold, we will say that systems $A$ and $B$ are entangled. 

This formal exposition sits well with the above account of the Bell pair. Here it was claimed that the spin of the particles were entangled, when we could not describe the state of the full system in terms of the states of two particles as spanned by $\lbrace \ket{\uparrow}, \ket{\downarrow} \rbrace$. Particularly, the outcome of measurements could not be accounted for by such a product state.

Generally, if $\ket{\Psi}$ is not a product state of $\mathcal{H}_A$ and $\mathcal{H}_B$, it follows that a pure state, $\sum_i c_i \ket{\psi^A_i}$, cannot give the same outcome as $\ket{\Psi}$ for a measurement on subsystem $A$. However, the outcome of such measurements can be reproduced by some ensemble of orthogonal states in $\mathcal{H}_A$, where each of these states, $\lbrace \ket{\psi^A_i} \rbrace$, are associated with a (classical) probability, $p_i$, of finding the system in that state when making the measurement on $A$.

Thus, the expectation value for the measurement on state $\ket{\Psi}$ agrees with the ensemble average of that measurement on this ensemble. Expressing this in terms of an operator, $\mathcal{O}_A$, acting on subsystem $A$, this implies:
\begin{equation}
\expval{\mathcal{O}_A} = \sum_i p_i \bra{\psi^A_i} \mathcal{O}_A \ket{\psi^A_i}
\end{equation}
To prove that there is such a class, $\ket{\psi^A_i}$, of orthogonal states, consider the expectation value of $\mathcal{O}_A$ when the full state is $\ket{\Psi}$
\begin{equation}
\begin{split}\label{expvalO}
\expval{\mathcal{O}_A} & = \bra{\Psi} \mathcal{O}_A \otimes \mathbb{I} \ket{\Psi} \\& =\left( \sum_{i,j} c_{i,j}^* \bra{\psi^B_j} \otimes \bra{\psi^A_i} \right) \mathcal{O}_A \left( \sum_{k,l} c_{k,l} \ket{\psi^A_k} \otimes \ket{\psi^B_l} \right) 
\\& = \sum_{i,j} \sum_{k,l} \delta_{jl} \bra{\psi^A_i} c_{i,j}^* \, c_{k,l} \, \mathcal{O}_A \ket{\psi^A_k} 
\\&
 = \sum_{i,k} \sum_{j} c_{i,j}^* \, c_{k,j} \, \bra{\psi^A_i} \mathcal{O}_A \ket{\psi^A_k} 
 \\& = \sum_{i,k} \sum_{n,m} \sum_{j} c_{i,j}^* \, c_{k,j} \, \braket{\psi^A_i}{\psi^A_n} \bra{\psi^A_n} \mathcal{O}_A \ket{\psi^A_m} \braket{\psi^A_m}{\psi^A_k}
 \\& = \sum_{n,m} \bra{\psi^A_m} \left( \sum_{i,k} \sum_{j} c_{i,j}^* \, c_{k,j} \, \ket{\psi^A_k} \bra{\psi^A_i} \right) \ket{\psi^A_n} \bra{\psi^A_n} \mathcal{O}_A \ket{\psi^A_m}
 \\& = \sum_{m} \bra{\psi^A_m} \mathcal{O}_A \, \rho_A \ket{\psi^A_m}
 \\& = \tr(\mathcal{O}_A \, \rho_A)
\end{split}
\end{equation}
where the last line is obtained by defining the operator
\begin{equation}\label{rho_A}
\rho_A = \sum_{i,k} \sum_j c_{i,j}^* \, c_{k,j} \ket{\psi^A_k} \bra{\psi^A_i}
\end{equation}
and by the definition of the trace
\begin{equation}
\tr(\mathcal{O}) = \sum_i \bra{i} \mathcal{O} \ket{i}.
\end{equation}
\\

As seen, the operator $\rho_A$ also only acts on subsystem $A$ and it is known as the reduced density matrix. It may be obtained from the density matrix of the full system, $\rho \equiv \ket{\Psi} \bra{\Psi}$, by tracing out the degrees of freedom outside of the subsystem. In the case where the full system is divided into two subsystems $A$ and $B$ this is
\begin{equation}
\begin{split}
\tr_B(\rho) & = \tr_B(\ket{\Psi} \bra{\Psi}) 
\\& = \sum_m \bra{\psi^B_m} \left( \sum_{k,l} c_{k,l} \ket{\psi^A_k} \otimes \ket{\psi^B_l} \right) \left( \sum_{i,j} c_{i,j}^* \bra{\psi^B_j} \otimes \bra{\psi^A_i} \right) \ket{\psi^B_m}
\\& = \sum_m \sum_{k,l} \sum_{i,j} \delta_{ml} \delta_{jm} c_{k,l} c_{i,j}^* \ket{\psi^A_k} \bra{\psi^A_i}
\\& = \sum_m \sum_{i,k} c_{k,l} c_{i,j}^* \ket{\psi^A_k} \bra{\psi^A_i}
\end{split}
\end{equation}
which is equal to (\ref{rho_A}) above. 

The density matrix is an hermitian operator, i.e. $\rho = \rho^\dagger$, and therefore it has real eigenvalues. Furthermore, the eigenvalues of the density matrix, $p_i$, can be shown to be non-negative. Thus
\begin{equation}
p_i \in \mathbb{R}^+
\end{equation}
The eigenvalues, $p_i$, are also sometimes known as the entanglement spectrum of the density matrix for a state $\ket{\Psi}$.

Using the associated eigenvectors we can express the reduced density matrix as
\begin{equation}
\rho_A = \sum p_i \ket{\phi^A_i} \bra{\phi^A_i}
\end{equation}
where $\ket{\phi^A_i} $ are the eigenvectors of $\rho_A$.

Inserting this into the expression for the expectation value of the operator $\mathcal{O}_A$, we find
\begin{equation}
\begin{split}
\expval{\mathcal{O}_A} & = \tr(\mathcal{O}_A \, \rho_A)
\\& = \tr \left( \mathcal{O}_A \sum p_i \ket{\phi^A_i} \bra{\phi^A_i} \right)
\\& = \sum_{i,j} p_i \bra{\phi^A_j} \left( \mathcal{O}_A \ket{\phi^A_i} \bra{\phi^A_i} \right) \ket{\phi^A_j}
\\& = \sum_{i,j} p_i \bra{\phi^A_j} \mathcal{O}_A \ket{\phi^A_i} \delta_{ij}
\\& = \sum_i p_i \bra{\phi^A_i} \mathcal{O}_A \ket{\phi^A_i}
\end{split}
\end{equation}
where we have used that the trace is the same in any orthonormal basis. This proves the claim that for the full system in a state $\ket{\Psi}$ there is an ensemble of orthogonal states in $\mathcal{H}_A$, $\lbrace p_i, \phi^A_i \rbrace$ that reproduces the expectation value of an operator, $\mathcal{O}_A$, that only acts on subsystem $A$.

From the last line of (\ref{expvalO}), we notice that the reduced density matrix for subsystem $A$, $\rho_A$, contains all information about the state, $\ket{\Psi}$, in subsystem $A$. Concerning our current interest -- entanglement -- it is interesting to see how $\rho_A$ includes information about the entanglement between subsystem $A$ and the full system. Particularly, we find that if $\rho_A$ only has one non-zero eigenvalue then the subsystem $A$ is in a pure state and thereby not entangled with the rest of the system.

To see this, let us assume that only the eigenvalue $p_1$ of $\rho_A$ is non-zero. We then have
\begin{equation}
\begin{split}
\expval{\mathcal{O}_A} &  = \bra{\Psi} \mathcal{O}_A \otimes \mathbb{I} \ket{\Psi} 
\\& = \tr(\mathcal{O}_A \, \rho_A)
\\& = \sum_i p_i \bra{\phi^A_i} \mathcal{O}_A \ket{\phi^A_i}
\\& = p_1 \bra{\phi^A_1} \mathcal{O}_A \ket{\phi^A_1}
\end{split}
\end{equation}
This implies that there is a single state of the Hilbert space for the subsystem -- in this case $\ket{\phi^A_1} \in \mathcal{H}_A$ -- that includes all the information of full state in that subsystem; there is no classical uncertainty about the state of the subsystem. If the full system consists of two subsystems $A$ and $B$ and $A$ is a pure state -- in this case $\ket{\phi^A_1}$ -- then it follows that $B$ is a pure state. Assuming that this pure state of $B$ is $\ket{\theta^B_1}$, we can thereby express the full state, $\ket{\Psi}$, as
\begin{equation}
\ket{\Psi} = \ket{\phi^A_1} \otimes \ket{\theta^B_1} 
\end{equation}
This is recognized as a product state, and the two subsystems is therefore \textit{not} entangled. In summary, if the reduced density matrix for some subsystem has only one non-zero eigenvalue, then the subsystem is \textit{not} entangled with the rest of the full system. The reduced density matrix contains the information about the entanglement between the subsystem and the rest of the full system.

\section{Entanglement entropy}
These properties makes it natural to use the reduced density matrix to quantify the entanglement between a subsystem and the rest of the system. For this purpose, we will define the entanglement entropy, $S$, as the the von Neumann entropy associated with a density matrix $\rho$ :
\begin{equation}\label{EE}
S = - \tr( \rho \log( \rho )).
\end{equation}

A number of properties makes the von Neumann entropy a good choice when quantifying entanglement. First, we observe that $S \geq 0$ which sits well with the fact that the notion of ``anti-entanglement" is incomprehensible. To see that $S \geq 0$, remember that $\rho$ must be diagonalized in the basis of its eigenvectors with diagonal elements equal to its eigenvalues, $p_i$. Since $0 \leq p_i \leq 1$, it follows that $\log(p_i) \leq 0$. Therefore, $\tr( \rho \log( \rho )) \leq 0$ which implies that $S \geq 0$. 

Another nice property is that the von Neumann entropy vanishes for a pure state. Thus, if there is no entanglement between a subsystem under consideration and the full system, then the entanglement entropy of that system is zero. This follows, since it holds for the density matrix of a pure state that $\rho^2 = \rho$. Thus
\begin{equation}
\begin{split}
& \rho \log(\rho) = \rho \log(\rho^2) = 2 \rho_B \log(\rho) \\& \Rightarrow \rho \log(\rho_B)=0
\\& \Rightarrow S(\rho) = -\tr( \rho \log( \rho )) = 0.
\end{split}
\end{equation}
The definiti

The von Neumann entropy also nicely reproduces the usual entropy formulas of well known ensembles. Taking the example of the microcanonical ensemble given by the combination of states and probabilities
\begin{equation}
\lbrace \ket{E_i}, p_i = \frac{1}{n} \rbrace
\end{equation}
where $E_i \in \langle E, E + dE \rangle$ and $n$ is the number of energy eigenstates in this interval. The entropy for the microcanonical ensemble is given as
\begin{equation}
S_{micro} = \log(n) = \sum_i p_i \log(n) 
\end{equation}
where the last equality is obtained by the identity 
\begin{equation}
\sum_{i=1}^n p_i = \sum_{i=1}^n \frac{1}{n} = 1
\end{equation}
which implies thinking of the total entropy as getting a contribution, $p_i \log(n)$, from each state, $E_i$, in the ensemble.

Now using the equality $\log(n) = - \log \left( \frac{1}{n} \right)$ and that $p_i = \frac{1}{n}$, we find
\begin{equation}
S_{micro} = \sum_i p_i \log(n) = - \sum_i p_i \log(p_i) = -\tr( \rho \log ( \rho ))
\end{equation}
where the last line follows since since $p_i = \frac{1}{n}$ are the eigenvalues of $\rho$ such that
\begin{equation}
\rho = 
\begin{pmatrix}
  \frac{1}{n} & 0 & \cdots & 0 \\
  0 & \frac{1}{n} & \cdots & 0 \\
  \vdots  & \vdots  & \ddots & \vdots  \\
  0 & 0 & \cdots & \frac{1}{n} 
 \end{pmatrix}
\end{equation}
such that that taking the trace equals a sum over $p_i$.

This is exactly the von Neumann entropy. However, it is worth noticing at this point that contrary to thermodynamic entropy which only is defined for equilibrium states, entanglement entropy is defined for any state and we can therefore also consider dynamical changes in entanglement entropy. Something that will become relevant later on.
\\

The von Neumann entropy also quantifies entanglement in agreement with our immediate expectations. To see this, consider the previous example of a bell pair system, but now with variable entanglement
\begin{equation}
\ket{\Psi} = \alpha \ket{\uparrow}_A \ket{\downarrow}_B + \beta \ket{\downarrow}_A \ket{\uparrow}_B
\end{equation}
where $\ket{\uparrow}_A$ is the spin state of subsystem $A$ and $\abs{\alpha}^2 + \abs{\beta}^2 = 1$ due to normalisation.

To compute the entanglement entropy, i.e. the von Neumann entropy, for one of the subsystem $A$, one first has to find the reduced density matrix
\begin{equation}
\begin{split}
\rho_A & = \tr_B (\ket{\Psi} \bra{\Psi})
\\& = \sum_{i=\uparrow, \downarrow} \bra{i}_B \left( \alpha \ket{\uparrow}_A \ket{\downarrow}_B + \beta \ket{\downarrow}_A \ket{\uparrow}_B \right) \left( \alpha^* \bra{\uparrow}_A \bra{\downarrow}_B + \beta^* \bra{\downarrow}_A \bra{\uparrow}_B \right) \ket{i}_B
\\& 
\begin{split} 
=\bra{\uparrow}_B \left( \beta \ket{\downarrow}_A \ket{\uparrow}_B \right) & \left(\beta^* \bra{\downarrow}_A \bra{\uparrow}_B \right) \ket{\uparrow}_B \\& \ + \bra{\downarrow}_B \left( \alpha \ket{\uparrow}_A \ket{\downarrow}_B \right) \left( \alpha^* \bra{\uparrow}_A \bra{\downarrow}_B + \right) \ket{\downarrow}_B
\end{split}
\\& = \abs{\beta}^2 \ket{\downarrow}_A \bra{\downarrow}_A + \abs{\alpha}^2 \ket{\uparrow}_A \bra{\uparrow}_A
\end{split}
\end{equation}

Having found the reduced density matrix, we can find an expression for the entanglement entropy
\begin{equation}
\begin{split}
S_A & = - \tr ( \rho_A \log{\rho_A} )
\\& = - \left( \abs{\beta}^2 \log(\abs{\beta}^2) \right) - \left( \abs{\alpha}^2 \log(\abs{\alpha}^2) \right)
\\& = - \left( (1 - \abs{\alpha}^2) \log(1 - \abs{\alpha}^2) \right) - \left( \abs{\alpha}^2 \log(\abs{\alpha}^2) \right)
\end{split}
\end{equation}
where the last line in obtained from the constraint $\abs{\alpha}^2 + \abs{\beta}^2 = 1$.

\begin{figure}\label{plotSA}
\begin{tikzpicture}
\begin{axis}[
    axis lines = left,
    xlabel = $\abs{\alpha}^2$,
    ylabel = {$S_A$},
]
%Below the red parabola is defined
\addplot [
    domain=0:1, 
    samples=50, 
    color=red,
]
{-(1-x)*ln(1-x)-x*ln(x)};
\end{axis}
\end{tikzpicture}
\caption{Plot of the entropy as a function of $\abs{\alpha}^2$. As expected the entanglement entropy is maximal for $\abs{\alpha}^2=\abs{\beta}^2=0.5$ and the entanglement entropy vanishes for $\abs{\alpha}^2=1$ and $\abs{\alpha}^2=0$ which corresponds to pure states of the subsystem $A$.}
\end{figure}
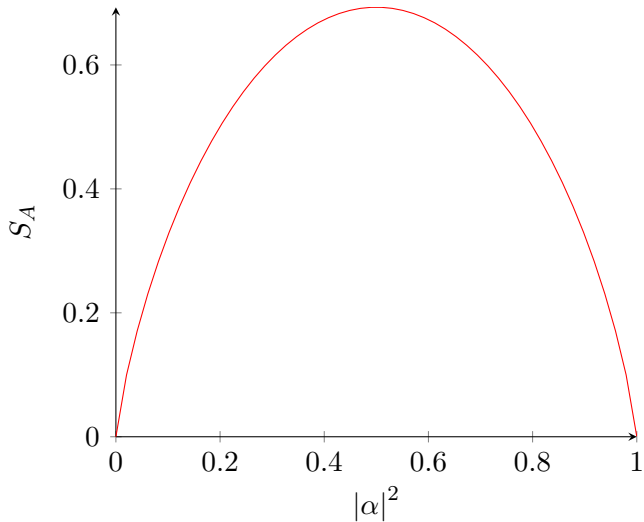

As seen from figure \ref{plotSA} the entanglement entropy is maximal for $\abs{\alpha}^2=\abs{\beta}^2=0.5$. The entanglement entropy vanishes for $\abs{\alpha}^2=1$ and $\abs{\alpha}^2=0$ which corresponds to pure states of the subsystem $A$. This is exactly what is expected for a Bell pair state.

In general we find that if $S_A = 0$ then $\ket{\psi_A}$ is a pure state. If $S_A \neq 0$ then $\ket{\psi_A}$ is in a mixed state; the degrees of freedom in $A$ are entangled with the degrees of freedom in in the rest of the full system.

\section{Entanglement in CFTs}
\label{Entanglement in CFTs}
Going to quantum field theories, some subtleties arise. A quantum field theory is defined on a $d$-dimensional Lorentzian spacetime, $\partial M$, thus to define a state of the system one must first choose a Cauchy (spacelike) slice, $ \Sigma_{\partial M}$, of the spacetime that defines a moment of simultaneity.\footnote{The notation $\partial M$ and $\Sigma_{\partial M}$ is chosen for comply with later notation.} With respect to $ \Sigma_{\partial M}$ we may define the Hilbert space of states $\mathcal{H}_{\Sigma_{\partial M}}$. For any state of the system, $\ket{\Psi}$, we have $\ket{\Psi} \in \mathcal{H}_{\Sigma_{\partial M}}$.

We may conceive of $\mathcal{H}_{\Sigma_{\partial M}}$ as a tensor product of the Hilbert spaces associated with each point and therefore as comprising of all the local degrees freedom in the quantum field theory. This construction is perhaps better understood as the continuum limit of a lattice system on $\Sigma_{\partial M}$ with some lattice spacing $\epsilon$. The degrees of freedom are located at the lattice sites and associated with the Hilbert space $\mathcal{H}_i$, where $i$ is some numbering of the lattice sites. The full Hilbert space is then given as a tensor product of the Hilbert space at each lattice site
\begin{equation}
\mathcal{H} = \otimes_i \mathcal{H}_i.
\end{equation}
The continuum limit is then obtained by sending $\epsilon \rightarrow 0$ and we can conceive of $\mathcal{H}_{\Sigma_{\partial M}}$ as constructed in this way. 

We want to interrogate about the entanglement entropy of some subregion $B \subset \Sigma_{\partial M}$, i.e. about some quantitative measure of the entanglement between $B$ and its compliment $\overline{B}$. Here, the compliment of $B$ is defined to satisfy $B \cup \overline{B} = \Sigma_{\partial M}$. The entangling surface separating $B$ and $\overline{B}$, we denote $\partial B$. Since a CFT is a local quantum field theory, there are specific degrees of freedom associated with specific spatial regions; again the analogy to a continuum limit of a lattice system is helpful. We can therefore regard the full quantum system as composed of two subsystems, $Q_B$ and $Q_{\overline{B}}$, associated with two spatially separated regions $B$ and $\overline{B}$. As a consequence we can decompose the Hilbert space of states of the full system as we did in the section above.\footnote{Some complications are involved in making such a decomposition in a gauge invariant way, but these will not be considered here.}

Having made this construction, one may obtain an expression for the density matrix operator $\rho_B$. To evaluate the entanglement entropy one also needs to obtain an expression for the operator $\log(\rho_B)$. However, taking the logarithm of a continuum operator -- such as $\rho_B$ in the context of a quantum field theory -- proves to be technically complicated. In fact, it proves to be more tractable to evaluate the Rényi entropy
\begin{equation}
S^{(q)}_B = \frac{1}{1-q} \log(\tr( \rho_{B}^q ))
\end{equation}
and then take the limit
\begin{equation}
\lim_{q \rightarrow 1} S^{(q)}_B = S_B.
\end{equation}
The Rényi entropy may be evaluated using the so-called replica trick. It is beyond the scope of the current project to go into details of this procedure. The procedure is reviewed in \citep{calabrese_entanglement_2004}, where it is also employed to obtain the entanglement entropy for 2-dimensional conformal field theories. These prove to be a special case, since they possess a Virasoro algebra with central charge $c$. This simplifies the evaluation of entanglement entropy using the replica trick. Indeed, the evaluation of entanglement entropy proves to be impossible in all but the simplest higher dimensional cases.

In 2-dimensional conformal field theories, any spatial slice takes the form of a line. Define $B$ to be the interval $-a < x < a$ centred around the origin on. As shown in \citep{calabrese_entanglement_2004}, the entanglement entropy for such a single  interval is found to diverge, however, this is as expected since we, in a continuum CFT, will have unregulated short distance entanglement over the entangling surface $\partial B$. To regulate this, we introduce the cutoff $\epsilon$. With this cutoff, the entanglement entropy of a single interval centred around the origin is found to be
\begin{equation}
S_B = \frac{c}{3} \log \left( \frac{2a}{\epsilon} \right).
\end{equation}
The simple expression for the entanglement entropy is a consequence of the Virasoro algebra, which is not realized in higher dimensional CFTs. In chapter \ref{The Ryu-Takayanagi formula}, we will compare this result to an expression for the entanglement entropy of a single interval in a 2-dimensional conformal field theories as obtained by bulk-calculation, i.e. to an expression for $S_B$ obtained via the AdS side of the AdS/CFT correspondence. It is this correspondence -- also known as the gauge/gravity duality -- that we now turn to.

\chapter{A Brief Introduction to Gauge/Gravity Duality}\label{A brief introduction to gauge/gravity duality}
Ever since its conception two decades ago \citep{maldacena_large-n_1999}, the gauge/gravity duality has intrigued researchers in many branches of physics. The conjecture is a that certain local field theories on a fixed spacetime background -- i.e. without gravity -- are dynamically equivalent to certain theories that include a gravitational interaction; they describe the same physics.

The most well known example of this duality is the one between conformal field theories\footnote{Conformal field theories are quantum field theories that are conformally invariant i.e. invariant under dillations, $x^{\mu} \rightarrow \lambda x^{\mu}$ where $\lambda$ is a constant , and special conformal transformation, $x^{\mu} \rightarrow \frac{x^{\mu} - a^{\mu} x^2}{1-2a}$.} in $d$ dimensions ($CFT_d$) and anti-de Sitter spacetimes in $d+1$ dimensions ($AdS_{d+1}$), where the $CFT_d$ is defined on a fixed spacetime that is identical to the $d$-dimensional asymptotic boundary of $AdS_{d+1}$. For this reason, the gauge/gravity duality is also known as the AdS/CFT correspondence.

Initially, the duality may seem unlikely since one would generally expect the number of degrees of freedom in $d+1$ dimensions to be larger than the number of degrees of freedom in $d$ dimensions. Thus, if the duality obtains, it suggests that the degrees of freedom in theories of gravity behave differently from those in local quantum field theories.

The first piece of evidence suggesting this to be the case is the holographic principle known from black hole thermodynamics. From statitical mechanics we know that entropy can be regarded as a measure for the degrees of freedom of a system. As shown in \citep{bekenstein_black_1973}, for general relativity to be consistent with the laws of thermodynamics, the entropy of a black hole has to scale with the area of its event horizon, $A_{BH}$
\begin{equation}
S_{BH} = \frac{A_{BH}}{4 G_N}
\end{equation}
Indeed, this can be shown to be the upper bound of the entropy in $d+1$ dimensional Einstein gravity (the Bekenstein Bound) for a spatial volume bounded by the spatial surface with area $A$. Thus, in general relativity the entropy $S_{GR}$ for the d-dimensional spatial slice $\Sigma_{M_{d+1}}$ of the manifold $M_{d+1}$ scales as
\begin{equation}
S_{GR} \propto \text{Area}(\Sigma_{M_{d+1}}) \propto \text{Vol}(\Sigma_{\partial M_{d+1}})
\end{equation}
where $\Sigma_{\partial M_{d+1}}$ is the $d-1$ dimensional boundary of $\Sigma_{M_{d+1}}$. Thus, gravity realises the holographic principle that the degrees of freedom in a volume scales with the surface area of that volume.

For a $d$-dimensional local quantum field theory, the entropy of the $d-1$ dimensional spatial slice of a system scales with the volume of that spatial slice. Following the duality, a CFT state with the spacetime dual $M_{d+1}$ is defined on a spacetime identical to the asymptotic boundary of $M_{d+1}$, $\partial M_{d+1}$. Thus, the entropy, $S_{CFT}$ scales as
\begin{equation}
S_{CFT} \propto \text{Vol}(\Sigma_{\partial M_{d+1}})
\end{equation}
where $\Sigma_{\partial M_{d+1}}$ is a spacelike co-dimension two surface. This is in agreement with the result obtained for a gravitational theory under the assumption of the holographic principle. This, of course, is no proof of the conjectured duality between $CFT_d$ and $AdS_{d+1}$, however, it suggests that it is not absurd that a local quantum field theory in $d$ dimensions can have the same number of degrees of freedom as a gravitational theory in $d+1$ dimensions. 

\section{$\mathcal{N}=4$ Super Yang-Mills theory and type IIB string theory on $AdS_5 \times S^5$}
Originally, the gauge/gravity duality was conceived by comparing an open string and a closed string perspective of a stack of $N$ D3-branes \citep{maldacena_large-n_1999}. This lead to the conjecture that $\mathcal{N}=4$ Super Yang-Mills theory (SYM) in four dimensional Minkowski spacetime is dual to type IIB string theory on the poincaré patch of the spacetime $AdS_5 \times S^5$.\footnote{For global AdS, the CFT dual is defined on $\mathbf{R} \times S^3$.} $\mathcal{N}=4$ SYM is a conformally invariant quantum field theory with gauge group $SU(N)$ and is characterised by the two free parameters $N$ -- the rank of the gauge group -- and Yang-Mills coupling constant, $g_{YM}$. This is referred to as the `CFT side' of the duality. The string theory on $AdS_5 \times S^5$ is referred to as the `AdS side' and is characterised by the string length $l_s = \sqrt{\alpha'}$, the string coupling constant $g_s$, and the curvature radius of $AdS_5$, $L$, which is also the curvature radius of $S^5$. As a free parameter, the latter two only occur as the ratio $l_s/L$ and are related -- according to the AdS/CFT correspondence -- to the free parameters on the CFT side in the following way
\begin{equation}
2 g^2_{YM} N = 2 \lambda = \frac{L^4}{l_s^4}
\end{equation}
where $\lambda = g^2_{YM} N$ is known as the 't Hooft coupling. The string coupling, $g_s$, and the Yang-Mills coupling, $g_{YM}$, are related by
\begin{equation}
g^2_{YM} = 2 \pi g_s.
\end{equation}

Thus stated, the conjectured duality is one between a four and a ten dimensional theory. The holographic principle asserts that the number of degrees of freedom in a $d+1$ dimensional gravity theory may be the same as the number of degrees of freedom in a $d$ dimensional gauge theory. This, however, still leaves five dimensions unaccounted for in the duality between four dimensional $\mathcal{N}=4$ SYM and type IIB string theory on $AdS_5 \times S^5$. To explain how the number of degrees of freedom can match serves as a crude consistency check of the duality. 

The details of the duality is such that the $CFT_4$ is defined on a spacetime identical to the asymptotic boundary of the dual spacetime; in this case the flat asymptotic boundary of the Poincaré patch of $AdS_5$. Therefore, the five dimensional version of the AdS side can be obtained by a Kaluza-Klein \citep{schwarz_covariant_1983} reduction of the string theory on $S^5$. More precisely, $S^5$ can be expanded in a complete set of spherical harmonics such that any field on $AdS_5 \times S^5$ can be reduced to a tower of massive fields on $AdS_5$, where the masses are due to the origin of these fields in the expansion in spherical harmonics of $S^5$. If the holographic principle holds for this reduced theory in five dimensions, then it is consistent with it having the same number of degrees of freedom as a four dimensional gauge theory like $\mathcal{N}=4$ Super Yang-Mills theory.

Again, this is no proof of the duality. Indeed, it turns out to be very hard to prove the duality for generic $g_{YM}$ and $\lambda$ (and thereby generic $g_s$ and $L/l_s$). A proof would require a full non-perturbative definition of quantized type IIB string theory on curved backgrounds, which we currently lack. This restricts most explicit calculations on the AdS side to the regime where string perturbation theory can be employed; the regime where $g_s \ll 1$. If $l_s/L$ is kept fixed, this reduces the AdS side to classical string theory in the sense that only tree level are taken into account, while the higher string genus expansions are disregarded. On the CFT side, this is equivalent to the limit where $g_{YM} \ll 1$ while $\lambda = g_{YM}^2 N$ is kept fixed. Thus, $N \rightarrow \infty$. This is the well known 't Hooft limit, where only planar diagrams in the gauge theory contribute to leading order in $1/N$. Generally, for fixed $\lambda$
\begin{equation}\label{gsl/N}
\lambda = g_{YM}^2 N \Rightarrow \lambda = 2 \pi g_s N \Rightarrow g_s \propto \frac{1}{N}
\end{equation}
Thus, an expansion in $1/N$ on the CFT side is equivalent to the string theory genus expansion in $g_s$. 

If $l_s/L \rightarrow 0$, the AdS side further simplifies. In this limit, the string length, $l_s$, is much smaller than the curvature radius, $L$ and therefore the strings may be approximated by point particles in the form of type IIB supergravity. This supergravity is defined on a weakly curved $AdS_5 \times S^5$ (since $l_s \ll L$) and is dual to $\mathcal{N}=4$ SYM in its strong coupling limit, i.e. $\lambda \gg 1$. This limit is referred to as the weak form of the duality as opposed to the strong form, where $l_s/L$ and therefore $\lambda$ are kept fixed.

Even though a proof of the duality is not currently possible, a number of non-trivial tests and checks have been carried out. These have their outset in either the weak or the strong form of the duality and evaluates the same quantities on both sides; hitherto with perfect agreement. For instance, the symmetries of $\mathcal{N}=4$ SYM form the supergroup $PSU(2,2|4)$ and so does the type IIB string theory on $AdS_5 \times S^5$. Also, scattering amplitudes which are protected by supersymmetry as one goes from weak to strong coupling can be compared between
the two sides and shown to be equal.

Assuming the weak form of the duality, tests are complicated by it being a weak/strong duality. This entails that one of the sides must be evaluated at strong coupling, which cannot be done at full generality on either side. However, observables independent of the coupling may still be compared and can be shown to be equal. Particularly, the classical fields of type IIB supergravity on $AdS_5 \times S^5$ maps to a subset of the operators in $\mathcal{N}=4$ SYM.\footnote{See \citep{howe_complete_1984} for a complete summary of the field content of $\mathcal{N}=2$ supergravity in 10 dimensions.}

Generally, there should be a one-to-one map between operators on the CFT side and fields on the AdS side which entails that the duality may be formulated as a correspondence between the generating functional for $\mathcal{N}=4$ SYM and partition function of type IIB string theory. The generating functional for correlation functions of gauge theory operators, $\mathcal{O}(x)$, is obtained by perturbing the Lagrangian, $\mathcal{L}$, by a source field, $\phi_{(0)}(x)$
\begin{equation}
\mathcal{L} \rightarrow \mathcal{L} + \phi_{(0)}(x) \mathcal{O}(x).
\end{equation}
The generating functional in Euclidean signature on the CFT side, $Z[\phi_{(0)}]_{CFT}$, thereby takes the form
\begin{equation}
Z[\phi_{(0)}]_{CFT} = \mathtt{N} \int \mathcal{D} \mathcal{O} e^{- \int d^d x (\mathcal{L} + \phi_{(0)}(x) \mathcal{O}(x)} \equiv \expval{ \exp \left( \int d^d x  \mathcal{O}(x) \phi_{0}(x) \right) }
\end{equation}
where $\mathtt{N}$ is some normalization. Assume that $\phi_{(0)}(x)$ is the boundary value of the fluctuating bulk field $\phi(z,0)$ such that
\begin{equation}
\lim_{z\rightarrow 0} z^{\Delta - d} \phi(x,z) = \phi_0 (x)
\end{equation}
where $\Delta$ is the dimension of $\mathcal{O}(x)$.\footnote{Observe that $\phi_0 (x) \neq \phi(z=0,x)$. Generally, $\phi(z=0,x)$ is divergent.} One then finds that
\begin{equation}\label{ZCFT=Zstring}
\expval{ \exp\left( \int d^d x \mathcal{O}(x) \phi_{0}(x) \right)} = Z_{string} \big\vert_{ \lim\limits_{z \rightarrow 0} z^{\Delta - d} \phi(x,z) = \phi_0 (x)}
\end{equation}
where $Z_{string} \big\vert_{ \lim\limits_{z \rightarrow 0} z^{\Delta - d} \phi(x,z) = \phi_0 (x)}$ is the classical string theory partition function in Euclidean signature that takes the path integral over all field configurations for $\phi(z,x)$ which takes the value $\phi_0 (x)$ at the asymptotic boundary of AdS. 

One complication is that $Z_{string}$ is not explicitly known. Therefore, it is often necessary to consider only the limit where the string length is much smaller than the curvature radius, i.e. the weak form introduced above as opposed to the strong form in (\ref{ZCFT=Zstring}). In this limit, the leading contribution to $Z_{string}$ comes from the field $\tilde{\phi}(x,z)$ that solves the supergravity equations of motion subject to the condition
\begin{equation}
\lim_{z\rightarrow 0} z^{\Delta - d} \tilde{\phi}(x,z) = \phi_0 (x).
\end{equation}
One thereby obtains the relation
\begin{equation}\label{ZCFT=Zsugra}
\expval{ \exp\left( \int d^d x \mathcal{O}(x) \phi_{0}(x) \right)} = e^{-\mathcal{S}_{sugra}} \big\vert_{ \lim\limits_{z \rightarrow 0} z^{\Delta - d} \tilde{\phi}(x,z) = \phi_0 (x)}
\end{equation}
where $\mathcal{S}_{SUGRA} \big\vert_{ \lim\limits_{z \rightarrow 0} z^{\Delta - d} \tilde{\phi}(x,z) = \phi_0 (x)}$ is the (renormalized) supergravity action.

If the source field of an operator in the CFT is known, then the relation (\ref{ZCFT=Zstring}) tells how operators (and their correlators) on the CFT side relate to fields (and field correlators) on the AdS side.

%How to use it: Page 197 GGD

\section{Gauge/gravity duality: The weak form}\label{The weak form}
As the subsequent chapters will deal with the weak form of the gauge/gravity duality, we will develop that in more detail here. To obtain the weak form, two limits were taken on the AdS side: $g_s \rightarrow 0$ and $l_s/L \rightarrow 0$. On the CFT side, this corresponded to the limit $N \rightarrow \infty$ and large $\lambda$, where $N$ is the rank of the gauge group of the theory and $\lambda$ is the t'Hooft coupling. These limits were taken to avoid having quantum gravity and stringy corrections on the AdS side. This allows the approximation of the AdS side with the Einstein Hilbert action in AdS spacetime
\begin{equation}\label{SEH}
S_{EH} = \int d^{d+1} x \sqrt{g} \left[ \frac{1}{16 \pi G_N^{d+1}} \left( R + \frac{d(d-1)}{L^2} \right) + \mathcal{L}_M \right]
\end{equation}
where $- \frac{d(d-1)}{L^2}$ is the cosmological constant in AdS spacetime and $\mathcal{L}_M$ are matter terms. With this approximation, the gauge/gravity duality can serve to relate general relativity in $d+1$ dimensions to a $d$-dimensional conformal field theory. To motivate the approximation of the AdS side by general relativity, we will first argue how supergravity obtains as an approximation on the AdS side and then show how the Einstein Hilbert action is a special case of supergravity action. 

Supergravity is an effective low-energy limit of superstring theory where only tree level interactions are included. Since energy -- in units $c=\hbar=1$ -- scales like inverse length, this is the limit where the energy is much smaller than the inverse of characteristic length scale of the theory, which in the case of string theory is the string length, $l_s$.  Thus, the low energy limit is equivalent to the limit where $l_s \rightarrow 0$. In supergravity, therefore, strings are approximated by point particles; stringy correction are ignored. This entails that only the massless closed string particles are included in the mode content of supergravity. To see this, remember that the mass spectrum of the particles in the closed string theory in Minkowski background is given by
\begin{equation}
M^2 = \frac{4 ( \mathbf{N} - 1)}{l_s^2}.
\end{equation}
where $\mathbf{N} = \sum_{n=1}^\infty \alpha_{-n} \alpha_n$ and $\alpha_{-n}$ and $\alpha_n$ are the annihilation and creation operators, respectively, of the closed string mode expansion. Thus, for Minkowski background, the massless modes -- for which $\mathbf{N}=1$ -- are independent of the string length, $l_s$, whereas the massive modes scale as $l_s^{-2}$. When $l_s \rightarrow 0$ the massive modes decouple such that only the massless modes contribute. Consequently, the low energy limit, and therefore supergravity on Minkowski background, only contains the massless modes of the closed superstring. The argument generalizes to closed string theory of $AdS$ background (for more details, see \citep{berenstein_strings_2002}).

Among the field content of the closed string is a graviton, which we with some foresight will denote $g_{A B}(X)$,\footnote{The $\mathbf{35}$ representation of $SO(8)$} and a dilaton, $\Phi(X)$.\footnote{The $\mathbf{1}$ representation of $SO(8)$.} Here, $X^A$ is the embedding function from the string world sheet to the target space. The graviton of course is interesting since it perturbs the background spacetime in which the string moves, and should therefore relate to the gravitational dynamics. However, also the fluctuations of the dilaton, it turns out, contribute to the metric perturbation as seen from the point of view of general relativity. To indicate how this obtains as well as how to derive the supergravity action, it is instructive to consider the graviton and dilaton parts of the string action. For convenience, we will work with the action for the closed bosonic string,\footnote{More presicely, the bosonic string action that only includes the dilaton and the graviton may be regarded as the low energy limit of the bosonic string (including only the massless fields) where the last massless field, the Kalb-Ramond field $B_{A B}$, is assumed to be zero.} however, the procedure generalizes to the full closed superstring action.

We will take $\alpha, \beta = 1,2$ to be worldsheet indexes and $A, B = 0,1,...D-1$ to be target space indexes, where $D$ is the dimension of the target space. We will further assume that the metric induced on the worldsheet by the background metric is $\gamma_{\alpha \beta}$, i.e. the background is not necessarily flat.\footnote{Since spacetime curvature in string theory originates from the closed string graviton, a curved background should be conceived off as consisting by a large enough number of closed strings such that spacetime curves even in the classical limit.} The world sheet can be parametrized by the timelike coordinate $\xi^0$ and spacelike coordinate $\xi^1$. Since we will work in Euclidean signature, it is convenient to define the coordinate $\xi^2 = i \xi^0$ such that $\xi^2$ is the Wick rotation of the worldsheet time parameter. As usual, $\xi^1$ is the parameter for string extension; for the closed string $0 \leq \xi^1 \leq 2\pi$. The Euclidean, bosonic string action on the world sheet for the graviton and dilaton then takes the form
\begin{equation}\label{SGPhi}
S_{g \Phi} = \frac{1}{4 \pi} \int d^2 \xi \sqrt{\gamma} \left[ \frac{1}{l_s^2} \gamma^{\alpha \beta} g_{A B}(X) \partial_\alpha X^A \partial_\beta X^B + R^{(2)} \Phi(X) \right]
\end{equation}
where $R^{(2)}$ is the 2-dimensional Ricci scalar on the world sheet. 
%As seen, this is the Polyakov action in general spacetime background plus a term associated with the dilaton field. 

The dilaton field may be decomposed into a constant part, $\expval{\Phi}$ (the vacuum expectation value of the dilation field) and fluctuations, $\tilde{\Phi}$, around $\expval{\Phi}$
\begin{equation}
\tilde{\Phi} = \Phi - \expval{\Phi}.
\end{equation}
We then find
\begin{equation}
\frac{1}{4 \pi} \int d^2 \xi \sqrt{\gamma} R^{(2)} \Phi(X) = \frac{1}{4 \pi} \int d^2 \xi \sqrt{\gamma} R^{(2)} \tilde{\Phi}(X) +  \chi \, \expval{\Phi}.
\end{equation}
where $\chi = \frac{1}{4 \pi} \int d^2 \xi \sqrt{\gamma} R^{(2)}$ such that $\chi$ is identical to the Euler characteristic of the world sheet. As a consequence, it only depends on the topology of the world sheet
\begin{equation}
\chi = -2 (q - 1)
\end{equation}
where $q$ is the genus of the world sheet. One finds that $\chi = 2$ for a sphere ($q=0$), $\chi = 0$ for a torus ($q=1$) and that $\chi$ is negative for higher genus, $q \geq 2$, surfaces. Separating the constant and fluctuating parts of the dilaton field, the action thereby takes the form
\begin{equation}
S_{g \Phi} = S_{g \tilde{\Phi}} + \expval{\Phi} \chi
\end{equation}
where the last term only depends on the world sheet topology.

The Euclidean generating functional associated with $S_{g \Phi}$ takes the form as a path integral over world sheet metrics, $\gamma_{\alpha \beta}$, of manifolds, $\mathcal{M}$, and embedding functions, $X^A$
\begin{equation}
Z_{g \Phi} = \int_{\mathcal{M}} \mathcal{D} \gamma_{\alpha \beta} \int \mathcal{D} X^{A} e^{-S_{g \Phi}}.
\end{equation}
Since $\expval{\Phi} \chi$ only depends on the topology, one may rewrite the generating functional as
\begin{equation}
\begin{split}
Z_{g \Phi} & =  \sum_{q=0}^{\infty} \left( e^{- \expval{\Phi}} \right)^{- 2 (q - 1)} \int_{\mathcal{M}^q} \mathcal{D} \gamma_{\alpha \beta} \int \mathcal{D} X^{A} e^{-S_{g \tilde{\Phi}}} \\& = \frac{1}{g_s^2} \sum_{q=0}^{\infty} g_s^{2 g} \int_{\mathcal{M}^q} \mathcal{D} \gamma_{\alpha \beta} \int \mathcal{D} X^{A} e^{-S_{g \tilde{\Phi}}}
\end{split}
\end{equation}
where $g_s \equiv e^{\expval{\Phi}}$; the constant part of the dilaton field serves as the string coupling constant. For $g_s \ll 1$, the first term in the sum dominates. In this limit, therefore, one only has to consider the path integral over world sheets that are topological spheres. This gives rise to the tree level diagrams known as Witten diagrams. In subsequent chapters, we will suppose this approximation to be valid.\footnote{In fact the effective coupling is $e^{\Phi(X)}$ which means that it must be small for all $X$ in order for perturbation theory to be valid.}

It proves to be convenient initially to keep $\Phi$ and only later decompose it in $\expval{\Phi}$ and $\tilde{\Phi}$. An effective Lagrangian obtained from perturbation theory in $\Phi$ will take the form
\begin{equation}\label{Leff}
\mathcal{L}_{Eff} = \sum_{q=0}^{\infty} \order{e^{-2(q-1) \Phi}}
\end{equation}
where terms $\propto e^{-2 \Phi}$ are denoted as tree level, and higher order terms are string loop corrections.\footnote{These should not be confused with loop corrections due to an expansion in the string length, $l_s$.} 

For general backgrounds, an observable calculated using $Z_{g \Phi}$ will diverge even at tree level. To regulate this divergence, we must introduce a cut-off, $\mu$, and the background fields will generally depend on this cut-off. To keep track of this dependence we introduce the functionals
\begin{equation}
\beta^{g}_{A B} = \mu \frac{\partial g_{AB}}{\partial \mu}, \; \; \beta^{\Phi} = \mu \frac{\partial \Phi}{\partial \mu}.
\end{equation}
An obvious problem with the background fields being dependent on the cut-off is that this threatens to break conformal invariance. Just like gauge invariance is a defining symmetry of Yang-Mills theory as a quantum theory, so is conformal invariance in string theory. 

A good indicator as to whether conformal invariance is broken is to look at the expectation value of the trace of the energy momentum tensor, $\expval{T^{\alpha}_{\alpha}}$, that must vanish. At each order of the expansion in $g_s$, the trace may be expressed in terms of the beta-functions. Not going into the details, one finds for the regulated action at tree level\footnote{For higher order string loop corrections see \citep{callan_string_1987}.}
\begin{equation}
\expval{T^{\alpha}_{\alpha}}_{Tree} = \frac{\beta^{\Phi}}{12} R^{(2)} + \frac{1}{2 l_s^2} \beta^{g}_{AB} g^{\alpha \beta} \partial_\alpha X^A \partial_\beta X^B.
\end{equation}
For this trace to vanish, we see that
\begin{equation}\label{beta=0}
\beta^{g}_{AB} = \beta^{\Phi} = 0
\end{equation}
i.e. neither of the background fields can depend on the regulator, $\mu$. This proves to impose strict constraints on these background fields.

Even if string loop corrections are disregarded, $S_{g \Phi}$ is a non-linear two-dimensional quantum field theory action, it can generally only be studied in perturbation theory. The true dimensionless coupling for the action is $l_s/L$, where $L$ is the characteristic length of the background field $g_{A B}$; for instance the AdS curvature radius in $AdS$ background. As seen, only the dimensionfull coupling, $l_s^2$, occurs explicitly in the action, while $L$ is hidden in $g_{A B}$. We may therefore expand $S_{G \Phi}$ in powers of $l_s^2$ just remembering that perturbation theory is only valid for $l_s/L \ll 1$ and not just $l_s \ll 1$. 

Using this expansion, one finds
\begin{equation}
\begin{split}
\beta^{\Phi} & = D-26 + \frac{3}{2}l_s^2 \left[ 4 (\nabla \Phi)^2 - 4 \nabla_A \nabla^{A} \Phi - R^{(D)} \right] + \order{l_s^4} + \order{e^{2 \Phi}} \\
\beta^{g}_{A B} & = l_s^2 \left( R^{(D)}_{A B} + 2 \nabla_A \nabla_B \Phi \right) + \order{l_s^4} + \order{e^{2 \Phi}}
\end{split}
\end{equation}
where we note that we are now considering the $D$-dimensional Ricci scalar and tensor. For $l_s/L \ll 1$ the terms $\order{l_s^4} \rightarrow 0$ and for $g_s \ll 1$ the string loop order corrections vanish. Imposing the constraint $\beta^i = 0$ (\ref{beta=0}), one finds in this limit (for $D=26$) the following constraints in the background fields
\begin{equation}\label{beta=0}
\begin{split}
0 & = 4 (\nabla \Phi)^2 - 4 \nabla_A \nabla^{A} \Phi - R^{(26)} \\
0 & = R^{(26)}_{A B} + 2 \nabla_A \nabla_B \Phi.
\end{split}
\end{equation}
Notably, for constant $\Phi$ this reduces to
\begin{equation}\label{0=R}
0 = R^{(D)}_{A B}.
\end{equation}
Thus as desired, in the limit $l_s/L \ll 1$ (corresponding to the low energy limit) which we consider here one obtains the Einstein equations without matter (\ref{0=R}), when all other fields vanish (expect the dilaton field that can be a non-zero constant). 

While the beta functions was obtained as the renormalization group flow for the regulated world sheet action, they can also be regarded as equations of motion for the background fields that the string propagates through. We may therefore change the perspective and search instead for a target space (and therefore 26-dimensional) action that has (\ref{beta=0}) as its equations of motion. 
%Thus, we must find an action which satisfies:
%\begin{equation}
%\begin{split}
%\frac{\delta}{ \delta \Phi} S_{Tree} = - \sqrt{g} e^{-2 \phi} 
%\end{split}
%\end{equation}
Without going into the details, this action is
\begin{equation}
S_{Tree} = \frac{1}{2 \kappa_0^2} \int d^{26} X \sqrt{-g} e^{-2 \Phi} \left( R^{(26)} + 4 \partial_B \Phi \partial^B \Phi \right)
\end{equation}
where $\kappa_0$ is some constant. As expected from  according to (\ref{Leff}), we see an overall factor $e^{-2 \Phi}$ which is consistent with the fact that we only included leading order in $e^{2 \Phi}$ in the beta-functions, i.e. no string loop corrections. 

$S_{Tree}$ is the low-energy effective action for the target space in the presence of a closed bosonic string. It is the low-energy action since (\ref{beta=0}) is valid only for in the limit $l_s \ll 1$ where higher order corrections vanishes. Loop corrections to this tree level action will be additional terms $\order{l_s^2}$.

Comparing $S_{Tree}$ to the Einstein Hilbert action there are some similarities, but also two important differences. First is the presence of the factor $e^{-2 \Phi}$ and second is the positive sign of the kinetic term $4 \partial_B \Phi \partial^B \Phi$. These, however, can be accommodated by combining the metric field, $g_{A B}$ with the fluctuating part of the dilaton field, $\tilde{\Phi}$
\begin{equation}
\tilde{g}_{A B} = e^{-4 \tilde{\Phi}/(D-2)} g_{A B}.
\end{equation}
Associating this new metric with Ricci scalar, $\tilde{R}$ and taking $D=26$ we can then rewrite the metric as
\begin{equation}
S_{Tree}^E = \frac{1}{2 \kappa_0^2 \left( e^{\expval{\Phi}} \right)^2} \int d^{26} X \sqrt{-\tilde{g}} \left( \tilde{R}^{(26)} - \frac{1}{6} \partial_B \tilde{\Phi} \partial^B \tilde{\Phi} \right).
\end{equation}
This gives the same form as the Einstein Hilbert action in the presence of a field with kinetic term $\frac{1}{6} \partial_B \tilde{\Phi} \partial^B \tilde{\Phi}$. Expressing the action like this is known as the Einstein perspective as opposed to the string perspective in $S_{Tree}$. The constant factor can then be determined by comparison to the Einstein Hilbert action
\begin{equation}
2 \kappa_0^2 \left( e^{\expval{\Phi}} \right)^2 = 16 \pi G_N^{(26)}
\end{equation}
where $G_N^{(26)}$ is Newton's constant in 26 dimensions. Identifying, as above, $g_s \equiv e^{\expval{\Phi}}$, one obtains
\begin{equation}\label{gsGN}
\kappa_0^2 g_s^2 = 8 \pi G_N^{(26)} \Rightarrow \kappa^2 = 8 \pi G_N^{(26)}
\end{equation}
where $\kappa^2 = \kappa_0^2 g_s^2$. Interestingly, $G_N^{(26)} \propto g_s^2$, which is the parameter of the genus expansion of the generating functional for the closed string, $Z_{g \Phi}$. Defining the reduced Planck length in 26 dimensions $l_p = 8 \pi G_N^{(26)}$ and observing from dimensional analysis that $\kappa^2 \propto l_s^{24} g_s^2$, we find, for the weak coupling regime, that $g_s \ll 1 \Rightarrow l_p \ll l_s$. This support the previous assertion that the regime $g_s \ll 1$ is the regime of classical gravity, if we assume that effects from quantum gravity only become significant when the length scale of the system approaches the Planck length.

Starting instead from the 10 dimensional type IIB string action, we can analogously obtain 10 dimensional type IIB supergravity as the low energy effective action if all string loop corrections are ignored. Since it is tree level in string loop corrections it involves an overall factor $e^{-2 \Phi}$ and includes the same terms derived above -- the Ricci scalar and a kinetic term for the dilaton -- as well as additional fields due to the additional field content of the type IIB superstring action; among them a self dual four form field. None of these, however, lead to terms in the action identifiable as a cosmological constant. Remarkably, the equations of motion for 10 dimensional supergravity still permits solutions where the non-compact spacetime is an anti-de Sitter; a spacetime with constant negative cosmological constant. This obtains, if the metric is $AdS_5 \times S^5$ and the field strength for the four form field adequately relates to the AdS curvature radius. Also, the dilaton field has to be constant, while the dynamics of the remaining supergravity fields decouple.

As asserted above, a Kaluza-Klein reduction of $S^5$ leads to a tower of massive modes in the 5 dimensional compactified theory on $AdS_5$. Thus, AdS spacetime can appear as a solution for higher dimensional supergravity after a Kaluza-Klein reduction of a compact internal space even if the higher dimensional supergravity contains no cosmological constant. A Kaluza-Klein reduction may therefore reduce this higher dimensional supergravity theory to a lower dimensional gravity theory with a cosmological constant. Indeed, supergravity on $AdS_5 \times S^5$ may be compactified by a Kaluza-Klein reduction of $S^5$ to some gauged, 5 dimensional supergravity theory \citep{schwarz_covariant_1983}.\footnote{In gauged supergravity, the gravitinos couple to a gauge field with the consequence that the transformation law for the gravitinos includes a constant term. For the action to remain invariant under this transformation, a constant term -- the cosmological constant -- must be added to the action.} Such gauged supergravity theories does contain a cosmological constant and has $AdS$ spacetime as its (supersymmetric) vacuum \citep{gunaydin_compact_1986}.

The implications of this for the AdS/CFT correspondence is that in the low energy limit we may approximate the AdS side by the $d+1$-dimensional Einstein Hilbert action -- and therefore the Einstein equations -- with cosmological constant $\Lambda = - \frac{d(d-1)}{L^2}$, though we should note that this disregards contributions from the field strength of the four form field that, as argued above, does not decouple. Supposing the validity of this approximation, we can then finally provide holographic picture that will be the outset of the remainder of the thesis. 

%$S_{Tree}^E$ does not contain any cosmological constant. Since out interest is in AdS spacetimes with negative cosmological constant, it is relevant indicating how a cosmological constant might occur in the scheme above. As the name suggests, it should take the form of a constant term in the effective action, i.e. the terms can neither depend on $G_{\mu \nu}$ nor on $\tilde{\Phi}$, while it may depend on $\expval{\Phi}$ which it turns out to do (implicitly through $G_N$) in the Einstein Hilbert action with cosmological constant. That this constant term cannot depend on $\tilde{\Phi}$ is of particular interest, since this naively entails that the term must arise as a first order $(g=1)$ string loop correction, $\order{e^{-2(g-1) \Phi}} = \order{1}$. Indeed, it is shown that such a term does arise for flat background in \citep{fischler_dilaton_1986} and for general backgrounds in \citep{callan_string_1987}. 

%Do we need cosmological constant? If not argue why? $AdS_5 x S^5$ is a solution to supergravity without. What happens once we disregard $S^5$ (we are probably making a Kaluza-Klein reduction). Do we then need an argument to the origin of the negative cosmological constant?

\section{$CFT_d$ and $AdS_{d+1}$}\label{CFT_d and AdS_{d+1}}
The overall assumption is that Einstein gravity on $AdS_{d+1}$ is dual to some large $N$, large $\lambda$ CFT in d dimensions, where $\lambda = 2 g_{YM}^2 N$. %According to the AdS/CFT dictionary
%\begin{equation}
%g_{YM}^2 = 2 \pi g_s, \; \; 2 g_{YM}^2 N = \frac{L^4}{l_s^4}.
%\end{equation} 
%Following the above exposition of the low energy effective theory on the AdS side, we can verify that this regime of the CFT parameters is indeed dual to Einstein Gravity on the AdS side. From the relation between $\lambda$ and the ratio $L/l_s$, we see that large $\lambda \rightarrow l_s/L \ll 1$. The supergravity approximation (and thereby the Einstein Hilbert action) is valid for large $\lambda$. Extending the result obtained for fixed $\lambda$, $g_s \propto 1/N$, and the relation found above, $G_N^{(d+1)} \propto g_s^2$, we see, as desired, that in the large $N$ limit we only have to consider leading order contributions in perturbation theory of the Einstein Hilbert action.

%\textsc{Consider to include remarks about the Einstein equations and how there are two solutions without matter depending on boundary conditions. See Takayanagi (2016) p. 36-37.}

We will consider a one parameter family of CFT states, $\ket{\Psi (\zeta)}$ where $\ket{\Psi (0)}$ represents the CFT vacuum. We assume that $\ket{\Psi (\zeta)}$ lives in a $d$-dimensional Minkowski spacetime, $R^{d-1,1}$, such that $\ket{\Psi (0)}$ is dual to the $(d+1)$-dimensional, Poincaré patch of pure Anti-de Sitter spacetime with metric 
\begin{equation}
ds^2 = \frac{L^2}{z^2}(dz^2+dx_{\mu} dx^{\mu} )
\end{equation}
where $L$ is a length scale that controls the negative curvature of bulk spacetime. The spacetime has a Minkowski boundary for $z \rightarrow 0$ which is in accordance with the general result of the AdS/CFT correspondence that the quantum state $\ket{\Psi}$ with a spacetime dual $M_{\Psi}$ is defined on the boundary of the dual spacetime,  $\partial M_{\Psi}$. Often, we will be interested in arbitrary Cauchy surfaces $\Sigma_{\partial M_{\Psi}} \subset \partial M_{\Psi}$.

Generally, the spacetime dual of excited states $\ket{\Psi}$ can be represented by the metric
\begin{equation}
ds^2 = \frac{L^2}{z^2}(dz^2 + \Gamma_{\mu \nu}(x, z) dx^\mu dx^\nu )
\end{equation}
where $\Gamma_{\mu \nu}$ is a parameter that controls the perturbation, i.e. how much the spacetime deviates from pure AdS. As seen, pure AdS corresponds to a spacetime where $\Gamma_{\mu \nu} = 0$. Thus, $\Gamma_{\mu \nu} \rightarrow 0$ for $\zeta \rightarrow 0$ whereas $\Gamma_{\mu \nu} \neq 0$ when $\zeta \neq 0$. 

For small enough $z$, one finds that $\Gamma_{\mu \nu}(x,z) = \eta_{\mu \nu} + z^d \bar{h}_{\mu \nu}(x, z)$ and we may therefore express the metric as
\begin{equation}\label{FefGra}
ds^2 = \frac{L^2}{z^2}(dz^2 + dx_\mu dx^\mu + z^d \bar{h}_{\mu \nu}(x, z) dx^\mu dx^\nu ).
\end{equation}
This is known as the Fefferman-Graham coordinates. The metric can be used when the gravitational excitations are near the asymptotic boundary, $z \rightarrow 0$. On the CFT side, this corresponds to the small excitations, i.e. to states $\ket{\Psi (\zeta)}$ with sufficiently small $\zeta$. These states are small perturbations of the CFT vacuum and the spacetime dual, $M(\zeta)$, can be represented as perturbations of pure AdS spacetime and will take the form of asymptotically AdS spacetimes described by (\ref{FefGra}). 

Higher excited states will on the other hand be dual to AdS spacetimes with very different geometry that may even have different topology. For such spacetimes, we cannot asumme that $l_s/L \ll 1$ which entails that the Einstein Hilbert action is no longer a valid description of the AdS side, but needs corrections of order $l_s^2$ and higher. This, therefore, takes us beyond the regime of classical (super)gravity. Intuitively, this follows since one has to take into account the effects of the massive string modes in high energy states. 

An example of an excited CFT state with a non-trivial dual classical spacetime is the thermal state of a CFT on $d$-dimensional Minkowski spacetime that is dual to the planar Schwarzschild-AdS black hole in $d+1$ dimensions
\begin{equation}
ds^2 = - f_M(r) dt^2 + \frac{dr^2}{f_M(r)} + \frac{r^2}{L^2} d\Omega_{d-2}^2
\end{equation}
where $f_M(r) = \frac{r^2}{L^2} - \frac{\mu}{r^{d-2}}$. CFT states dual to black hole spacetimes will play an important role in chapter \ref{Entanglement and Spacetime}, however, our interest will be in the eternal black hole in $d+1$ introduced in section \ref{Asymptotically AdS black holes}. The eternal black hole is conjectured to be dual to a thermal double state of two CFTs defined on $\mathbb{R} \times S^{d-1}$ \citep{maldacena_eternal_2003}.\footnote{A number of authors have recently questioned this conjectured duality (see \citep{marolf_eternal_2013,avery_no_2013,mathur_what_2014}). Thus, in the assessment of the conclusions drawn from this duality, it is therefore worth keeping in mind that the duality remains disputed. The general claims about the relation between entanglement and spacetime is not directly disproved if this duality turns out to be false, the intuitive appeal of the qualitative argument will be somewhat damaged.} 

The two identical regions I and III of the eternal black hole (see figure \ref{PMSASBH}) have asymptotic boundaries -- denoted $A$ and $B$ in figure \ref{PMSASBH} -- with spacetime $\mathbb{R} \times S^{d-1}$.\footnote{This follows since the eternal black hole spacetime is asymptotically pure global AdS.} These are also causally disconnected. In accord with the general result of the AdS/CFT correspondence, the CFT dual to the eternal black hole is defined on a spacetime identical to this asymptotic boundary of the AdS spacetime, $A \cup B$.\footnote{Note that for the eternal black hole this is not a contiguous spacetime, but instead a spacetime that consists of two disjoint copies of the same spacetime.} Thus, the full quantum system is comprised of two identical quantum subsystems, $Q_A$ and $Q_B$, that covers the local degrees of freedom of the CFTs living on the spacetimes identical to $A$ and $B$ respectively. Locality of the CFT requires that $Q_A$ and $Q_B$ cannot interact, i.e. their Hamiltonians must be uncoupled. However, they can still be correlated via entanglement.

More precisely, the two subsystems $Q_A$ and $Q_B$ can be associated with two identical Hilbert spaces of states, $\mathcal{H}_A$ and $\mathcal{H}_B$, that may be spanned by an orthogonal basis consisting of (again identical) energy eigenstates $\mathsmaller{ \big\lbrace \ket{E_i^A}\big\rbrace}$ and $\mathsmaller{ \big\lbrace \ket{E_i^B}\big\rbrace}$. The thermofield double state, $\ket{\Psi}$, that is the CFT state dual to the eternal black hole can then be expressed as
\begin{equation}\label{double}
\ket{\Psi}=\frac{1}{\sqrt{Z(\beta)}}\sum_i e^{-\beta E_i / 2} \ket{E_i^A} \otimes \ket{E_i^B}
\end{equation}
where $\beta$ is the inverse temperature of one of the subsystems\footnote{Since they are identical, the inverse temperature is the same in both subsystems.} and $Z = \sum_i e^{- \beta E_i}$. Notably, the expression of the thermofield double state, (\ref{double}), in terms of energy eigenstates explicitly unveils the local degrees of freedom of the two subsystems, $Q_A$ and $Q_B$, to be entangled through the weighted sum over states $\sum_i e^{-\beta E_i / 2} \ket{E_i^A} \otimes \ket{E_i^B}$. This will be of interest later.

%\textsc{How do we see more explicitly that such higher excitations goes beyond the scope of the Einstein Hilbert action? Does higher energy imply stronger 't Hooft-coupling? Not immediately, because conformal invariance ensures vanishing beta functions...}

%\textsc{Are there highly excited states that are well behaved such that they may still be accounted for using the Einstein Hilbert action?}

%\textsc{Does conformal invariance imply that the path integral over all world sheet with a particular genus all give the same result (due to scale invariance)? How else is the result obtained that we only include the on shell fields in the weak form?}

%\textsc{Will observables calculated using $Z_{G \Phi}$ converge if $\Phi$ is constant and $g_s \ll 1$?}

%\textsc{Are the fluctuations $\tilde{\Phi}$ quantum fluctuations such that $\Phi$ is constant in the classical regime and only varies in the quantum theory?}

%\textsc{What about the saddle point approximation? Where does it enter?}

\chapter{The Ryu-Takayanagi Formula}\label{The Ryu-Takayanagi formula}
So far, we have only provided very superficial remarks about the quantitative relation between $d$-dimensional CFT and the $d+1$-dimensional AdS spacetime. The reason for this is that the subsequent study solely relies on a single entry from the AdS/CFT dictionary; the Ryu-Takayanagi formula \citep{ryu_holographic_2006} and its covariant generalization \citep{hubeny_covariant_2007}. This formula relates entanglement entropy on the CFT side to areas of extremal co-dimension two surfaces on the AdS side, thereby relating an intrinsically quantum mechanical phenomena on the CFT side to the geometry of the bulk in the spacetime dual. Using this relation alone, it can be shown that if the first law of entanglement entropy is satisfied in a CFT, then Einstein equations to linear order are satisfied in the spacetime dual.

\section{Relation to holography}
Above, we already mentioned the remarkable result from black hole thermodynamics due to Bekenstein and Hawking that relates the entropy of a black hole to its horizon area
\begin{equation}
S_{BH} = \frac{A_{BH}}{4 G_N}
\end{equation}
This gave rise to the holographic principle, which we used as a consistency check for the gauge/gravity duality. This entails a particular way of conceiving of the Bekenstein Hawking formula as a relation between some entropy of a CFT subsystem and the area of the black hole horizon; that is the area of a particular co-dimension two surface in the spacetime dual to that CFT. Thus, rather than being a statement about black hole thermodynamics, this holographic interpretation of the Bekenstein Hawking formula regards it as a piece of the AdS/CFT dictionary. As a consequence, it is natural to speculate whether the Bekenstein Hawking formula signifies a more general relation between entropies of CFT subsystems and co-dimension two surfaces in the bulk of the spacetime dual of that CFT. 

\begin{figure}\label{cylinder}
\begin{center}
\includegraphics[scale=0.5]{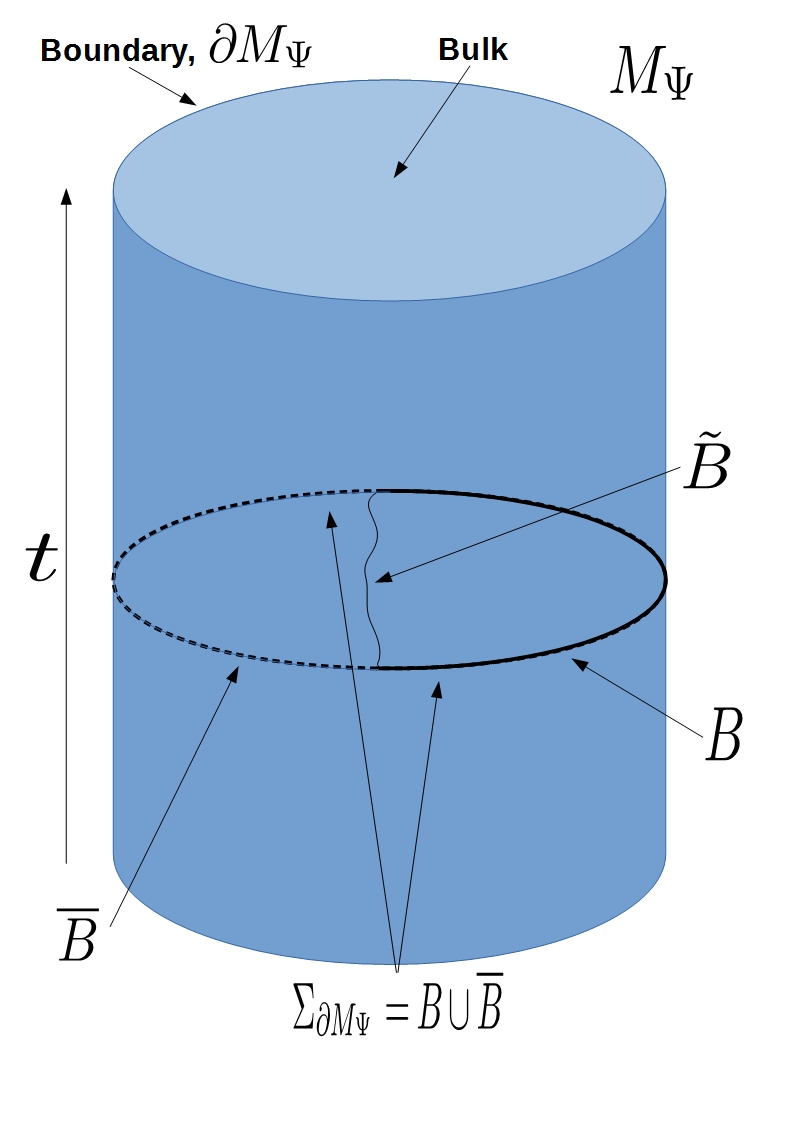}
\caption{For the purpose of illustration we here depict spherical space with time, $S^d \times R$, that is conformally equivalent to global AdS.}
\end{center}
\end{figure}

As it turns out, such a generalization does indeed exist. For spacetimes described by Einstein gravity with a dual CFT state, Ryu and Takayanagi conjectured that the entanglement entropy (defined as in (\ref{EE})) of some spatial region on the CFT side is proportional to the area of the co-dimension two surface in the bulk of the spacetime dual that minimizes the area functional under certain conditions.

To put this more precisely, let us again consider quantum state, $\ket{\Psi}$, with the dual spacetime, $M_\Psi$. In accordance with the gauge/gravity duality, $\ket{\Psi}$ is defined in the asymptotic boundary of $M_\Psi$, $\partial M_\Psi$. Let $\Sigma_{\partial M_\Psi}$ be a Cauchy surface of $\partial M_\Psi$ and let $B$ be a subregion of the Cauchy slice such that $B \cup \overline{B} = \Sigma_{\partial M_\Psi}$, where $\overline{B}$ is the compliment of $B$. The entanglement entropy, $S_B$, of the quantum system living on $B$ is the entropy associated with entanglement between the quantum systems $Q_B$ and $Q_{\overline{B}}$ over entangling surface that coincides with the boundary of $B$, $\partial B$, i.e. the von Neumann entropy of the reduced density matrix
\begin{equation}
\rho_B = \tr_{\overline{B}} \left( \ket{\Psi} \bra{\Psi} \right)
\end{equation} 
where the degrees of freedom in $Q_{\overline{B}}$ are traced out of the full density matrix. Ryu and Takayangi's conjecture is that $S_B$ is proportional to the area of the co-dimension two surface, $\tilde{B}$, whose boundary is $\partial \tilde{B} \equiv \tilde{B} \vert_{\partial M_{\Psi}} = \partial B$ and which extremizes the area functional 
\begin{equation}\label{Area}
A = \int d^d \sigma \sqrt{\det(g_{ab})}
\end{equation} 
where $g_{ab}$ is the metric of $M_\Psi$ induced on the surface $\tilde{B}$. If more surfaces extremizes (\ref{Area}), then the further requirement is imposed that $\tilde{B}$ must be the surface of least area that is homologous to $B$. Thus, $\tilde{B}$ must be the surface smoothly retractable to $B$ that minimizes the area functional. The Ryu-Takayanagi formula can then be expressed as
\begin{equation}\label{RyuTakayanagi}
S_B = \frac{A(\tilde{B})}{4 G_N^{(d+1)}}.
\end{equation}
It is worth noting that the area of $\tilde{B}$ generally will be divergent. This is due to the infinite proper distance from any point in AdS spacetime to the boundary. This, however, is not a problem. As remarked in section \ref{Entanglement in CFTs}, short distance entanglement over the entangling surface in continuum quantum field theories renders the entanglement entropy for any quantum subsystem divergent. A regulation of this short distance entanglement on the CFT side exactly corresponds to introducing a boundary cutoff, $\epsilon$, such that $z > \epsilon$. This cutoff then serves to regulate the area as given by (\ref{Area}).

%where  $\tilde{B}$ is the co-dimension two surface in the bulk of $M_\Psi$, that minimizes the area functional of $M_\Psi$ and which separates a spatial slice of the bulk with boundary, $\Sigma_{\partial M_\Psi}$, into two parts such that $\Sigma_{\partial M_\Psi}$ is separated into two regions $B$ and its compliment, $\overline{B}$ by the asymptotic boundary, $\partial \tilde{B}$, of $\tilde{B}$ (see figure \ref{cylinder}). 

Comparing the Ryu-Takayanagi formula with the Bekenstein-Hawking formula, the similarity is striking. This is as expected if the Ryu-Takayanagi formula is a generalization of the Bekenstein-Hawking formula in its holographic interpretation. Thus, to establish the Bekenstein-Hawking formula as a special case can serve as a first test of the Ryu-Takayanagi formula. If the entropy, $S_{BH}$, in the Bekenstein Hawking formula is interpreted as an entanglement entropy, $S_B$, over some entangling surface, $\partial S_B$, then the black hole horizon must be interpreted as a co-dimension two surface, $\tilde{B}$, that minimizes the area functional which has $\partial B = \partial \tilde{B}$ and which is homologous to $B$. The problem, therefore, consists in checking whether black hole horizons in general are extremal co-dimension two surfaces that are homologous to a region on the asymptotic boundary of the black hole spacetime. 

For spacetimes with black holes situated in the bulk, the black hole horizon cannot coincide with a surface that ends on the boundary, since $\partial \tilde{B}_{BH} = \emptyset$. Thus, if such bulk black hole horizons are to be identified with the extremal surfaces from the Ryu-Takayanagi formula, it must be for the special case where $\partial B = \emptyset$. This is the limit where $B$ covers the full spacetime boundary, i.e. $\overline{B} = \emptyset$. Given these conditions, it is tempting to suppose that $\tilde{B}$ is the empty set as well, however, since a black hole spacetime is not simply connected -- it contains a singularity -- the empty set is not homologous with $B$. Instead, $\tilde{B}$ must be a surface that enclose the singularity. Considering all such surfaces, the one with minimal area is the black hole horizon. Obviously, this immediately entails that the dual quantum state to a bulk black hole must be a thermal state. If $B$ is such that its compliment, $\overline{B}$, is empty, then the quantum state on $B$ must be thermal in order for it to have non-zero entropy and thereby for it to have a black hole spacetime dual with non-zero horizon area. Thus, the Bekenstein-Hawking formula in its holographic interpretation may be viewed be a special case of the Ryu-Takayanagi formula. 

Claiming this, however, it is necessary to make a qualifying remark about the type of entropy involved. If the full system is in a thermal state, it has an entropy and is therefore not a pure state. But knowing only this, the entropy cannot be designated as entanglement entropy; one simply does not know the origin of the entropy. For this reason, the type of entropy was left unqualified above. However for the eternal black hole, the thermal properties explicitly originate from entanglement effects and the entropy, therefore, is entanglement entropy. As stated in section \ref{CFT_d and AdS_{d+1}}, the eternal black hole is dual to a thermofield double state
\begin{equation}
\ket{\Psi}=\frac{1}{\sqrt{Z(\beta)}}\sum_i e^{-\beta E_i / 2} \ket{E_i} \otimes \ket{E_i}.
\end{equation}
This is a pure state, i.e. the von Neumann entropy of the state $\ket{\Psi}$ is zero, but expressed as an entangled state of two subsystems. Tracing out the degrees of freedom of one of the subsystems, the other subsystem is in a thermal state. Knowing the full state, we can identify the thermal entropy of one of the subsystems with the entanglement entropy between the two subsystems. Either subsystem simply is in a thermal state, and consequently the extremal bulk co-dimension two surface whose area is proportional to this entanglement entropy must be identical with the black hole horizon following the argument above. Thus, when we consider the entropy of one of the subsystems in the eternal black hole and notes its origin, then this entropy can be identified as entanglement entropy. Therefore, the Bekenstein-Hawking formula applies to the eternal black hole as a special case of the Ryu-Takayangi formula. 
\\
\section{Uses of the Ryu-Takayanagi formula}
As a practical tool, the Ryu-Takayanagi formula allows us to evaluate entanglement entropies in strongly coupled field theories by evaluating the area of $\tilde{B}$ in the weakly coupled dual gravity theory. However, we could also work the other way around. Knowing the entanglement entropy, we can determine the area of the co-dimension two surface, $\tilde{B}$. In itself, this area is not of much interest. It would be interesting, if one could reconstruct the entire dual spacetime from the entanglement entropy of the field theory. Knowing a single area of a co-dimension two surface in the bulk of this spacetime will not get us very far; many spacetimes will be compatible with this particular area. However, subdividing the Cauchy surface, $\Sigma_{\partial M_{\Psi}}$, into different subsystems will give other entanglement entropies that can be related to different bulk surface areas.

In fact, the problem is highly overconstrained ensuring that a single spacetime at most -- the spacetime dual of that CFT state -- is compatible with all the areas in the bulk that corresponds to entanglement entropies on the CFT side.  Thus, according to the Ryu-Takayanagi formula it is in principle possible to find the spacetime dual of any CFT simply by calculating entanglement entropies. It is worth noting that the entanglement entropies overconstrain the dual spacetime such that there, for a general CFT, will be no spacetime that is compatible with all the entropies. Consequently, only a smaller subset of CFTs are expected to have dual descriptions as geometric spacetime theories.

However, finding the spacetime dual even for these CFTs faces severe technical difficulties both due to problems associated with the evaluation of entanglement entropies on the CFT side, and due to problems in solving the overconstrained problem on the gravity side. Also, the Ryu-Takayanagi formula is limitated to first order in the bulk, and therfore correction will occur from quantum gravity if we go deeper into the bulk.

However, setting aside these technical difficulties and considering only first order pertubations in the bulk where the gravitational theory is well-described by classical Einstein gravity coupled to matter (without curvature couplings), the Ryu-Takayanagi formula suggests an intimate relation between entanglement in $d$ dimensions -- an inherently quantum mechanical phenomena -- and then the entire bulk spacetime in $d+1$ dimensions in the (rare) cases where gauge/gravity duality obtains.

It is exactly this relation that will be explored subsequently. However, prior to this some evidence for the Ryu-Takayanagi formula will be provided.
\section{Evidence for the Ryu-Takayanagi formula}
%\textsc{If the cosmological constant is only non-zero if first order string loop correction are non-zero, then we should perhaps worry that this entails entanglement in bulk and therefore corrections to the Ryu-Takayanagi formula may occur already in pure AdS. Problem: $AdS_5 x S^5$ is a solution to Supergravity without cosmological constant.}

At its conception, the evidence of the Ryu-Takayanagi formula consisted in the a number of successful tests, where the extremal surface approach was employed to reproduce known results for entanglement entropies of CFT states. These quantitative comparisons are not very numerous due to the strong coupling of the CFT for $d > 2$. Thus, most of these comparisons are between $CFT_2$ and $AdS_3$. 

One such example is the matching of the entanglement entropy for a single interval in a $CFT_2$ on $\mathbb{R}^{1,1}$. As we did in section \ref{Entanglement in CFTs}, we will consider an interval with length $2a$ centred around the origin on the spacelike slice, $t=0$, such that $B=\lbrace x \in \mathbb{R} \vert -a < x < a \rbrace$ and $\partial B = \lbrace -a,a \rbrace$. For a $CFT_2$ on $\mathbb{R}^{1,1}$, the spacetime dual is the Poincaré patch of pure $AdS_3$. Since the background is Poincaré-AdS, the metric induced on the spatial slice, $t=0$, is $ds^2 = \frac{L^2}{z^2} \left( dx^2 + dz^2 \right)$. Following the prescription of the Ryu-Takayanagi formula, the entaglement entropy of $B$ is equal to the area of a co-dimension two bulk surface, $\tilde{B}$, in the $t=0$ plane parametrized by coordinates $x$ and $z$; see the metric (\ref{Poincaremetric}). $\tilde{B}$ must be a curve that has $\tilde{B}\vert_{\partial M} = \partial B$, i.e. a curve whose endpoints at $z=0$ are $x=-a$ and $x=a$. Furthermore, it must minimize the area functional (\ref{Area}). 

We will choose the parametrize $B$ in terms of the boundary coordinate $x$. The metric induced on $B$ -- with the single component $g_{xx}$ -- may then be expressed as
\begin{equation}
g_{xx} = \frac{L^2}{z^2} \left( \frac{\partial x}{\partial x}\frac{\partial x}{\partial x} + \frac{\partial z}{\partial x} \frac{\partial z}{\partial x} \right) = \frac{L^2}{z^2} \left( 1 + z'(x)^2 \right).
\end{equation}
Since induced metric has only one component, we have $\det(g) = g_{xx}$. The area functional therefore takes the form
\begin{equation}
A = \int_{-a}^a dx \sqrt{g_{xx}} = \int_{-a}^a dx \frac{L}{z} \sqrt{1 + z'(x)^2}.
\end{equation}
Defining $A = \frac{l}{z} \sqrt{1 + z'(x)^2}$, we may then find the extremum of the area functional using the Euler-Lagrange equation
\begin{equation}
\begin{split}
0 & = \frac{\partial A}{\partial z} - \frac{d}{dx} \frac{\partial A}{\partial z'} 
\\& = - \frac{L \sqrt{1 + z'(x)^2}}{z^2} - \frac{d}{dx} \frac{L z'(x)}{z \sqrt{1 + z'(x)^2} }
\\& = - L \frac{ \left( z \, z''(x) + z'(x)^2 + 1 \right) }{z^2 \left( z'(x)^2 + 1 \right)^{3/2}}.
\end{split}
\end{equation}
The solution is $z = \sqrt{a^2 - x^2}$ -- a semi-circle around the origin in the $zx$ place -- as may be verified by inserting back into the numerator of Euler-Lagrange equation:
\begin{equation}
z \, z''(x) + z'(x)^2 + 1 = - \frac{a^2}{z^2} + \frac{x^2}{z^2} + 1 = 0
\end{equation}
where we have used $1 = \frac{a^2 - x^2}{z^2}$. Thus, the co-dimension two surface, $\tilde{B}$, must satisfy $z^2 + x^2 = a^2$.

Inserting this into the expression for the area, we find 
\begin{equation}
A = \int_{-a}^a dx \frac{L}{a^2-x^2} \sqrt{1 + \frac{x^2}{a^2 - x^2}}.
\end{equation}
As expected, this integration diverges. To regulate it, we require that $z > \epsilon$ for some small $\epsilon$. The integration then yields
\begin{equation}
A = 2L \ln\left( \frac{2a}{\epsilon} \right).
\end{equation}
Inserting this into the Ryu-Takayanagi we find
\begin{equation}
S_B = \frac{A}{4 G_N} = \frac{L}{2 G_N} \ln\left( \frac{2a}{\epsilon} \right).
\end{equation}
If we identify $c = \frac{3}{2} \frac{L}{G_N}$ in accord with the $AdS_3/CFT_2$ dictionary, we get
\begin{equation}
S_B = \frac{c}{3} \ln\left( \frac{2a}{\epsilon} \right).
\end{equation}
The same result as obtained by direct calculation in section \ref{Entanglement in CFTs}.

Besides direct agreement between the entanglement entropies in cases where the calculations are tractable on both the AdS and the CFT side, the Ryu-Takayanagi formula has also been shown to satisfy various known properties of entanglement entropy such as strong subadditivity. A partial derivation of the Ryu-Takayanagi formula for general spherical entangling surfaces was proposed in \citep{casini_towards_2011} and based on this a full derivation from the AdS/CFT dictionary was developed in \citep{lewkowycz_generalized_2013}. Both derivations are beyond the scope of the current project. However, elements of \citep{casini_towards_2011} will be explored in later chapters since a number of aspects from this partial derivation of the Ryu-Takayanagi formula will prove useful for other purposes.

\chapter{Linearized Gravity in AdS Background}\label{Linearized gravity in AdS background}
In this section we introduce linearized gravity gravity first in a generic background and then in the specific case of an anti-de Sitter background. 
\section{Perturbing spacetime}
Linearized gravity is a type of approximation scheme for Einstein gravity where the spacetime is characterized by some background metric, $g_{ab}^0$, plus a weak perturbation, $\delta g_{ab}$, that generates only small corrections to the background field such that the full metric field, $g_{ab}$, is given by
\begin{equation}
g_{ab} = g^0_{ab}+\delta g_{ab}
\end{equation}
where $\delta g_{ab}$ is assumed to be small compared to $g^0_{ab}$. Linearized gravity is then obtained by expanding Einsteins field equations to linear order in $\delta g_{ab}$. Since the perturbation, $\delta g_{ab}$, is small, we will assume that we can use the background metric to raise and lower indices.

For convenience, we will consider a one paramater family of metrics, $g_{ab}(\zeta)$, such that for each value of $\zeta$, $g_{ab}(\zeta)$ solves the Einstein equations and $g_{ab}(0) = g_{ab}^0$. Thus, the first order correction to the background metric -- the perturbation $\delta g_{ab}$ -- is given by
\begin{equation}
\frac{\partial g_{ab}(\zeta)}{\partial \zeta}\vert_{\zeta=0} = \delta g_{ab}.
\end{equation}
The linearized Einstein equations can then similarly be obtained as
\begin{equation}
\frac{\partial G^E_{ab}(\zeta)}{\partial \zeta}\vert_{\zeta=0} = 8 \pi G_N \frac{\partial T_{ab}(\zeta)}{\partial \zeta}\vert_{\zeta=0}.
\end{equation}
These equations imposes constraints on $\delta g_{ab}$. To find them, we must establish what change is induced to linear order on the Einstein tensor by the linear order perturbation $\delta g_{ab}$. For convenience, we will introduce the notation
\begin{equation}
\frac{\partial G^E_{ab}(\zeta)}{\partial \zeta}\vert_{\zeta=0} \equiv \dot{G}^E_{ab}
\end{equation}
and may thereby rephrase the above as the task of finding an expression for $\dot{G}^E_{ab}$ in terms of $\delta g_{ab}$.

The Einstein tensor with cosmological constant is given by
\begin{equation}
G^E_{ab} = R_{ab} - \frac{1}{2} R g_{ab} + \Lambda g_{ab} = \left( \delta^c_a \delta^d_b - \frac{1}{2} g^{cd} g_{ab} \right) R_{cd} + \Lambda g_{ab}.
\end{equation}
where $R_{ab}$ is the Ricci tensor, $R = g^{ab} R_{ab}$ is the Ricci scalar and $\Lambda$ is the cosmological constant. Thus,
\begin{equation}\label{GEdot}
\begin{split}
\dot{G}^E_{ab} & = \frac{\partial}{\partial \zeta}\vert_{\zeta=0} \left[ \left( \delta^c_a \delta^d_b - \frac{1}{2} g^{cd} g_{ab} \right) R_{cd} + \Lambda g_{ab} \right] 
\\& = \left( \delta^c_a \delta^d_b - \frac{1}{2} g^{cd} g_{ab} \right) \dot{R}_{cd} - \frac{1}{2} \left( g^{cd} \delta g_{ab} - \delta g^{cd} g_{ab}  \right) R_{cd} + \Lambda \delta g_{ab}
\end{split}
\end{equation}
where we have defined $\delta g^{ab} \equiv g^{ac} g^{bd} \delta g_{bd}$ from which it follows that $\frac{\partial g^{ab}}{\partial \zeta}\vert_{\zeta=0} \equiv - \delta g^{ab}$.\footnote{To see this, consider
\begin{equation}
\begin{split}
\frac{\partial}{\partial \zeta} \delta^a_c & = \frac{\partial}{\partial \zeta} g^{ab} g_{bc} 
\\ 0 & = g_{bc} \frac{\partial g^{ab}}{\partial \zeta} + g^{ab}  \frac{\partial g_{bc}}{\partial \zeta}
\\ 0 & = g^{cd} g_{bc} \frac{\partial g^{ab}}{\partial \zeta} + g^{cd} g^{ab}  \delta g_{bc}
\\ 0 & = \frac{\partial g^{ad}}{\partial \zeta} + \delta g^{ad}
\end{split}
\end{equation}}
To find an expression for $\dot{G}^E_{ab}$, one must therefore first find an expression for the perturbation of the Ricci tensor $\dot{R}_{ab}$. The Ricci tensor is given as
\begin{equation}
R_{ab} = R_{acb}^{\; \; \; \; c}
\end{equation}
where $R_{acb}^{\; \; \; \; c}$ is the Riemann tensor. The Riemann tensor, in turn, depends on the metric connections:
\begin{equation}
R_{abc}^{\; \; \; \; d} = - \partial_a \Gamma_{bc}^d + \partial_b \Gamma_{ac}^d - \Gamma_{ae}^d \Gamma_{bc}^e + \Gamma_{be}^d \Gamma_{ac}^e
\end{equation}
The linear order change in the Riemann tensor induced by the linear order perturbation, $\delta g_{ab}$ may be expanded as
\begin{equation}
\begin{split}
\frac{\partial}{\partial \zeta} R_{abc}^{\; \; \; \; d} & = - \frac{\partial}{\partial \zeta} \left( \partial_a \Gamma_{bc}^d - \partial_b \Gamma_{ac}^d + \Gamma_{ae}^d \Gamma_{bc}^e - \Gamma_{be}^d \Gamma_{ac}^e \right)
\\& = - \partial_a \dot{\Gamma}_{bc}^d + \partial_b \dot{\Gamma}_{ac}^d - \dot{\Gamma}_{ae}^d \Gamma_{bc}^e - \Gamma_{ae}^d \dot{\Gamma}_{bc}^e + \dot{\Gamma}_{be}^d \Gamma_{ac}^e + \Gamma_{be}^d \dot{\Gamma}_{ac}^e
\\& = - \nabla_a \dot{\Gamma}_{bc}^d + \nabla_b \dot{\Gamma}_{ac}^d
\end{split}
\end{equation}
where we have used that the partial derivative is independent of $\zeta$. The last line follows since
\begin{equation}
\begin{split}
- \nabla_a \dot{\Gamma}_{bc}^d + \nabla_b \dot{\Gamma}_{ac}^d &= - \partial_a \dot{\Gamma}_{bc}^d + \Gamma_{ab}^e \dot{\Gamma}_{ec}^d + \Gamma_{ac}^e \dot{\Gamma}_{be}^d - \Gamma_{ae}^d \dot{\Gamma}_{bc}^e \\& \; \; \; \; + \partial_b \dot{\Gamma}_{ac}^d - \Gamma_{ba}^e \dot{\Gamma}_{ec}^d - \Gamma_{bc}^e \dot{\Gamma}_{ae}^d + \Gamma_{be}^d \dot{\Gamma}_{ac}^e
\\& = - \partial_a \dot{\Gamma}_{bc}^d + \Gamma_{ac}^e \dot{\Gamma}_{be}^d - \Gamma_{ae}^d \dot{\Gamma}_{bc}^e + \partial_b \dot{\Gamma}_{ac}^d - \Gamma_{bc}^e \dot{\Gamma}_{ae}^d + \Gamma_{be}^d \dot{\Gamma}_{ac}^e.
\end{split}
\end{equation}
Here we have used that the metric connection is torsion free: $\Gamma_{ab}^e=\Gamma_{ba}^e$.

What remains is to find perturbation of the metric connection. To find an expression for this, consider initially the derivative with respect to $\zeta$ of the vanishing covariant derivative of the metric
\begin{equation}
\begin{split}
0 & = \frac{\partial}{\partial \zeta} \nabla_a(\zeta) g_{bc}(\zeta) 
\\& = \frac{\partial}{\partial \zeta}  \left( \partial_a g_{bc} - \Gamma_{ab}^d g_{dc} - \Gamma_{ac}^d g_{ad} \right)
\\& = - \dot{\Gamma}_{ab}^d g_{dc} - \dot{\Gamma}_{ac}^d g_{ad} + \nabla_a \delta g_{bc}.
\end{split}
\end{equation}
Reorganizing this, we find
\begin{equation}\label{ndg}
\begin{split}
\nabla_a \delta g_{bc} & = \dot{\Gamma}_{ab}^d g_{dc} + \dot{\Gamma}_{ac}^d g_{ad} 
\\& = g^{de} \dot{\Gamma}_{abe} g_{dc} + g^{de} \dot{\Gamma}_{ace} g_{ad}
\\& = \delta^{e}_c \dot{\Gamma}_{abe} + \delta^{e}_b \dot{\Gamma}_{ace}
\\& = \dot{\Gamma}_{abc} + \dot{\Gamma}_{acb}.
\end{split}
\end{equation}
Another constraint follows, since the metric connection must be torsion free in the first two indices. Thus,
\begin{equation}
\dot{\Gamma}_{(ab)}^c = \dot{\Gamma}_{ab}^c.
\end{equation}
Now, using first (\ref{ndg}) and then that the metric connection must be torsion free, we find
\begin{equation}
\begin{split}
\dot{\Gamma}_{abc} = \nabla_a \delta g_{bc} - \dot{\Gamma}_{acb} = \nabla_a \delta g_{bc} - \dot{\Gamma}_{cab}.
\end{split}
\end{equation}
Applying this two more times, we get
\begin{equation}
\begin{split}
\dot{\Gamma}_{abc} & = \nabla_a \delta g_{bc} - \dot{\Gamma}_{cab}
\\& = \nabla_a \delta g_{bc} - \left( \nabla_c \delta g_{ab} - \dot{\Gamma}_{bca} \right)
\\& = \nabla_a \delta g_{bc} - \nabla_c \delta g_{ab} + \left( \nabla_b \delta g_{ca} - \dot{\Gamma}_{abc} \right).
\end{split}
\end{equation}
The metric perturbation now occurs both on the left and on the right hand side above. Collecting these, dividing by two and raising the last index, we finally get
\begin{equation}
\begin{split}\label{dotGamma}
\dot{\Gamma}_{ab}^c & = \frac{1}{2} g^{cd} \left( \nabla_a \delta g_{bd} - \nabla_d \delta g_{ab} + \nabla_b \delta g_{da} \right).
\end{split}
\end{equation}
As seen, this expression only depends on metric perturbation and covariant derivatives taken with respect to the background metric.

Inserting this expression, the perturbation of the Ricci tensor becomes
\begin{equation}
\begin{split}
\dot{R}_{ab} & = \dot{R}_{acb}^{\; \; \; \; c} = - \nabla_a \dot{\Gamma}_{cb}^c + \nabla_c \dot{\Gamma}_{ab}^c
\\& = - \frac{1}{2} g^{cd} \nabla_a \left( \nabla_c \delta g_{bd} - \nabla_d \delta g_{cb} + \nabla_{b} \delta g_{cd} \right) + \frac{1}{2} g^{cd} \nabla_c \left( \nabla_a \delta g_{bd} - \nabla_d \delta g_{ab} + \nabla_b \delta g_{da} \right) 
\\& = - \frac{1}{2} g^{cd} \left( R_{acb}^{\; \; \; \; e} \delta g_{ed} + R_{acd}^{\; \; \; \; e} \delta g_{be} - \nabla_a \nabla_d \delta g_{cb} + \nabla_a  \nabla_{b} \delta g_{cd} + \nabla_c \nabla_d \delta g_{ab} - \nabla_c \nabla_b \delta g_{da} \right)
\end{split}
\end{equation}
where the last line follows from the identity
\begin{equation}
\nabla_a \nabla_c \delta g_{bd} - \nabla_c \nabla_a \delta g_{bd} = R_{acb}^{\; \; \; \; \; e} \delta g_{ed} + R_{acd}^{\; \; \; \; \; e} \delta g_{be}.
\end{equation}
Using the analogous identity
\begin{equation}
\nabla_c \nabla_b \delta g_{da} - \nabla_b \nabla_c \delta g_{da} = R_{cbd}^{\; \; \; \; \; e} \delta g_{ea} + R_{cba}^{\; \; \; \; \; e} \delta g_{de}
\end{equation}
we can substitute for the last term above and act with $g^{cd}$ on all terms
\begin{equation}
\begin{split}
\dot{R}_{ab} & = - \frac{1}{2} g^{cd} \bigg( R_{acb}^e \delta g_{ed} + R_{acd}^e \delta g_{be} - \nabla_a \nabla_d \delta g_{cb} + \nabla_a  \nabla_{b} \delta g_{cd} + \nabla_c \nabla_d \delta g_{ab} \\& \; \; \; \; \; \; \; \; \; \; \; \; \; \; \; - \nabla_b \nabla_c \delta g_{da} - R_{cbd}^e \delta g_{ea} - R_{cba}^e \delta g_{de} \bigg)
\\& = - \frac{1}{2} \bigg( R_{a \; b}^{\; d \; e} \delta g_{ed} + R_{ac}^{\; \; \; ce} \delta g_{be} - \nabla_a \nabla^c \delta g_{cb} + \nabla_a \nabla_{b} \delta g_{c}^c + \nabla_c \nabla^c \delta g_{ab} \\& \; \; \; \; \; \; \; \; \; \; \; - \nabla_b \nabla^d \delta g_{da} - R_{cb}^{\; \; \; ce} \delta g_{ea} - R_{\; ba}^{d \; \; e} \delta g_{de} \bigg)
\\& = - \frac{1}{2} \bigg( \nabla_c \nabla^c \delta g_{ab} - \nabla_a \nabla^c \delta g_{cb} - \nabla_b \nabla^c \delta g_{ca}  \\& \; \; \; \; \; \; \; \; \; \; \;  + \nabla_a \nabla_{b} \delta g_{c}^c - R_{a}^{\; d} \delta g_{bd} - R_{b}^{\; d} \delta g_{da} + 2 R_{a \; b}^{\; c \; d} \delta g_{dc} \bigg)
\end{split}
\end{equation}
where we have used $R_{cb}^{\; \; \; ce} = R_{b}^{\; e}$, $R_{ac}^{\; \; \; ce} = - R_{ca}^{\; \; \; ce} = - R_{a}^{\; e}$, $- R_{\; ba}^{d \; \; e} = R_{b \; a}^{\; d \; e}$, and also $R_{a \; b}^{\; c \; d} \delta g_{dc} =  R_{b \; a}^{\; c \; d} \delta g_{cd}$.

\section{Perturbing the Poincaré patch}\label{Perturbing the Poincare Patch}
The expression in terms of the Riemann and Ricci tensor is convenient since we will be interested in the special case where the background is the Poincaré patch of anti-de Sitter spacetime; a solution to Einsteins equations with negative cosmological constant
\begin{equation}
\Lambda = - \frac{d (d-1)}{2L^2}.
\end{equation}
This is a maximally symmetric spacetime and the Riemann tensor may therefore be expressed as
\begin{equation}
R_{abcd} = \frac{-1}{L^2} (g_{ac} g_{bd} - g_{ad} g_{bc})
\end{equation}
and the Ricci tensor takes the form
\begin{equation}
R_{ab} = \frac{-d}{L^2} g_{ab}.
\end{equation}
Using the Poincaré coordinates, the metric tensor is diagonal, such that $g_{ab}$ and $R_{ab}$ vanishes for $a \neq b$. Consequently, the Riemann tensor contracted with the metric perturbation becomes:
\begin{equation}
\begin{split}
R_{a\;b}^{\; c \; d} \delta g_{cd} & = g^{ce} g^{df} R_{aebf} \, \delta g_{cd}
\\& = g^{ce} g^{df} (g_{ab} g_{ef} - g_{af} g_{eb}) \frac{-1}{L^2} \, \delta g_{cd}
\\& = (\delta^{c}_f g^{df} g_{ab} - \delta^{c}_b \delta^{d}_a) \frac{-1}{L^2} \, \delta g_{cd}
\\& = (g^{cd} g_{ab} - \delta^{c}_b \delta^{d}_a) \frac{-1}{L^2} \, \delta g_{cd}
\end{split}
\end{equation}
and the Ricci tensor contracted with the metric perturbation becomes
\begin{equation}
\begin{split}
R_a^{\; c} \delta g_{bc} & = g^{cd} R_ {ad} \delta g_{bc}
\\& = g^{cd} \frac{-d}{L^2} g_{ad} \delta g_{bc}
\\& = \frac{-d}{L^2} \delta^c_a \delta g_{bc}
\\& = \frac{-d}{L^2} \delta g_{ba}.
\end{split}
\end{equation}
Subsequently, we will primarily be interested in $G^E_{tt}$. Using the above we find
\begin{equation}
\begin{split}
\dot{G}^E_{tt} & = - \frac{1}{2} (- \delta g^{cd} g_{tt} + g^{cd} \delta g_{tt} ) R_{cd} + ( \delta^c_t \delta^d_t - \frac{1}{2} g^{cd} g_{tt} ) \dot{R}_{cd} - \frac{d (d-1)}{2L^2} \delta g_{tt}
\\& = \frac{1}{2} (- g^{ce} g^{df} \delta g_{ef} g_{tt} + g^{cd} \delta g_{tt} ) \frac{d}{L^2} g_{cd} \\& \; \; \, \, + \dot{R}_{tt} - \frac{1}{2} \left( \dot{R}_{tt} - \dot{R}_{zz} - \delta^{ij} \dot{R}_{ij} \right) - \frac{d }{2L^2} \delta g_{tt}
\\& = \frac{d}{2 L^2} (- \delta^{e}_d g^{df} \delta g_{ef} g_{tt} + \left[ \delta^c_c - (d - 1) \right] \delta g_{tt} ) + \frac{1}{2} ( \dot{R}_{tt} + \dot{R}_{zz} + \delta^{ij} \dot{R}_{ij} ) - \frac{d (d-1)}{2L^2} \delta g_{tt}
\\& = \frac{d}{2 L^2} (\delta^{ij} \delta g_{ij} - g^{tt} \delta g_{tt} g_{tt} + \left[ (d+1) - (d-1) \right] \delta g_{tt} ) + \frac{1}{2} \delta^{ab} \dot{R}_{ab} - \frac{d (d-1)}{2L^2} \delta g_{tt}
\\& = \frac{d}{2 L^2} (\delta^{ij} \delta g_{ij} + \delta g_{tt} ) + \frac{1}{2} \delta^{ab} \dot{R}_{ab} 
\end{split}
\end{equation}
where we have assumed radial gauge ($\delta g_{zz} = 0$) and used that $\delta^a_a = d+1$. 

What remains is to evaluate $\delta^{ab} \dot{R}_{ab}$:
\begin{equation}
\begin{split}
\delta^{ab} \dot{R}_{ab} & = \delta^{ab} \bigg( - \frac{1}{2} \nabla_c \nabla^c \delta g_{ab} + \frac{1}{2} \nabla_a \nabla^c ( \delta g_{bc} - \frac{1}{2} \delta g_d^{\; d} g_{bc} ) 
\\& \; \; \; \; \; \; \; \; \; \; + \frac{1}{2} \nabla_b \nabla^c ( \delta g_{ac} - \frac{1}{2} \delta g_d^{\; d} g_{ac} ) +   \frac{1}{2} R_a^{\; c} \delta g_{bc} +  \frac{1}{2} R_b^{\; c} \delta g_{ac} - R_{a \; b}^{\; c \; d} \delta g_{cd} \bigg)
\\& = \delta^{ab} \bigg( - \frac{1}{2} \nabla_c \nabla^c \delta g_{ab} + \nabla_a \nabla^c ( \delta g_{bc} - \frac{1}{2} \delta g_d^{\; d} g_{bc} ) +   R_a^{\; c} \delta g_{bc} - R_{a \; b}^{\; c \; d} \delta g_{cd} \bigg)
\\& = \frac{z^2}{L^2} \bigg[  - \frac{d}{z^2} \delta g_{tt} - \left( \partial_z^2 + \partial_i \partial^i - \frac{(d-5)}{z} \partial_z + \frac{d-4}{z^2} \right) \delta^{kl} \delta g_{kl} + \partial^i \partial^j \delta g_{ij} \bigg]
\end{split}
\end{equation}
where we have defined $\partial_i \equiv \delta_{ij} \partial^j$.

We will suppose that $\zeta$ is small such that the dual spacetimes can be represented using the Fefferman-Graham coordinates by the metric
\begin{equation}\label{FG}
\begin{split}
ds^2 & = \frac{L^2}{z^2}(dz^2 + dx_\mu dx^\mu + z^d \tilde{h}_{\mu \nu}(x, z) dx^\mu dx^\nu )
\\& = \frac{L^2}{z^2}(dz^2 + dx_\mu dx^\mu) + z^{d-2} h_{\mu \nu}(x, z) dx^\mu dx^\nu
\end{split}
\end{equation}
where we, in the last equality, have defined $h_{\mu \nu} \equiv \tilde{h}_{\mu \nu} L^2$. This is convenient since it permits us to write the perturbed metric on the form
\begin{equation}
g_{a b} = g_{ab}^{0} + \delta g_{ab} = g_{ab}^{AdS} + z^{d-2} h_{ab}.
\end{equation}
Here we have identified $\delta g_{ab} = z^{d-2} h_{ab}$ where $h_{ab}$ is equal to $h_{\mu \nu}$ with the additional components $h_{zz}= 0$ and $h_{az}=0$. From this, we readily see that the Fefferman-Graham coordinates presumes radial gauge for the perturbation $z^{d-2} h_{ab}$.\footnote{This gauge freedom follows since there are $d+1$ coordinates. One can therefore gauge fix $d+1$ parameters of $H$: $h_{z a}$ has $d$ components and $h_{zz}$ is another, which gives $d+1$ gauge fixed components in total.} Also, we have assumed that the background metric is the Poincaré patch of AdS: \begin{equation}
g^0_{ab} = 
\begin{pmatrix}
- \frac{L^2}{z^2} & & & \\
& \frac{L^2}{z^2} & &   \\
& & \ddots & \\
& & & \frac{L^2}{z^2}.
\end{pmatrix}
\end{equation}

Inserting this into the expression for $\delta^{ab} \dot{R}_{ab}$, one finds
\begin{equation}
\begin{split}
\delta^{ab} \dot{R}_{ab} & = \frac{z^2}{L^2} \bigg[ - \frac{d}{z^2} z^{d-2} h_{tt} - \left( \partial_z^2 + \partial_i \partial^i - \frac{(d-5)}{z} \partial_z + \frac{d-4}{z^2} \right) \delta^{kl} z^{d-2} h_{kl} \\& \; \; \; \; \; \; \; \; \; \; \; + \partial^i \partial^j z^{d-2} h_{ij} \bigg]
\\& = \frac{z^d}{L^2} \bigg[ - \frac{d}{z^2} h_{tt} - \bigg( \frac{(d-2)^2}{z^2} - \frac{d-2}{z^2} + 2 \frac{d-2}{z} \partial_z + \partial_z^2 + \partial_i \partial^i \\& \; \; \; \; \; \; \; \; \; \; \;  - \frac{(d-5)}{z} \frac{d-2}{z} - \frac{(d-5)}{z} \partial_z - \frac{d-4}{z^2} \bigg) \delta^{kl} h_{kl} + \partial^i \partial^j h_{ij} \bigg]
\\& = \frac{z^d}{L^2} \bigg[ - \frac{d}{z^2} h_{tt}  - \left( \partial_z^2 + \frac{d+1}{z}\partial_z + \frac{d}{z^2} + \partial_i \partial^i \right) \delta^{kl} h_{kl} + \partial^i \partial^j h_{ij} \bigg].
\end{split}
\end{equation}
Reinserting this into the expression for $\dot{G}^E_{tt}$ and again substituting $\delta g_{ab}$ with $z^{d-2}h_{ab}$, one finally gets
\begin{equation}
\begin{split}
\dot{G}^E_{tt} & = \frac{d}{2 L^2} (\delta^{ij} z^{d-2} h_{ij} + z^{d-2} h_{tt} ) \\& + \frac{z^d}{2 L^2} \bigg[ - \frac{d}{z^2} h_{tt} - \left( \partial_z^2 + \frac{d+1}{z}\partial_z + \frac{d}{z^2} + \partial_i \partial^i \right) \delta^{kl} h_{kl} + \partial^i \partial^j h_{ij} \bigg]
\\& = \frac{z^d}{2 L^2} \bigg[  - \left( \partial_z^2 + \frac{d+1}{z}\partial_z + \partial_i \partial^i \right) \delta^{kl} h_{kl} + \partial^i \partial^j h_{ij}\bigg]
\end{split}
\end{equation}
The equation 
\begin{equation}
\dot{G}^E_{tt} = 0
\end{equation}
is the tt-component of the linearized Einstein equations in vacuum in the presence of a negative cosmological constant $- \frac{d (d-1)}{2L^2}$.

As one may expect, $\dot{G}^E_{ab} = 0$ is also the equations of motion obtained to leading order for an expansion in $G_N$ of the Einstein-Hilbert action. The first subleading correction in $G_N$ turns on the coupling to matter fields with the inclusion of the source term $8 \pi G_N \delta \expval{T_{ab}}$.

\chapter{Entanglement and Spacetime}\label{Entanglement and Spacetime}
An initial qualitative indication of the relation between gravity on the AdS side and entanglement on the CFT side may be obtained by considering the dual pair: the eternal black hole and the thermofield double state. Here the Bekenstein-Hawking formula interpreted holographically, i.e. as a special case of the Ryu-Takayanagi formula, can be used to monitor what happens to the black hole horizon in the eternal black hole when entanglement is added or removed between the double states in the thermofield double state. Thus, it serves as a probe of the relation between entanglement on the CFT side and some spacetime surface on the AdS side. While this special case may be studied employing a holographic interpretation of the Bekenstein-Hawking formula, general, holographic, classical spacetimes may similarly be studied using the Ryu-Takayanagi formula. Below, the special case of the thermofield double state/eternal black hole will be examined first, and we will then subsequently consider the generalization to arbitrary quantum states with a classical spacetime dual.

\section{Entanglement and the eternal black hole}\label{Entanglement and Black Holes}
The thermofield double state is a particular state, $\ket{\Psi} \in \mathcal{H}_A \otimes \mathcal{H}_B$, in which we can find the quantum system comprised of the subsystems $Q_A$ and $Q_B$. Interestingly, the expression of the thermofield double state, (\ref{double}), in terms of energy eigenstates explicitly unveils how the degrees of freedom of the two subsystems are entangled through the weighted sum over states $\sum_i e^{-\beta E_i / 2} \ket{E_i^A} \otimes \ket{E_i^B}$. More explicitly, we find that the density matrix for one of the subsystems is given by
\begin{equation}
\begin{split}
\rho_A & = \tr_B(\ket{\Psi} \bra{\Psi}) 
\\& = \frac{1}{Z(\beta)} \sum_m \bra{E_m^B} \left( \sum_i e^{-\beta E_i / 2} \ket{E_i^A} \otimes \ket{E_i^B} \right) \left( \sum_j e^{-\beta E_j / 2} \bra{E_j^A} \otimes \bra{E_j^B} \right) \ket{E_m^B}
\\& = \frac{1}{Z(\beta)} \sum_m \sum_i \sum_j e^{-\beta (E_j+E_i) / 2} \delta_{jm} \delta_{im} \ket{E_i^A} \bra{E_j^A}
\\& = \frac{1}{Z(\beta)} \sum_m e^{-\beta E_m} \ket{E_m^A} \bra{E_m^A}
\end{split}
\end{equation}
from which it follows that
\begin{equation}
\begin{split}
S_A & = - \frac{1}{Z(\beta)} \sum_m e^{-\beta E_m} \log\left( \frac{1}{Z(\beta)} e^{-\beta E_m}\right) 
\\& =  \frac{1}{\sum_i e^{-\beta E_i}} \sum_m e^{-\beta E_m} \left( \beta E_m + \log(\sum_j e^{-\beta E_j}) \right) 
\\& = \frac{1}{\sum_i e^{-\beta E_i}} \sum_m e^{-\beta E_m} \beta E_m + \sum_j \log(e^{-\beta E_j})
\end{split}
\end{equation}
where we have used $Z(\beta) = \sum_i e^{-\beta E_i}$. This will generally be non-zero except for the limit where $\beta \rightarrow \infty$, i.e. the limit where the temperature goes to zero. It is evident from this, how the inverse temperature $\beta$ controls the entanglement entropy and therefore the amount of entanglement between the degrees of freedom in the two subsystems.

From the reduced density matrix, $\rho_A$, we see that in the limit $\beta \rightarrow \infty$ only the ground state of the subsystem, $Q_A$, is occupied. This follows since all the energy eigenvalues other than the vacuum state, $E_{i \neq 0}$, are non-zero and therefore vanish in the limit $\beta \rightarrow \infty$ as a consequence of the factor $e^{-\beta E_i}$. Due to the symmetry of the thermofield double state, the same holds for the subsystem, $Q_B$. Thus, in the limit $\beta \rightarrow \infty$ the full system is the product of these two ground states
\begin{equation}
\ket{\Phi} = \ket{E_0^A} \otimes \ket{E_0^B}
\end{equation}
i.e. $\ket{\Psi} \stackrel{\beta \rightarrow \inf}{=} \ket{\Phi} $. 

Manifestly, $\ket{\Phi}$ is a product state and does not, as expected from the expression for $S_A$, contain entanglement between the degrees of freedom in $Q_A$ and $Q_B$. In this state, therefore, the two systems are completely uncorrelated. What is argued in \citep{van_raamsdonk_building_2010} is that a first indication of the relation between entanglement and spacetime may be obtained from the differences between the spacetime dual of a product state like $\Phi$ and an entangled state like $\Psi$.

The state $\Phi$ where there is no entanglement between $Q_A$ and $Q_B$ consists of the tensor product of two identical pure states; more specifically a product of the identical vacuum states of these subsystems, $\ket{E_0}$. If we suppose that a pure state is dual to some spacetime, then the product of two pure states is dual to the product of two such spacetimes. Thus, the full spacetime consists of the product of two completely uncorrelated spacetimes. It seems, therefore, reasonable to suppose that these two spacetimes are disconnected, i.e. no light signal travelling from one spacetime can intersect a light signal travelling from the other. From the conjectured duality between the thermofield double state and the eternal black hole, it follows that the spacetime dual of the thermofield double state, where $Q_A$ and $Q_B$ are entangled, is a connected spacetime, i.e. in the eternal black hole a light signal travelling from $A$ can intersect a light signal travelling from $B$. The comparison between $\Psi$ and $\Phi$ in the context of the thermofield double states suggests that entanglement between $Q_A$ and $Q_B$ is a necessary condition for connectivity between $A$ and $B$ in the dual eternal black hole.

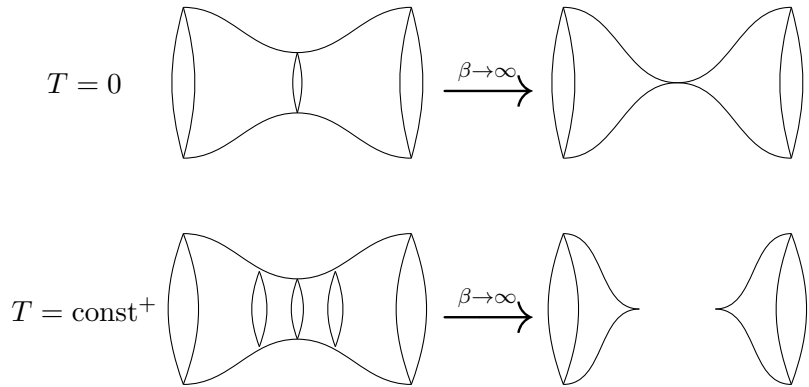
\begin{figure}[h]
\begin{center}
\begin{tikzpicture}

     \draw (-3,3) to[out=255,in=105] (-3,1);
     \draw (-3,3) to[out=285,in=75] (-3,1);
     \draw (-3,3) to[out=0,in=180] (-1.5,2.4);
     \draw (-3,1) to[out=0,in=180] (-1.5,1.6);
     \draw (-1.5,2.4) to[out=255,in=105] (-1.5,1.6);
     \draw (-1.5,2.4) to[out=285,in=75] (-1.5,1.6);
     \draw (-1.5,2.4) to[out=0,in=180] (0,3);
     \draw (-1.5,1.6) to[out=0,in=180] (0,1);
     \draw (0,3) to[out=255,in=105] (0,1);
     \draw (0,3) to[out=285,in=75] (0,1);
     
     %Lower left hand figure
     
\draw (-3,0) to[out=250,in=110] (-3,-2);
     \draw (-3,0) to[out=290,in=70] (-3,-2);
     \draw (-3,0) to[out=0,in=180] (-1.5,-0.6);
     \draw (-3,-2) to[out=0,in=180] (-1.5,-1.4);
     \draw (-1.5,-0.6) to[out=250,in=110] (-1.5,-1.4);
     \draw (-1.5,-0.6) to[out=290,in=70] (-1.5,-1.4);
     \draw (-2,-0.5) to[out=250,in=110] (-2,-1.5);
     \draw (-2,-0.5) to[out=290,in=70] (-2,-1.5);
     \draw (-1,-0.5) to[out=250,in=110] (-1,-1.5);
     \draw (-1,-0.5) to[out=290,in=70] (-1,-1.5);
     \draw (-1.5,-0.6) to[out=0,in=180] (0,0);
     \draw (-1.5,-1.4) to[out=0,in=180] (0,-2);
     \draw (0,0) to[out=250,in=110] (0,-2);
     \draw (0,0) to[out=290,in=70] (0,-2);   
     
     \node (V)    at (1,2)   {$\stackrel{\beta \rightarrow \infty}{\mathlarger{\mathlarger{\mathlarger{\mathlarger{\mathlarger{\longrightarrow}}}}}}$};
     \node (V)    at (1,-1)   {$\stackrel{\beta \rightarrow \infty}{\mathlarger{\mathlarger{\mathlarger{\mathlarger{\mathlarger{\longrightarrow}}}}}}$};
     
     \draw (2,3) to[out=255,in=105] (2,1);
     \draw (2,3) to[out=285,in=75] (2,1);
     \draw (2,3) to[out=0,in=180] (3.5,2);
     \draw (2,1) to[out=0,in=180] (3.5,2);
     \draw (3.5,2) to[out=0,in=180] (5,3);
     \draw (3.5,2) to[out=0,in=180] (5,1);
     \draw (5,3) to[out=255,in=105] (5,1);
     \draw (5,3) to[out=285,in=75] (5,1);
     \node at (-4.3,2) {$T=0$};
     
     %Lower right hand figure
     
     \draw (2,0) to[out=250,in=110] (2,-2);
     \draw (2,0) to[out=290,in=70] (2,-2);
     \draw (2,0) to[out=0,in=180] (3,-1);
     \draw (2,-2) to[out=0,in=180] (3,-1);
     \draw (4,-1) to[out=0,in=180] (5,0);
     \draw (4,-1) to[out=0,in=180] (5,-2);
     \draw (5,0) to[out=250,in=110] (5,-2);
     \draw (5,0) to[out=290,in=70] (5,-2);
     \node  at (-4.3,-1) {$T=\mathrm{const^+}$};
\end{tikzpicture}
\caption{\label{thermopinch} 
Depiction of the behavior of the spatial slices $T=0$ and $T$ equals a constant of the eternal black hole when when $\beta \rightarrow \infty$ in the dual quantum state $\ket{\Psi}$.}
\end{center}
\end{figure}

This conclusion sits well with the Bekenstein-Hawking formula that relates the entropy of a black hole, $S_{BH}$, with its horizon area, $A_{BH}$ \citep{bekenstein_black_1973}.
\begin{equation}\label{Bekenstein}
S_{BH} = \frac{A_{BH}}{4G}.
\end{equation}
The duality between the eternal black hole and the thermofield double state entails that the black hole entropy -- regardless of its origin on the AdS side -- is equal to the entanglement entropy of one of the subsystems in the thermofield double state \citep{emparan_black_2006}. Employing this relation, it is possible to monitor the horizon area as seen from either side of the eternal black hole, when entanglement between $Q_A$ and $Q_B$ is removed in the dual thermofield double state. Since the horizon area, $A_{BH}$, and entanglement entropy, $S_A$, is proportional, it follows that when entanglement is removed between $Q_A$ and $Q_B$, then the black hole horizon area decreases. Since the entanglement entropy is controlled by $\beta$, it follows that the black hole horizon area decreases when $\beta$ increases. Apparently, in the limit where all entanglement is removed between $Q_A$ and $Q_B$ (i.e. where $\beta \rightarrow \infty$), the entanglement entropy goes to zero and so does the horizon area. This limit is depicted in figure \ref{thermopinch} for spatial slices $T=0$ and $T$ equal to a positive constant, where $T$ is the timelike Kruskal-Szekeres coordinate.

The depiction suggests that the spacetime pinches such that it is only contiguous at $T = 0$. Thus, in the limit where $\beta \rightarrow \infty$ in the thermofield double state, the spacetime dual is connected by a single point. This does not exactly reproduce the initial assumption that the product of two pure states, such as $\ket{E_0} \otimes \ket{E_0}$, are dual to two completely disconnected spacetimes. A singularity remains, whose interpretation is uncertain. This, however, should perhaps not worry us too much. Due to the critical minimum temperature for AdS-Schwarzschild black holes below which they cannot exist, it is unclear whether the above account of the limit $\beta \rightarrow \infty$ is correct. When $\beta \rightarrow \infty$, the temperature of the thermal state of $Q_A$ and $Q_B$ goes to zero, i.e. below the critical temperature. We should therefore be careful when taking the limit $\beta \rightarrow \infty$. Despite this complication, it is clear how removing (some) entanglement between the two subsystems of the thermofield double state, $Q_A$ and $Q_B$, decreases the connectivity in the dual eternal black hole spacetime. From this latter observation, van Raamsdonk -- not considering the complication related the minimum temperature -- concludes: ``In this example, \textit{classical connectivity arises by entangling the degrees of freedom in the two components}" \citep[2325, emphasis in original]{van_raamsdonk_building_2010}. Entanglement between $Q_A$ and $Q_B$ is a necessary condition for spacetime connectivity between regions A and B in the eternal black hole. 

\section{Beyond the eternal black hole}\label{Beyond the Black Hole}
While conceived in the context of the duality between the eternal black hole and the thermofield doublestate, the relation between entanglement and spacetime connectivity generalises to any quantum state with a classical spacetime dual.

Suppose again that $\ket{\Psi}$ is a generic quantum state with a classical spacetime dual $M_{\Psi}$ with boundary $\partial M_\Psi$ and $\ket{\Psi}$ is a state in the Hilbert space, $\mathcal{H}_{\partial M_{\Psi}}$, for a CFT defined on (a spatial slice of) $\partial M_\Psi$. Now, divide the boundary $\partial M_\Psi$ into two regions $B$ and $\overline{B}$, such that $B \cup \overline{B} = \partial M_\Psi$ (see figure \ref{cylinder}). Since a CFT is a local quantum field theory, there are specific degrees of freedom associated with specific spatial regions. We can therefore regard the full quantum system as composed of two subsystems, $Q_B$ and $Q_{\overline{B}}$, associated with two spatially separated regions $B$ and $\overline{B}$. As a consequence, we can decompose the Hilbert space of states of the full system as a tensor product of the Hilbert spaces of states of $Q_B$ and $Q_{\overline{B}}$:\footnote{Some complications are involved in making such a decomposition in a gauge invariant way, but these will not be considered here.}
\begin{equation}
\mathcal{H}_{\partial M_\Psi}=\mathcal{H}_B \otimes \mathcal{H}_{\overline{B}}
\end{equation}
$\ket{\Psi}$ can therefore be expressed as a sum over products of states $\ket{\psi^{\overline{B}}_i} \in \mathcal{H}_{\overline{B}}$ and $\ket{\psi^B_i} \in \mathcal{H}_B$:
\begin{equation}
\ket{\Psi}=\sum_{i,j}p_{i,j} \ket{\psi^B_i} \otimes \ket{\psi^{\overline{B}}_j}
\end{equation}
This will generally not be a product state, i.e. a product of a state in $\mathcal{H}_{\overline{B}}$ and $\mathcal{H}_{\overline{B}}$. Thus, the local degrees of freedom in $Q_B$ and $Q_{\overline{B}}$ will generally be entangled.

Again, assume that the spacetime dual of a product state
\begin{equation}
\ket{\Phi} = \left( \sum_i c_i \ket{\psi^B_i} \right) \otimes  \left( \sum_j d_j \ket{\psi^{\overline{B}}_j} \right)
\end{equation}
is dual to two disconnected spacetimes. One then obtains the result that entanglement between $Q_B$ and $Q_{\overline{B}}$ in $\ket{\Psi}$ is a necessary condition for the dual spacetime $M_\Psi$ to be a connected spacetime. The duality between the thermofield double state and the eternal black hole is just a particular example of this.

We may monitor this more closely using the Ryu-Takayanagi formula:
\begin{equation}
S_B = \frac{A( \tilde{B})}{4G_N}
\end{equation}
Again, it follows that changing the entanglement between $Q_B$ and $Q_{\overline{B}}$ changes the connectivity between the two corresponding regions in the spacetime dual.

\begin{figure}[h]
\begin{center}
\begin{tikzpicture}
  % Four corners of left diamond
     \draw (-4,0) to[out=90,in=180] (-2,2);
     \draw (-4,0) to[out=-90,in=180] (-2,-2);
     \draw[dotted,thick] (0,0) to[out=90,in=0] (-2,2);
     \draw[dotted, thick] (0,0) to[out=-90,in=0] (-2,-2);
     \node at (-0.2,1.4) {$\overline{B}$};
     \node at (-3.8,-1.4) {$B$};
     \draw[dashed,rounded corners=5mm] (-2,2) -- (-2.3,1) -- (-1.6,0) -- (-2.2,-1.7) -- (-2,-2);
     \node at (-2,0) {$\tilde{B}$};

     \node (V)    at (0.75,0)   {$\stackrel{\mathsmaller{\ket{\Psi} \rightarrow \ket{\Phi}} \; \; \;}{\mathlarger{\mathlarger{\mathlarger{\mathlarger{\mathlarger{\longrightarrow}}}}}}$};
     
     \draw (1.5,0) to[out=90,in=180] (2.5,1.5);
     \draw (2.5,1.5) to[out=0,in=135] (4,0);
     \draw (1.5,0) to[out=-90,in=180] (2.5,-1.5);
     \draw (2.5,-1.5) to[out=0,in=-135] (4,0);
     \draw[dotted,thick] (4,0) to[out=45,in=180] (5.5,1.5);
     \draw[dotted,thick] (5.5,1.5) to[out=0,in=90] (6.5,0);
     \draw[dotted,thick] (4,0) to[out=-45,in=180] (5.5,-1.5);
     \draw[dotted,thick] (5.5,-1.5) to[out=0,in=-90] (6.5,0);
     \node at (6.3,1.4) {$\overline{B}$};
     \node at (1.6,-1.4) {$B$};

\end{tikzpicture}
\caption{\label{genpinch} Depiction of the behavior of the spatial slice $\Sigma_{\partial M_{\Psi}}$ when all entanglement is removed between $Q_B$ and $Q_{\overline{B}}$. Note that the quantum state is defined on a fixed spacetime identical to the asymptotic boundary of the spacetime dual. Thus, the change in $\tilde{B}$ is solely due to changes in the bulk metric despite the appearance to the contrary.}
\end{center}
\end{figure}
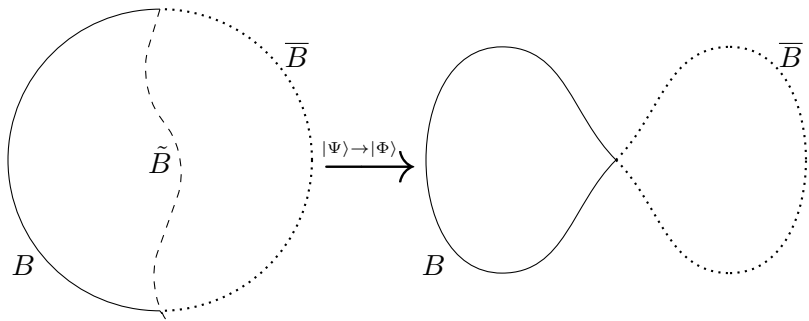

When $\ket{\Psi} \rightarrow \ket{\Phi}$, the state of the full quantum system becomes a product of two pure states such that there is no entanglement between the local degrees of freedom in $Q_B$ and $Q_{\overline{B}}$. Thus, in this limit the entanglement entropy, $S_B$, goes to zero and so does the area of $\tilde{B}$ according to the Ryu-Takayanagi formula. More explicitly stated, in this limit the bulk metric changes such that the minimal area dividing the two asymptotic regions $B$ and $\overline{B}$ in the spacetime goes to zero; the spacetime dual of the quantum state pinches when $\ket{\Psi} \rightarrow \ket{\Phi}$. For the spatial surface $\Sigma_{\partial M_{\Psi}}$, figure \ref{genpinch} depicts the limit where all entanglement is removed between $Q_B$ and $Q_{\overline{B}}$.

Again, the limit where all entanglement is removed between the quantum subsystems does not exactly reproduce the expectation that the tensor product of two pure states is dual to a disconnected spacetime. The two regions $B$ and $\overline{B}$ remain connected by a single singular point that has no clear interpretation. Nevertheless, it evident how entanglement between $Q_B$ and $Q_{\overline{B}}$ is related to the connectivity between $B$ and $\overline{B}$ in the spacetime dual. Further support for this is found in an argument from the mutual information\footnote{Schematically, the mutual information, $I(A,B)$, can be defined as $I(A,B) = S(A) + S(B) - S(A \cup B)$ where $S(M)$ is the entanglement entropy between a region $M$ and the rest of the system \citep{van_raamsdonk_building_2010}.} between a point in $B$ and one in $\overline{B}$ to the effect that the proper distance in the dual spacetime between any two such points goes to infinity when the entanglement between $Q_B$ and $Q_{\overline{B}}$ goes to zero \citep{wolf_area_2008}. As summarised by van Raamsdonk, ``the two regions of spacetime pull apart and pinch off from each other" \citep[2327]{van_raamsdonk_building_2010}. In other words, the conclusion from section \ref{Entanglement and Black Holes} extends even to spacetimes with a contiguous boundary. 

If the AdS side and the CFT side in the AdS/CFT correspondence are regarded as two different ways to encode the same information about a physical system, then the above qualitative investigation of the relation between spacetime and entanglement in the AdS/CFT correspondence offers an indication of how some of this information is encoded in the two different descriptions. The aspect of the physical system described on the AdS side as spacetime geometry is encoded on the CFT side as entanglement structure. It is, however, important to notice that the argument merely demonstrates how entanglement on the CFT side is a necessary condition for spacetime connectivity on the AdS side. Indeed, bulk entanglement corrections to the Ryu-Takayanagi formula entails that entanglement is not sufficient for spacetime connectivity.
\chapter{Linearized Gravity from Entanglement}\label{Entropy and Field Equations}
If the entanglement structure indeed can represent the same physics as bulk spacetime, there must be constraints on entanglement that corresponds to Einstein's field equations. Finding such correspondences would provide more quantitative justification for the claim that the same physical phenomena can indeed be encoded either as entanglement structure on the CFT side or as bulk spacetime. 

In the following, it will be shown that if perturbations of the CFT vacuum state satisfies a particular constraint on entanglement entropy then the spacetime dual of the perturbed state satisfies Einstein's field equations expanded to linear order around pure AdS. Thus, there is a general constraint on entanglement entropy in the CFT state that corresponds to imposing Einstein's field equations in the bulk of the spacetime dual.

The constraint imposed on the CFT side is the following:
\begin{equation}\label{dS=dE}
dS = dE
\end{equation}
where $S$ is the entanglement entropy of a region in a CFT and $E$ is some energy associated with that region. 

(\ref{dS=dE}) is similar to the first law of thermodynamics and a derivation of Einstein equations from this relation sits well with previous attempts to derive these equations from thermodynamics \citep{jacobson_thermodynamics_1995}. However, contrary to the first law of thermodynamics (\ref{dS=dE}) holds for arbitrary perturbations of the vacuum state and not only for thermal/equilibrium states as it is the case for the first law of thermodynamics.

To derive Einstein equation from (\ref{dS=dE}), one must establish a translation between this relation in the holographic CFT to quantities in the spacetime dual. For the entanglement entropy (LHS of (\ref{dS=dE})), this translation is obtained by the Ryu-Takayanagi formula:
\begin{equation}\label{Ryu-Takayanagi}
S = \frac{A(\tilde{B})}{4 G_N}.
\end{equation}
The change in energy (RHS of (\ref{dS=dE})) can be translated as an energy associated with the asymptotic behavior of the metric. This, together with the Ryu-Takayanagi formula allows the translation of (\ref{dS=dE}) holding in the CFT into a relation that must hold in the spacetime dual, which turn out to be the linearized Einstein equations in vacuum.

The material of this chapter is based on \citep{lashkari_gravitational_2014,faulkner_gravitation_2014,van_raamsdonk_lectures_2016}. No new results are derived here, however, a number of details are filled in which are not explicit elsewhere in the literature.

As in section \ref{CFT_d and AdS_{d+1}}, we will consider a one parameter family of CFT states, $\ket{\Psi (\zeta)}$ and assume that $\ket{\Psi (\zeta)}$ lives in a $d$-dimensional Minkowski spacetime, $R^{d-1,1}$, such that $\ket{\Psi (0)}$ has as its spacetime dual the $(d+1)$-dimensional pure Poincarè patch of Anti-de-Sitter spacetime. The metric of the dual spacetime therefore takes the form
\begin{equation}
ds^2 = \frac{L^2}{z^2}(dz^2+dx_{\mu} dx^{\mu} )
\end{equation}
Again, the dual spacetime of the quantum state $\ket{\Psi}$ will be doneted $M_{\Psi}$. Its boundary will be denoted  $\partial M_{\Psi}$ and arbitrary Cauchy surfaces on this boundary will be denoted  $\Sigma_{\partial M_{\Psi}} \subset \partial M_{\Psi}$.

Within this framework, we first want to demonstrate that
\begin{equation}\label{dSA=dEA}
\frac{d}{d\zeta} S_B = \frac{d}{d\zeta} E^{Hyp}_B
\end{equation}
holds for all small perturbations of the CFT vacuum state i.e. for all $\ket{\Psi (\zeta)}$ with small $\zeta$. Here, $S_B$ is the entanglement entropy of $\ket{\Psi (\zeta)}$ for a ball shaped region $B \subset \Sigma_{\partial \mathcal{M}_{\Psi}}$ and $\frac{d}{d\zeta} S_B$ is the variation of this entropy compared to $\ket{\Psi (0)}$. $E^{Hyp}_B$ is the so-called hyperbolic energy of the region $B$ and $\frac{d}{d\zeta} E^{Hyp}_B$ is the variation of this energy with respect to $\ket{\Psi (0)}$.

Secondly, we want to demonstrate that if (\ref{dSA=dEA})  holds for a quantum state then linearized Einstein equations hold in the spacetime dual of this quantum state. To do this, we will provide a holographic interpretation of $S_B$ and $E^{Hyp}_B$ which both translate as integrations over (parts of) $M(\zeta)$ in terms of the metric perturbation $h(x,z)$. Requiring an equality to first order between the holographic expressions of $\left.\frac{d}{d \zeta} \right\vert_{\zeta = 0}S_B$ and $\left.\frac{d}{d \zeta} \right\vert_{\zeta = 0}E^{Hyp}_B$ will then be shown to be equivalent to imposing Einstein's equations to first order on $h(z,x)$.

In the following, it will be assumed that all states, $\ket{\Psi(\zeta)}$, are small perturbation of $\ket{\Psi(0)}$ and that their spacetime duals, $\mathcal{M}(\zeta)$, consequently are small perturbations of $M(0)$, i.e. of the Poincarè patch of pure AdS.
\section{Entropy and the modular hamiltonian}\label{Entropy and the Modular Hamiltonian}
The entanglement entropy of a spatial region in a CFT is defined via the reduced density matrix. Again, we define for some CFT state $\ket{\Psi}$ the reduced density matrix associated with an arbitrary spatial region $B$ as:
\begin{equation}
\rho_B = \tr_{\overline{B}} ( \ket{\Psi} \bra{\Psi} ) 
\end{equation}
where $\overline{B}$ is the compliment of $B$. The von Neumann entropy is
\begin{equation}
S_B = - \tr ( \rho_B \log ( \rho_B ) )
\end{equation}

Now, consider the perturbed state $\ket{\Psi(\zeta)}$ such that $S_B$ is the entanglement entropy in the region $B$. Varying this with respect to $\zeta$, we find (to any order in $\zeta$)
\begin{equation}\label{dSB=dtr}
\begin{split}
\frac{d}{d \zeta} S_B & = \frac{d}{d \zeta} - \tr ( \rho_B \log ( \rho_B ) ) ) \\& = - \tr ( \log ( \rho_B ) \frac{d}{d \zeta} \rho_B ) - \tr( \rho_B \frac{d}{d \zeta}  \log(\rho_B))
\\& = - \tr ( \log ( \rho_B ) \frac{d}{d \zeta} \rho_B ) - \tr( \rho_B \frac{1}{\rho_B} \frac{d}{d \zeta} \rho_B)
\\& = - \tr ( \log ( \rho_B ) \frac{d}{d \zeta} \rho_B ) - \tr(\frac{d}{d \zeta} \rho_B)
\\& = - \tr ( \log ( \rho_B ) \frac{d}{d \zeta} \rho_B )
\end{split}
\end{equation}
where last line follows since the trace of density matrix must be unity.\footnote{It has fixed normalization and therefore $\tr( \frac{d}{d \zeta} \rho_B) = \frac{d}{d \zeta} \tr( \rho_B) = 0$.} Defining the modular Hamiltonian of the region $B$: $H_B \equiv - \log(\rho_B(\zeta=0)  )$ we get
\begin{equation}
\frac{d}{d \zeta} S_B = \tr ( H_B \frac{d}{d \zeta} \rho_B )
\end{equation}
Here $H_B$ is independent of $\zeta$ and therefore we can take the derivative outside the trace
\begin{equation}
\frac{d}{d \zeta} \tr ( H_B \rho_B ) = \frac{d}{d \zeta} \expval{H_B}
\end{equation}
where the equality follows from the definition of the expectation value $\expval{A} = \tr( \rho \, A )$.

Thereby, we obtain the general relation that holds for an arbitrary region $B \subset \Sigma_{\partial M_{\Psi}}$ and for any perturbation:
\begin{equation}\label{dSA=dHA}
\frac{d}{d\zeta} S_B = \frac{d}{d\zeta} \expval{H_B}
\end{equation}

Generally, the modular Hamiltonian, $H_B$, is not a local operator. Thus, it cannot in general be evaluated from the local fields of a quantum field theory. However, there are certain exception where $H_B$ is a local operator. One example is the special case where $B$ is a ball shaped region with radius $R$.\footnote{I.e. a particular spatial, co-dimension two surface of $R^{d-1,1}$.} The domain of dependence for this region is defined as usual as consisting of those points for which all causal curves that passes through the point and also passes through $B$. 

As shown by \citep{casini_towards_2011}, there is a conformal mapping of this domain of dependence for a ball shaped region in Minkowski spacetime to a Rindler wedge of Minkowski spacetime. This is the part of Minkowski spacetime that is accessible for a uniformly boosted observer. Identifying $x=r \cosh( \eta )$ and $t=r \sinh (\eta)$, the Rindler wedge has the metric
\begin{equation}
ds^2 = dr^2 - r^2 d \eta^2
\end{equation}
where $r>0$.

Generally, it holds that the density matrix for a half space (like the Rindler wedge) is thermal for the vacuum state of a Lorentz-invariant quantum field theory.\footnote{Unruh radiation is a well known effect of this.} To see this, consider first the Wick rotated (Euclidean) statement of the path integral for a quantum field theory where $t=i \tau$:
\begin{equation}
\matrixel{\phi_1(\tau_1)}{e^{- \tau H}}{ \phi_0(\tau_0)} = \mathtt{N} \int_{\phi(\tau_0)) = \phi_0}^{\phi(\tau_1)=\phi_1} \mathcal{D}\phi \; e^{- S^{Euc} (\phi(\tau))}
\end{equation}
where $\phi_1$ and $\phi_0$ are respectively final and initial field configurations. This may be identified as density matrix of a thermal state $e^{-\beta H}/Z$ under the assumption $Z = \mathtt{N}$ and $\tau_0 = 0$ and $\tau_1 = \beta$
\begin{equation}
\matrixel{\phi_1}{e^{- \beta H}}{ \phi_0} = \mathtt{N} \int_{\phi(0) = \phi_0}^{\phi(\beta)=\phi_1} \mathcal{D}\phi \; e^{- S^{Euc} (\phi(\tau))}.
\end{equation}
The partition function, i.e the trace of the density matrix $e^{- \beta H}$, can be expressed as
\begin{equation}
Z(\beta) = \tr(e^{- \beta H}) = \int_{\tau} \matrixel{\phi(\tau)}{e^{- \beta H}}{ \phi(\tau)}.
\end{equation}
This is equivalent to a path integral over all periodic paths with period $\beta$ which we express as
\begin{equation}
Z(\beta) = \oint_{\phi(0) = \phi(\beta)} \mathcal{D}\phi \; e^{- S^{Euc} (\phi(\tau))}.
\end{equation}
We can obtain the reduced density matrix for a region $B$ by tracing out the degrees of freedom in the compliment of $B$, $\overline{B}$,
\begin{equation}\label{thermalstate}
\matrixel{\phi^B_1}{e^{- \beta H}}{ \phi^B_0} = \frac{1}{Z} \int_{\phi^B(0) = \phi^B_0}^{\phi^B(\beta)=\phi^B_1} \oint_{\phi^{\overline{B}}(0) = \phi^{\overline{B}}(\beta)} \mathcal{D}\phi \; e^{- S^{Euc} (\phi(\tau))}.
\end{equation}
Now, note that for some state $\ket{\Psi}$,
\begin{equation}
\lim_{\beta \rightarrow \infty} e^{- \beta H} \ket{\Psi} = \mathtt{N} \ket{E_0}
\end{equation}
where $\ket{E_0}$ is the vacuum state. This follows since all energy eigenstates other than the vacuum state are suppressed by a factor $e^{- \beta (E-E_0)}$. This entails that 
\begin{equation}
\braket{\phi^B_0}{\mathtt{N} E_0} = \matrixel{\phi_0}{\lim_{\beta \rightarrow \infty} e^{- \beta H}}{ \Psi} = \mathtt{N} \int_{\phi^B(\infty)}^{\phi^B(0)=\phi^B_0} \mathcal{D}\phi \; e^{- S^{Euc} (\phi(\tau))}
\end{equation}
where we have suppressed the integral over the degrees of freedom in $\overline{B}$. From this, we can obtain an expression for the vacuum density matrix
\begin{equation}
\begin{split}
\matrixel{\phi^B_1}{\rho_B^{vac}}{\phi^B_0} = \braket{\phi^B_1}{\mathtt{N} E_0} \braket{\phi^B_0}{\mathtt{N} E_0}^* = \mathtt{N}^2 \int_{\phi^B(0^+)=\phi^B_0}^{\phi^B(0^-)=\phi^B_1} \mathcal{D}\phi \; e^{- S^{Euc} (\phi(\tau))}
\end{split}
\end{equation}
which follows since
\begin{equation}
\braket{\phi^B_0}{\mathtt{N} E_0}^* = \matrixel{ \Psi}{\lim_{\beta \rightarrow \infty} e^{\beta H}}{\phi_0}.
\end{equation}

Consider the special case where we are interested in the density matrix for the Rindler wedge of the Minkowski vacuum, i.e. a half-space of a particular QFT vacuum state. First, define the region $R$ of Minkowski spacetime by $x^1 > 0$. Thus $R$ is a half-space. We then have
\begin{equation}
\begin{split}
\matrixel{\phi^R_1}{\rho^R_{vac}}{\phi^R_0} = \mathtt{N}^2 \int_{\phi^R(0^+)=\phi^R_0}^{\phi^R(0^-)=\phi^R_1} \mathcal{D}\phi \; e^{- S^{Euc} (\phi(\tau))}
\end{split}
\end{equation}
where $\tau$ is still the wick rotated time coordinate. Change the coordinates to polar coordinates, $x^1 = r \cos(\theta)$ and $\tau = r \sin(\theta)$ such that the metric may be expressed as $ds^2 = dr^2 + r^2 d\theta^2$ (where we have suppressed the remaining direction $x^{i \neq 1}$). We then have
\begin{equation}\label{dmvac}
\begin{split}
\matrixel{\phi^R_1}{\rho_R^{vac}}{\phi^R_0} = \mathtt{N}^2 \int_{\phi^R(\theta = 0)=\phi^R_0}^{\phi^R(\theta = 2 \pi)=\phi^R_1} \mathcal{D}\phi \; e^{- S^{Euc} (\phi(\tau))}.
\end{split}
\end{equation}
Comparing to (\ref{thermalstate}), we identify this as a density matrix over $Z=\mathbb{N}$ with $\beta = 2 \pi$ and Hamiltonian, $H_{\eta}$, that generates evolution in the variable $\eta = i \theta$, i.e.
\begin{equation}\label{dmthermal}
\matrixel{\phi^R_1}{\rho_R^{vac}}{\phi^R_0} = %\mathtt{N}^2 \int_{\phi^R(\theta = 0)=\phi^R_0}^{\phi^R(\theta = 2 \pi)=\phi^R_1} \mathcal{D}\phi \; e^{- S^{Euc} (\phi(\tau))} = \frac{\matrixel{\phi^R_1}{e^{- 2 \pi H_{\eta}}}{\phi^R_0}}{Z}
\end{equation}
In the coordinates, $(r, \eta)$, the metric takes the form $ds^2 = dr^2 - r^2 d\eta^2$ which we recognize as the metric of the Rindler wedge in Minkowski spacetime. $H_{\eta}$, therefore, is the generator of Rindler time (the boost generator of Minkowski spacetime) and $e^{- 2 \pi H_{\eta}}$ is a thermal density matrix of the Rindler wedge. Thus, combining the results (\ref{dmvac}) and (\ref{dmthermal}), we find that reduced vacuum density matrix for the half space $R$ is equal to the thermal density matrix defined in terms of the boost generator with $\beta = 2 \pi$, i.e. \begin{equation}\label{dmvac=dmthermal}
\rho_R^{vac} = \frac{1}{Z} e^{- 2 \pi H_{\eta}}. 
\end{equation}

In the Rindler wedge, $\eta$ is the time coordinate and classical time translations are generated by the vector field
\begin{equation}\label{deta}
d_\eta = x d_t + t d_x.
\end{equation}
In Minkowski space, time translations are simply generated by the vector field $d_t$ and the Hamiltonian, therefore, can be obtained from the energy density operator for Minkowski spacetime, $T_{tt}$, by
\begin{equation}
H = \int d^{d-1} x \, T_{tt}.
\end{equation}
By the same procedure, one can obtain the Hamiltonian for the Rindler wedge, $H_\eta$, however, its expression as a sum of vector fields complicates things. The complication can be avoided by noting that the Hamiltonian must be the same for any Cauchy surface of the spacetime, we can therefore choose $t=0$ such that the second term in (\ref{deta}) is zero. Thereby, $H_\eta$ can be expressed in terms of the full Minkowski spacetime energy density operator $T_{tt}$,
\begin{equation}
H_\eta = \int_{x>0} d^{d-1} x \, [ x T_{tt} ]
\end{equation}
where $x>0$ is required to ensure that we only include the right Rindler wedge. With this, we can obtain an expression for the density matrix for the ball shaped region using the inverse mapping from the Rindler wedge to the domain of dependence for $B$. Since we are considering a CFT state, the energy eigenstates, including the vacuum
state, are invariant under this conformal mapping.  Associating with this mapping the transformation $\mathtt{U}$, on finds
\begin{equation}\label{rhoBH}
\begin{split}
\rho_B & = \mathtt{U}^\dagger \rho_\eta \mathtt{U}
\\&   =  \frac{1}{Z} e^{-2 \pi \mathtt{U}^\dagger H_{\eta} \mathtt{U}}
\end{split}
\end{equation}
where the last equality follows from (\ref{dmvac=dmthermal}). Now define $H_B^{vac} \equiv 2 \pi \mathtt{U}^\dagger H_{\eta} \mathtt{U} - \log(Z)$ such that $\rho_B = e^{- H_B^{vac}}$. From this it immediately follows that $H_B^{vac} = - \log(\rho_B)$ and we may therefore identity $H_B^{vac}$ as the modular Hamiltonian and observe that (\ref{rhoBH}) yields a direct expression for the modular Hamiltonian of a ball shaped region in terms of the known Hamiltonian for the Rindler wedge, $H_\eta$ and the transformation $\mathtt{U}$ from the Rindler wedge to the domain of dependence for a ball shaped region.

Using this, one finds
\begin{equation}\label{HB=T00}
H_B^{vac} = 2 \pi \int_B d^{d-1} x \frac{R^2-(\vec{x}-\vec{x}_0)^2}{2R} T^{tt}(x)
\end{equation}
where, $T^{tt}$ is the energy density operator for the CFT and $(\vec{x}-\vec{x}_0)^2$ serves as a squared radial coordinate centred in the center of the ball. For further details, see \citep{faulkner_gravitation_2014,casini_towards_2011}. 

We can then obtain the sought expression for the first order variation in $\zeta$ away from vacuum of the expectation value of modular Hamiltonian
\begin{equation}\label{dHB=dT00}
\left.\frac{d}{d \zeta} \right\vert_{\zeta = 0} \expval{H_B} = 2 \pi \int_B d^{d-1} x \frac{R^2-(\vec{x}-x_0)^2}{2 R} \left.\frac{d}{d \zeta} \right\vert_{\zeta = 0} \expval{T^{tt}(x)}
\end{equation}

Finally, making an identification between the expectation value of the modular Hamiltonian and the hyperbolic energy, $\expval{H_B} \equiv E_B^{hyp}$, the relation \begin{equation}
\frac{d}{d\zeta} S_B = \frac{d}{d\zeta} E^{Hyp}_B
\end{equation} is obtained by substitution into (\ref{dSA=dHA}).
\section{Holographic interpretation of $\delta S_B$}
To see what corresponds to 
\begin{equation}
\frac{d}{d\zeta} S_B = \frac{d}{d\zeta} E^{Hyp}_B
\end{equation}
in the spacetime dual, we must find an interpretation of $\frac{d}{d\zeta} S_B$ and $\frac{d}{d\zeta} E^{Hyp}_B$ on the AdS side.

A holographic interpretation of entanglement entropy is already provided by the Ryu-Takayanagi formula (\ref{Ryu-Takayanagi}). This involves the evaluation of the area of a minimal co-dimension two surface in the spacetime bulk that divides the boundary region $B$ and its compliment.

Generally, one can parametrize a co-dimension two surface as $X^{a}(\sigma)$ where $X$ is an embedding function. The area functional, $A(g, X)$, for such surfaces can then be expressed as
\begin{equation}\label{AgX}
A(g, X) = \int d^{d-1} \sigma \sqrt{\det(\gamma_{\mu \nu})}
\end{equation}
where $\det(\gamma_{\mu \nu})$ is the determinant of the metric induced on this surface by the embedding function $X^a$ such that
\begin{equation}
\gamma_{\mu \nu} = g_{a b}(x) \frac{\partial X^a}{\partial \sigma^\mu} \frac{\partial X^b}{\partial \sigma^\nu}
\end{equation}

We will use the notation $A(g, X_{ext} )$ for the extremal of the area functional, where $X_{ext}^{a}(\sigma)$ is the embedding function which extremizes the area.

Now, consider a small perturbation of the metric $g_{a b}(x)$
\begin{equation}\label{g=g0+dg}
g_{ab} = g_{ab}^0 + \delta g_{ab}
\end{equation}
that will entail a small variation of the spacetime. In the original spacetime the area, $A$, was extremized by $X^0_{ext}$ but with the variation in spacetime due to the variation in $g$ another surface with embedding function, $X_{ext}$, will extremize $A$.

The variation in extremal surface area is then
\begin{equation}\label{dAgX}
\delta A(g, X_{ext}) = \frac{\delta A(g, X_{ext}^0)}{\delta g} \delta g + \frac{\delta A(g^0, X_{ext})}{\delta X} \frac{ \delta X}{\delta g} \delta g
\end{equation}
Note that $X_{ext}^0$ extremizes the functional $A(g^0, X)$. This entails that 
\begin{equation}
A(g^0 , X_{ext} ) = A(g^0 , X_{ext}^0 + \delta X) = A(g^0 , X_{ext}^0) + \order{\delta X^2}
\end{equation}
i.e. the first order variations of the unperturbed extremal surface with respect to variation in the embedding function vanishes.\footnote{For an extremum, the first order variation always vanishes.} The second term in (\ref{dAgX}) can therefore be rewritten as
\begin{equation}
\frac{\delta A(g^0, X_{ext})}{\delta X} = \frac{\delta A(g^0, X_{ext}^0)}{\delta X} + \frac{\order{\delta X^2}}{\delta X} = \order{ \delta X}
\end{equation}
We then observe that $\delta X$ is of order $\delta g$ from which it follows
\begin{equation}
\order{ \delta X} \frac{ \delta X}{\delta g} \delta g = \order{\delta g^2}
\end{equation}
We therefore find that
\begin{equation}
\delta A(g, X_{ext}) = \frac{\delta A(g, X_{ext}^0)}{\delta g} \delta g + \order{\delta g^2}
\end{equation}
Thus, to linear order in $\delta g$, the variation in $A(g, X_{ext})$ depends on the variation in $g$ while the embedding function, $X$, is held fixed.

The variation of $A(g, X_{ext})$ is controlled by the variation in the induced metric. To first order in $\delta g$, the variation in the induced metric is proportional to the variation in the metric since the first order variation of the embedding function vanishes. Thus, reintroducing the perturbation parameter $\zeta$, we can express the perturbation of the induced metric as
\begin{equation}
\left.\frac{d}{d \zeta} \right\vert_{\zeta = 0} \gamma(\zeta) = \delta \gamma
\end{equation}
where we have suppressed the tensor notation and the dependence of $\gamma$ on $x$. 

The variation to first order in $\zeta$ (i.e. to first order in $\delta g$) of the square-root of the induced metric, $\det(\gamma(\zeta))$, entering in the area functional (\ref{AgX}) may then be expressed as
\begin{equation}
\left.\frac{d}{d \zeta} \right\vert_{\zeta = 0} \sqrt{\det(\gamma(\zeta))} = \frac{1}{2 \, \sqrt{\det(\gamma( \zeta))}} \left.\frac{d}{d \zeta} \right\vert_{\zeta = 0} e^{ \tr ( \ln ( \gamma( \zeta)))} 
\end{equation}
where we have used the identity $\det(M) = \exp(\tr(\ln(M)))$. 
The derivative becomes
\begin{equation}
\begin{split}
\left.\frac{d}{d \zeta} \right\vert_{\zeta = 0} e^{ \tr ( \ln ( \gamma( \zeta)))} & = \det(\gamma(0)) \left.\frac{d}{d \zeta} \right\vert_{\zeta = 0} \tr(\ln(\gamma( \zeta))) \\ & = \det(\gamma(0)) \tr( \left.\frac{d}{d \zeta} \right\vert_{\zeta = 0} \ln(\gamma( \zeta))) \\& = \det(\gamma(0)) \tr( \gamma(0) \left.\frac{d}{d \zeta} \right\vert_{\zeta = 0} \gamma( \zeta)) \\ & = \det(\gamma(0)) \tr( \gamma(0) \delta \gamma) \\& = \det(\gamma_{\mu \nu}^0) \gamma^{\rho \lambda}_0 \delta \gamma_{\rho \lambda} 
\end{split}
\end{equation}
where the last line reintroduces the tensor notation and replaces the trace with a sum over repeated indices. Thus we find the following expression for the variation of the square root 
\begin{equation}
\left.\frac{d}{d \zeta} \right\vert_{\zeta = 0} \sqrt{\det(\gamma(\zeta))}= \frac{1}{2 \, \sqrt{\det(\gamma_{\mu \nu}^0)}} \det(\gamma_{\mu \nu}^0) \gamma^{\rho \lambda}_0 \delta \gamma_{\rho \lambda} = \frac{1}{2} \sqrt{\det(\gamma_{\mu \nu}^0)} \gamma^{\rho \lambda}_0 \delta \gamma_{\rho \lambda}
\end{equation}
We can then insert this into the area functional, and get an explicit expression of the variation of the extremum of the area functional due to a first order variation in the metric
\begin{equation}\label{dA}
\delta A(g, X_{ext}) = \int d^{d-1} \sigma \frac{1}{2} \sqrt{\det(\gamma_{\mu \nu}^0)} \gamma^{\rho \lambda}_0 \delta \gamma_{\rho \lambda}.
\end{equation}
\\

Now, we want again to consider the particular example where the perturbed metric $g_{ab}$ is given by
\begin{equation}
g_{ab} = g_{ab}^{AdS} + z^{d-2} h_{ab}.
\end{equation}
The embedding function $X_{ext}$ of interest is the mapping from the co-dimension two surface, $\tilde{B}$, to the spacetime with metric $g_{ab}$, which extremizes the area of $\tilde{B}$, whose boundery at $z=0$ is a ball shaped region, $B \subset \Sigma_{\partial M}$, with radius $R$ and which is defined on the constant time slice $t=0$. In pure AdS, the bulk surface that extremizes the area of $\tilde{B}$ for a ball shaped boundary is
\begin{equation}\label{x2+z2=R2}
\vec{x}^2+z^2 = x_i x^i + z^2 = R^2
\end{equation}
where $x_i$ are the spatial Minkowski coordinates, i.e. the  coordinates on the boundary $\Sigma \partial M$. 

Parametrizing this extremal surface using these boundary coordinates, such that $(t,x^i)=\sigma^\mu$, one finds the embedding function
\begin{equation}
X_{ext}^{a}(t,x^ii): \mathbb{R}^{d-1} \rightarrow \mathbb{R}^{d+1}, (t,x^i) \mapsto \left\{
  \begin{array}{lr}
    X^0 = t_0 \\   
    X^1=x^1\\
    \vdots \\
    X^{d-1}=x^{d-1}\\
    X^{d}=Z=\sqrt{R^2-\vec{x}^2}
    \end{array}
\right\}
\end{equation}
Since we have chosen a constant time slice, we may disregard the time coordinate and consider only the spatial coordinates, $x^i$. The unperturbed induced metric, $\gamma_{ij}^0$, then takes the form
\begin{equation}
\gamma_{ij}^0 = g_{ab}^0 \frac{\partial X_{ext}^{a}}{\partial x^i} \frac{\partial X_{ext}^{b}}{\partial x^j} = \frac{L^2}{z^2} \sum_{k=1}^{d-1} \left(\frac{\partial X_{ext}^{k}}{\partial x^i} \frac{\partial X_{ext}^{k}}{\partial x^j}+\frac{\partial Z}{\partial x^i} \frac{\partial Z}{\partial x^j} \right).
\end{equation}
where the last equality follows since $\partial t_0/\partial t = 0$ and $g_{ab}^0 = g_{ab}^{AdS}$ and therefore is diagonal with Lorentzian signature and elements $L^2/z^2$. Now
\begin{equation}
\sum_{k=1}^{d-1} \frac{\partial X_{ext}^{k}}{\partial x^i} \frac{\partial X_{ext}^{k}}{\partial x^j} = \delta_{ij}
\end{equation}
and
\begin{equation}
\frac{\partial Z}{\partial x^i} \frac{\partial Z}{\partial x^j} = \frac{x_i x_j}{z^2}
\end{equation}
so we find that
\begin{equation}
\gamma_{ij}^0 = \frac{L^2}{z^2} \left( \delta_{ij} + \frac{x_i x_j}{z^2} \right)
\end{equation}
and 
\begin{equation}
\gamma^{ij}_0 = \frac{z^2}{L^2} \left( \delta^{ij} - \frac{x^i x^j}{R^2} \right).
\end{equation}
This can be verified using the relation
\begin{equation}\label{x2/z2-1=R2/x2}
\vec{x}^2 + z^2 = R^2 \Rightarrow \frac{\vec{x}^2}{z^2}+1 = \frac{R^2}{z^2}
\end{equation}
and the identity $\gamma_{ij} \gamma^{jk} = \delta_i^k$:
\begin{equation}
\begin{split}
\gamma_{ij}^0 \gamma^{jk}_0 & = \frac{L^2}{z^2} \left( \delta_{ij} + \frac{x_i x_j}{z^2} \right) \frac{z^2}{L^2} \left( \delta^{jk} - \frac{x^j x^k}{R^2} \right)
\\& = \delta_{ij} \delta^{jk} + \delta^{jk} \frac{x_i x_j}{z^2} - \delta_{ij} \frac{x^j x^k}{R^2} - \frac{x_i x_j}{z^2} \frac{x^j x^k}{R^2}
\\& = \delta_{i}^{k} + \frac{x_i x^k}{z^2} - \frac{x_i x^k}{R^2} - \frac{x_i x_j}{z^2} \frac{x^j x^k}{R^2}
\\& = \delta_{i}^{k} + \frac{x_i x^k}{z^2} - \frac{x_i x^k}{R^2} - \left( \frac{R^2}{z^2} - 1 \right) \frac{x_i x^k}{R^2}
\\& = \delta_{i}^{k} + \frac{x_i x^k}{z^2} - \frac{x_i x^k}{R^2} - \frac{x_i x^k}{z^2} - \frac{x_i x^k}{R^2}
\\& = \delta_{i}^{k}
\end{split}
\end{equation}

The square root of the determinant of the induced metric, $\sqrt{\gamma^0}$, becomes
\begin{equation}
\sqrt{\gamma^0} = \frac{L^{d-1}}{z^{d-1}} \sqrt{1 + \frac{\vec{x}^2}{z^2}} = \frac{L^{d-1}}{z^{d}} R
\end{equation}
where the last equality follows from (\ref{x2/z2-1=R2/x2}).
The first order variation of the induced metric, $\delta \gamma_{ij}$, is then given as
\begin{equation}
\begin{split}
\delta \gamma_{ij} & =  \delta g_{ab} \frac{\partial X_{ext}^{a}}{\partial x^i} \frac{\partial X_{ext}^{b}}{\partial x^j} = z^{d-2} h_{ab} \frac{\partial X_{ext}^{a}}{\partial x^i} \frac{\partial X_{ext}^{b}}{\partial x^j}
\\& = z^{d-2} \left( h_{mn} \frac{\partial X_{ext}^{m}}{\partial x^i} \frac{\partial X_{ext}^{n}}{\partial x^j} + h_{z \mu} \left( \ldots \right) + h_{z z} \left( \ldots \right) \right)
\\& = z^{d-2} h_{mn} \, \delta_{im} \, \delta_{jn} = z^{d-2} h_{ij}
\end{split}
\end{equation}
where we have used the assumption (radial gauge) $h_{z \mu}=h_{zz}=0$ and that the embedding function, $X_{ext}^{a}(x_i)$, is invariant under the first order variation of the the metric.

All of this can then be combined to obtain an expression for the first order variation around pure AdS in the area of the extremal surface with a ball shaped boundary
\begin{equation}
\begin{split}
\delta A_{AdS}^{B} & = \int d^{d-1} \sigma \frac{1}{2} \sqrt{\det(\gamma_{\mu \nu}^0)} \gamma^{\rho \lambda}_0 \delta \gamma_{\rho \lambda}
\\& = \int d^{d-1} x \frac{1}{2} \left[ \frac{L^{d-1}}{z^{d}} R \right] \left[ \frac{z^2}{L^2} \left( \delta^{ij} - \frac{x^i x^j}{R^2} \right) \right] \left[z^{d-2} h_{ij}\right]
\\& = \frac{L^{d-3} R}{2} \int d^{d-1} x \left( \delta^{ij} - \frac{x^i x^j}{R^2} \right) h_{ij}
\\& = \frac{L^{d-3} R}{2} \int_{\abs{\vec{x}-\vec{x_0}} \leq R} d^{d-1} x \, \left( \delta^{ij} - \frac{1}{R^2} (x^i-x_0^i)(x^j - x_0^j) \right) h_{ij}
\end{split}
\end{equation}
where the last line assumes that the center of the ball shaped region is $\vec{x_0}$. Using the Ryu-Takayanagi formula (\ref{RyuTakayanagi}) we find
\begin{equation}\label{dSB}
\begin{split}
\delta S_B = \frac{\delta A_{ads}}{4 G_N} & = \frac{L^{d-3}R}{8 G_N} \int_{\abs{\vec{x}-\vec{x}_0} \leq R} d^{d-1} x \, (\delta^{ij} - \frac{1}{R^2} (x^i-x_0^i)(x^j - x_0^j)) h_{ij}
\\& = \frac{L^{d-3}}{8 G_N R} \int_{\abs{\vec{x}-\vec{x}_0} \leq R} d^{d-1} x \, (R^2 \delta^{ij} - (x^i-x_0^i)(x^j - x_0^j)) h_{ij}
\end{split}
\end{equation}
where
\begin{equation}
\delta S_B \equiv \left.\frac{d}{d \zeta} \right\vert_{\zeta = 0} S_{B}.
\end{equation}
Thus, for CFT state $\ket{\Psi}$ that has pure AdS as its spacetime dual, a small variation in the entanglement entropy, $\delta S_B$, of a ball shaped region $B$ is proportional to the change in area of the extremal surface $\tilde{B}$, $\delta A_{AdS}$, due to the corresponding perturbation of the AdS metric.

\section{Holographic interpretation of $\delta E^{Hyp}_B$}\label{Holographic interpretation of dE}
From 
\begin{equation}
H_B^{vac} = 2 \pi \int_B d^{d-1} x \frac{R^2-(\vec{x}-\vec{x}_0)^2}{2R} T^{tt}(x)
\end{equation}
we see that a holographic interpretation of the hyperbolic energy, $E_B^{hyp}$, merely requires an interpretation of the energy density operator, $T^{tt}$, for the CFT.

To find this holographic interpretation, consider first an infinitesimal ball shaped region, $B_{inf}$, on the boundary $\Sigma_{\partial M}$ centred at $x_0$. For such a infinitesimal ball, the expectation value of the energy density operator can be assumed to be the same throughout the ball. To leading order, we can therefore replace the function $\left.\frac{d}{d \zeta} \right\vert_{\zeta = 0} \expval{T^{tt}(x)}$ with its central value $\left.\frac{d}{d \zeta} \right\vert_{\zeta = 0} \expval{T^{tt}(x_0)}$, where $x_0$ is the center. The variation in hyperbolic energy for such an infinitesimal ball, $\left.\frac{d}{d \zeta} \right\vert_{\zeta = 0} E_{B_{inf}}^{hyp}$, is given by (\ref{dHB=dT00}) when taking the limit $R \rightarrow 0$
\begin{equation}\label{ET}
\begin{split}
\left.\frac{d}{d \zeta} \right\vert_{\zeta = 0} E_{B_{inf}}^{hyp} & = \lim_{R \rightarrow 0} \bigg\lbrace 2 \pi \int_B d^{d-1} x \frac{R^2-(\vec{x}-\vec{x}_0)^2}{2R} \left.\frac{d}{d \zeta} \right\vert_{\zeta = 0} \expval{T^{tt}(x)} \bigg\rbrace
\\& = \pi \left.\frac{d}{d \zeta} \right\vert_{\zeta = 0} \expval{T^{tt}(x_0)} \int_{\abs{\vec{x}-\vec{x}_0} \leq R} d^{d-1} x \frac{R^2-(\vec{x}-\vec{x}_0)^2}{R} 
\\& = \frac{2 \pi R^d S_{d-2}}{d^2-1} \left.\frac{d}{d \zeta} \right\vert_{\zeta = 0} \expval{T^{tt}(x_0)}
\end{split}
\end{equation}
where $S_{d-2}$ is the area of the unit ball of dimension $d-2$ (surface area of the unit sphere in dimension $d-1$). From (\ref{dSA=dEA}) it then follows that 
\begin{equation}
\begin{split}
& \left.\frac{d}{d \zeta} \right\vert_{\zeta = 0} E_{B_{inf}}^{hyp} = \left.\frac{d}{d \zeta} \right\vert_{\zeta = 0} S_{B_{inf}} = \lim_{R \rightarrow 0} \delta S_B
\\& \Rightarrow \frac{2 \pi R^d S_{d-2}}{d^2-1} \left.\frac{d}{d \zeta} \right\vert_{\zeta = 0} \expval{T^{tt}(x_0)} = \lim_{R \rightarrow 0} \delta S_B
\\& \Rightarrow \left.\frac{d}{d \zeta} \right\vert_{\zeta = 0} \expval{T^{tt}(x_0)} = \frac{d^2-1}{2\pi S_{d-2}}\lim_{R \rightarrow 0} \left( \frac{1}{R^d} \delta S_B \right).
\end{split}
\end{equation}

Inserting the holographic interpretation for $\delta S_B$ into (\ref{dSB}) we find
\begin{equation}\label{TS}
\begin{split}
& \left.\frac{d}{d \zeta} \right\vert_{\zeta = 0} \expval{T^{tt}(x_0)} \\& \; = \frac{d^2-1}{2\pi S_{d-2}}\lim_{R \rightarrow 0} \left( \frac{1}{R^d} \frac{L^{d-3} R}{8 G_N} \int_{\abs{\vec{x}-\vec{x_0}} \leq R} d^{d-1} x \, (\delta^{ij} - \frac{1}{R^2} (x^i-x_0^i)(x^j - x_0^j)) h_{ij} \right)
\end{split}
\end{equation}
To investigate the limit $R \rightarrow 0$, it is convenient to define coordinates that are constant in this limit
\begin{equation}
\hat{x}=\frac{x^i - x_0^i}{R}, \; \hat{z}=\frac{z}{R}.
\end{equation}
With these coordinates we find the expression
\begin{equation}
\delta S_B = \frac{L^{d-3} R}{8 G_N} \int_{\vec{\hat{x}}^2 \leq 1} d^{d-1} \hat{x} R^{d-1} \, (\delta^{ij} - \hat{x}^i \hat{x}^j) h_{ij}(x,z)
\end{equation}
and
\begin{equation}
\left.\frac{d}{d \zeta} \right\vert_{\zeta = 0} \expval{T^{tt}(x_0)} = \frac{d^2-1}{2 \pi S_{d-2}} \frac{L^{d-3}}{8 G_N} \int_{\vec{\hat{x}}^2 \leq 1} d^{d-1} \hat{x} \, (\delta^{ij} - \hat{x}^i \hat{x}^j) h_{ij}(x,z).
\end{equation}
Since we consider the limit $R \rightarrow 0$, we assume that the metric perturbation is the same throughout the region enclosed by $\tilde{B}$, i.e. $h_{ij}(x,z) \vert_{R \rightarrow 0} = h_{ij}(x_0,z=0) \equiv h_{ij}$.
The integrals are over symmetric intervals and therefore all terms where $i \neq j$ will not contribute since
\begin{equation}
\int_{-a}^a dx \, \int_{-b}^b dy \; (\delta_{xy} - x y) h_{xy}(x_0,z=0) = h_{xy} \left(a^2-(-a)^2 \right) \left( b^2 - (-b)^2 \right) = 0.
\end{equation}
We therefore have
\begin{equation}
\left.\frac{d}{d \zeta} \right\vert_{\zeta = 0} \expval{T^{tt}(x_0)} = \frac{d^2-1}{2\pi S_{d-2}} \frac{L^{d-3}}{8 G_N}  h_{ii} \int_{\vec{\hat{x}}^2 \leq 1} d^{d-1} \hat{x} \, (1 - \hat{x}^i \hat{x}^i).
\end{equation}
Since there is nothing particular about the numbering of the coordinates $x^i$, the $d-1$ integrals above must be equal
\begin{equation}
\int_{\vec{\hat{x}}^2 \leq 1} d^{d-1} \hat{x} \, (1 - \hat{x}^1 \hat{x}^1) = \dots = \int_{\vec{\hat{x}}^2 \leq 1} d^{d-1} \hat{x} \, (1 - \hat{x}^{d-1} \hat{x}^{d-1}) = \frac{S_{d-2} \, d }{d^2-1}.
\end{equation}
One thereby finds\footnote{This is a rather remarkable result in its own right. Assuming only the Ryu-Takayanagi formula, we have derived the known relation \citep{balasubramanian_stress_1999,skenderis_asymptotically_2001} between the first order variation of the expectation value of the CFT stress energy tensor and the bulk metric perturbation $h$.}
\begin{equation}
\begin{split}
\left.\frac{d}{d \zeta} \right\vert_{\zeta = 0} \expval{T^{tt}(x_0)} & = \frac{d^2-1}{2\pi S_{d-2}} \frac{L^{d-3}}{8 G_N}  \left( \frac{S_{d-2} \, d }{d^2-1} h^i_{\; i} \right)
\\& = \frac{L^{d-3} \, d}{16 \pi G_N} \delta^{ij} h_{ij}
\end{split}.
\end{equation}
Now nothing was assumed for the point $x_0$, so this holds for all $x$. We can therefore substitute the first order variation of energy density, $T^{tt}(x)$, in the CFT state with the metric perturbation at the boundary, $\delta^{ij} h_{ij}(x)$ times some constant
\begin{equation}
\begin{split}
\left.\frac{d}{d \zeta} \right\vert_{\zeta = 0} E_{B}^{Hyp} & = 2 \pi \int_B d^{d-1} x \frac{R^2-(\vec{x}-\vec{x}_0)^2}{2R} \left.\frac{d}{d \zeta} \right\vert_{\zeta = 0} \expval{T^{tt}(x)} \\&
= \frac{L^{d-3} \, d}{16 G_N} \int_B d^{d-1} x \frac{R^2-(\vec{x}-\vec{x}_0)^2}{R} \delta^{ij} h_{ij}(x,z=0).
\end{split}
\end{equation}
This, thereby, provides a holographic interpretation of the hyperbolic energy for a ball shaped region on $\Sigma M_{\Psi}$. Below, we will define
\begin{equation}
\left.\frac{d}{d \zeta} \right\vert_{\zeta = 0} E_{B}^{Hyp} \equiv \delta E^{Hyp}_B
\end{equation}

\section{Combining the holographic results}\label{Combining the holographic results}
In the two previous section we found holographic interpretations (and expressions) for $\delta S_B$ and $\delta E_B^{hyp}$. The relation $\delta S_B = \delta E_B^{hyp}$ holds for any CFT quantum state. It will turn out that if and only if this equality holds for all disks in all Lorentz frames then the bulk metric satisfies the Einstein's field equations to linear order in $h$.

The two holographic interpretations are
\begin{equation}
\delta E^{Hyp}_B = \frac{L^{d-3} \, d}{16 G_N R} \int_B d^{d-1} x (R^2-(\vec{x}-\vec{x}_0)^2) \delta^{ij} h_{ij}(x,z=0) \equiv \delta E^{Grav}_B
\end{equation}
and
\begin{equation}
\delta S_B = \frac{L^{d-3}}{8 G_N R} \int_{\tilde{B}} d^{d-1} x \left( R^2 \delta^{ij} - (x^i x^j) \right) h_{ij} \equiv \delta S^{Grav}_B
\end{equation}
where we have introduced the shorthand $\delta E^{Grav}_B$ and $\delta S^{Grav}_B$ for the holographic interpretation of $\delta E^{Hyp}_B$ and $\delta S_B$, respectively.
Combining these and using the relation $\delta S_B = \delta E_B^{hyp}$, we find
\begin{equation}\label{dEG=dSG}
\begin{split}
& \delta E^{Grav}_B  = \delta S^{Grav}_B
\\& \Rightarrow 2 d \int_B d^{d-1} x \left( R^2-(\vec{x}-\vec{x}_0)^2 \right) \delta^{ij} h_ {ij}(x,z=0) = \int_{\tilde{B}} d^{d-1} x \left( R^2 \delta^{ij} - x^i x^j \right) h_{ij}(x,z)
\end{split}
\end{equation}
a relation between the metric perturbation on the boundary and in the perturbation in the bulk. This bulk constraint was first shown to be equivalent to linearized Einstein equation in vacuum in \citep{lashkari_gravitational_2014}, however, a more elegant derivation was developed in \citep{faulkner_gravitation_2014} inspired by the Iyer-Wald formalism \citep{iyer_properties_1994}. Only the details of the latter derivation will be recounted here.

In \citep{faulkner_gravitation_2014}, they show that the non-local constraint (\ref{dEG=dSG}) is equal to the linearized Einstein equations in vacuum if there is a form, $\chi_B(h)$, such that (off shell)
\begin{equation}\label{constr.int}
\int_B \chi_B(h) = \delta E^{Grav}_B, \; \; \int_{\tilde{B}} \chi_B(h) = \delta S^{Grav}_B
\end{equation}
and
\begin{equation}\label{dchi}
\text{d}\chi_B(h) = -\frac{1}{8 \pi G_N} \xi_B^a \dot{G}^{E}_{ab} \epsilon^b
\end{equation}
Here, $\text{d}$ is the exterior derivative, $\dot{G}^{E}_{ab}$ are linearized Einstein equations, $\xi_B^a$ is the Killing vector
\begin{equation}\label{xib}
\xi_B = - \frac{2 \pi}{R} (t-t_0) \left[z \partial_z + (x^i - x^i_0) \partial_i \right] + \frac{\pi}{R} \left[ R^2 - z^2 - (x^i - x^i_0)^2 - (t-t_0)^2 \right] \partial_t.
\end{equation}
We see that all but the time component vanishes at time, $t=t_0$. Further, on the extremal surface $\tilde{B}$ that satisfies $(\vec{x} - \vec{x}_0)^2 + z^2 = R^2$ and $t=t_0$ also the time component, $\xi^t_B$, vanishes. Thus, $\tilde{B}$ is a bifurcation surface of the killing vector $\xi_B$. Finally, $\epsilon^b$ is the volume form of co-dimension one surfaces in the bulk
\begin{equation}
\epsilon_b = \frac{1}{d!} \epsilon_{b c_2 \cdots c_{d+1}} dx^{c_2} \wedge \cdots \wedge dx^{c_{d+1}}
\end{equation}
where $\epsilon_{a_1 \cdots a_{d+1}}$ is the antisymmetric tensor in $d+1$ dimensions, for which we define $\epsilon_{z t x^1 \cdots x^{d-1}} = \sqrt{-g}$. For a surface of constant time, we have $d t = 0$ and all component but $\epsilon^t$ vanishes. This follows since all terms of $\epsilon_{b \neq t}$ either include the antisymmetric tensor with repeated indices or wedge products that involve $dt$.

Defining the volume $\Pi$ as the volume bounded by $B \cup \tilde{B} \equiv \partial \Pi$, where $B$ and $\tilde{B}$ as above are spacelike surfaces at $t=t_0$. From that it follows 
\begin{equation}\label{dchiS}
\mathbf{d}\chi_B(h)\vert_{\Pi} = -\frac{1}{8 \pi G_N} \xi_B^t \dot{G}^E_{tt} \epsilon^t.
\end{equation}
Furthermore, we require that 
\begin{equation}\label{dchi0}
\mathbf{d} \chi_B \vert_{\partial M} = 0
\end{equation}
where $\partial M$ is the boundary of AdS which entails that $d \chi_B \vert_{B} = 0$ since $B \subset \partial M$.

Assuming there is such a form, $\chi$, it can then be shown that if the non-local constraint $\delta S^{Grav}_B = \delta E_B^{Grav}$ is satisfied in the bulk, then Einstein's field equations to linear order are satisfied in the bulk, i.e. the constraint implies that $\dot{G}^E_{ab} = 0$. First, relation $\delta S^{Grav}_B = \delta E_B^{Grav}$ implies that
\begin{equation}\label{0=intX}
0 = \delta S^{Grav}_B - \delta E_B^{Grav} = \int_{\tilde{B}} \chi_B - \int_B \chi_B = \int_{B \cup \tilde{B}} \chi_B = \int_{\Pi} \mathbf{d}\chi_B
\end{equation}
where the last equality follows from Stokes theorem. Next, we multiply by $R$, take the derivative with respect to $R$ and expand using Leibniz' integration rule for forms \citep{flanders_differentiation_1973}
\begin{equation}
0=\frac{\partial}{\partial R} \int_{\Pi} R \mathbf{d}\chi_B = \int_{\Pi} \vec{v} \lrcorner (\mathbf{d}(R \mathbf{d} \chi_B)) + \int_{\partial \Pi} \vec{v} \lrcorner (R \mathbf{d}\chi_B) + \int_{\Pi} \frac{\partial}{\partial R} R \mathbf{d}\chi_B
\end{equation}
where $\lrcorner$ is the interior product and $\vec{v}$ is a vector orthogonal to $R$, such that $\vec{v} \lrcorner \mathbf{d}\chi_B(h) = \mathbf{d}\chi_B(h) \cdot \hat{r}$, where $\hat{r}$ is a unit vector in the radial direction. It is an axiom of exterior derivatives that $\mathbf{d}(\mathbf{d} \omega) = 0$ for any differential form $\mathbf{d} \omega$ of a smooth function $\omega$. Therefore, the first term in the expansion vanishes. Thus, inserting the expression (\ref{dchi}) for $\mathbf{d}\chi$ and using (\ref{dchiS}), we find
\begin{equation}
0 = -\frac{1}{8 \pi G_N} \left( \int_{\partial \Pi} R \xi_B^t \dot{G}^E_{tt} \epsilon^t \cdot \hat{r}+ \int_{\Pi} \frac{\partial}{\partial R} R \xi_B^t \dot{G}^E_{tt} \epsilon^t \right).
\end{equation}
The first term vanishes, which can be seen by splitting the surface integral over $\partial \Pi$ into an integral over $B$ and one over $\tilde{B}$ (valid since $\partial \Pi = B \cup \tilde{B}$). The integral over $B$ vanishes due the requirement $\mathbf{d} \chi \vert_{\partial M} = 0$. The integral over $\tilde{B}$ vanishes since $\tilde{B}$ is a bifurcation surface of $\xi_B^t$ ($\xi_B \vert_{\tilde{B}} = 0$). Thus, only the second term above is non-vanishing. The derivative, restricted to the volume $\Pi$ for which $t=t_0$, can be evaluated by inserting the time component of $\xi_B^t$ as given in (\ref{xib})
\begin{equation}
\begin{split}
\frac{\partial}{\partial R} R \xi_B^t \dot{G}^E_{tt} \epsilon^t & = \frac{\partial}{\partial R} R \frac{\pi}{R} \left[ R^2 - z^2 - (\vec{x}-\vec{x_0})^2 \right]  \dot{G}^E_{tt} \epsilon^t = 2 \pi R \dot{G}^E_{tt} \epsilon^t.
\end{split}
\end{equation}
We then finally get
\begin{equation}\label{0=intEe}
0 = -\frac{R}{4 G_N} \int_{\Pi} \dot{G}^E_{tt} \epsilon^t
\end{equation}
which is satisfied only if $\int_{\Pi} \dot{G}^E_{tt} \epsilon^t = 0$.

To see that this implies $\dot{G}^E_{tt} = 0$, observe that (\ref{0=intEe}) is of the form $\int_{\Pi(R)}  d^{d-1} x dz f(z,x)$. Then take the derivative of this with respect to $R$ and use Stokes theorem:
\begin{equation}
0 = \frac{\partial}{\partial R} \int_{\Pi(R)}  d^{d-1} x dz f(z,x) =  \int_{\partial \Pi(R)}  dA f(z,x)
\end{equation}
where $dA$ is the area element on $\partial \Pi(R)$. Expanding this integral, we find
\begin{equation}
0 = \int_{\partial \Pi(R)}  dA f(z,x) = \int_{B}  dA f(z,x) + \int_{\tilde{B}}  dA f(z,x) = \int_{\tilde{B}}  dA f(z,x)
\end{equation}
where the last equality follows from (\ref{dchi0}). Now, we identify $\int_{\tilde{B}}  dA f(z,x)$ as the average moment of the function $f(z,x)$ on $\tilde{B}$ and find that this vanishes. Next, consider the derivative of $\int_{\Pi(R)}  d^{d-1} x dz f(z,x)$ with respect to $x^i$:
\begin{equation}
\begin{split}
0 & = \frac{\partial}{\partial x^i} \int_{\Pi(R)}  d^{d-1} x dz f(z,x) 
\\& = \frac{\partial}{\partial R} \int_{\Pi(R)}  d^{d-1} x dz f(z,x) \frac{\partial R}{\partial x^i} 
\\& = \int_{\partial \Pi(R)}  dA \frac{x^i}{R}f(z,x) 
\\& = \int_{\tilde{B}}  dA \, x^i f(z,x)
\end{split}
\end{equation}
where the last equality follows since $1/R$ only contributes with a constant factor. Again, we used Stokes theorem. We identify this as the first moment of $f(x,z)$ on $\tilde{B}$ in the $x^i$ direction for vanishing average moment. As seen, this also vanishes.

Since the first order moment vanishes, it holds that
\begin{equation}
0 = \int_{\Pi(R)}  d^{d-1} x dz f(z,x) = \int_{\Pi(R)}  d^{d-1} x dz \, x^i f(z,x).
\end{equation}
Again, taking the derivative with respect to $x^i$ gives
\begin{equation}
\begin{split}
0 & = \frac{\partial}{\partial x^i} \int_{\Pi(R)}  d^{d-1} x dz \, x^i f(z,x)
\\& = \frac{\partial}{\partial R} \int_{\Pi(R)}  d^{d-1} x dz \, x^i f(z,x) \frac{\partial R}{\partial x^i}
\\& = \int_{\tilde{B}}  dA \, (x^i)^2 f(z,x).
\end{split}
\end{equation}
This, we recognize as the second moment of $f(x,z)$ on $\tilde{B}$ in the $x^i$ direction for vanishing average moment. Also this vanishes. We can continue this procedure and show that all moments of $f(x,z)$ on $\tilde{B}$ vanishes if $0 = \int_{\partial \Pi(R)}  dA f(z,x)$. Thus, $f(x,z)=0$. 

Going to our particular example $f(z,x) = -\frac{R}{4 G_N} \dot{G}^E_{tt} \epsilon^t$ with 
\begin{equation}
0 = -\frac{R}{4 G_N} \int_{\Pi} \dot{G}^E_{tt} \epsilon^t
\end{equation}
it follows that $\dot{G}^E_{tt} = 0$, since $R$, $G_N$ and $\epsilon^t$ are non-zero.

The argument so far has been carried assuming that the $B$ was a subset of a surface of constant time. Thus, we have assumed a frame of reference with four velocity, $u^{\mu}=(1,\vec{0})$. Going to a general frame of reference, one finds that the condition $\delta S_{B}^{Grav} = \delta E_{B}^{Grav}$ entails 
\begin{equation}\label{uuG=0}
u^\mu u^\nu \dot{G}^E_{\mu \nu} = 0.
\end{equation}
As a crude test, we see immediately that this reproduces the result above for the four velocity $u^{\mu}=(1,\vec{0})$. If we assume that (\ref{uuG=0}) holds for arbitrary four velocities $u^\mu$ then it follows that $\dot{G}^E_{\mu \nu} = 0$. Thus, the linearized Einstein equations are satisfied for the boundary coordinate directions. What remains is to show that $\dot{G}^E_{z \mu} = 0$ and $\dot{G}^E_{zz} = 0$. First note that $\dot{G}^E_{z \mu} = 0$ and $\dot{G}^E_{zz} = 0$ for $z=0$ which follows from (\ref{dchi0}). These are analogous to initial value constraints that must be satisfied by $\dot{G}^E_{ab}$. Consider the general result of the initial value formulation of general relativity that if the spatial components of the Einstein equations are satisfied everywhere, then the initial value constraints given on some spatial slice of the spacetime are satisfied everywhere. This result follows from the Bianchi identity $\nabla^a G^E_{ab} = 0$ \citep[252-267]{wald_general_1984}. However, this result may as well be applied to constraints at the surface $z=0$. If the remaining components, $G^E_{\mu \nu}$, of the Einstein equations are satisfied in all of spacetime as well as the constraints at the surface $z=0$ then these constraints are satisfied in all of spacetime. This entails that if $\dot{G}_{\mu \nu} = 0$ and $\dot{G}^E_{z \mu} = 0$ and $\dot{G}^E_{zz} = 0$ for $z=0$ then $\dot{G}^E_{z \mu} = 0$ and $\dot{G}^E_{zz} = 0$ for all $z$. Thus, if $\delta S^{Grav}_B = \delta E_B^{Grav}$ is satisfied in the bulk and there is a form $\chi_B(h)$ that satisfies the stated constraints then it follows that linearized Einstein equations are satisfied in the bulk. 

\section{The form $\chi_B(h)$}
What remains is to show that there is a form $\chi_B$ for which (\ref{constr.int}), (\ref{chi}), (\ref{dchi}), and (\ref{dchi0}) holds. Let us make the ansatz
\begin{equation}\label{chi}
\chi_B = - \frac{1}{16 \pi G_N} \left[ \left.\frac{d}{d \zeta} \right\vert_{\zeta = 0} \left( \nabla^a \xi^b_B \epsilon_{ab} \right) + \xi^b_B \epsilon_{ab} ( \nabla_c \delta g^{ac} - \nabla^a \delta g^{c}_{\, c} ) \right]
\end{equation}
where $\delta g^c_{\, c} = g^{bc} \delta g_{bc}$ and
\begin{equation}\label{epsilon}
\epsilon_{ab} = \frac{1}{(d-1)!} \epsilon_{a b c_3 \cdots c_{d+1}} dx^{c_3} \wedge \cdots \wedge dx^{c_{d+1}}.
\end{equation}
This entails $\nabla_c \epsilon_{ab} = 0$ since the covariant derivative of the Levi-Civita tensor vanishes.
We may therefore expand the variation of $\nabla^a \xi^b_B \epsilon_{ab}$ with respect to the metric as
\begin{equation}\label{vari}
\begin{split}
\left.\frac{d}{d \zeta} \right\vert_{\zeta = 0} \nabla^a \xi^b_B \epsilon_{ab} & = \left.\frac{d}{d \zeta} \right\vert_{\zeta = 0} \left( g^{ac} \nabla_c \xi^b_B \epsilon_{ab} \right)
\\& = \epsilon_{ab} g^{ac} \left.\frac{d  \nabla_c \xi^b_B }{d \zeta} \right\vert_{\zeta = 0} + g^{ac} \nabla_c \xi^b_B \left.\frac{d  \epsilon_{ab}}{d \zeta} \right\vert_{\zeta = 0}+ \left.\frac{d  g^{ac}}{d \zeta} \right\vert_{\zeta = 0}  \nabla_c \xi^b_B \epsilon_{ab}
\\& = \epsilon_{ab} g^{ac} \dot{\Gamma}_{cd}^b \xi^d_B + g^{ac} \nabla_c \xi^b_B \bar{\epsilon}_{ab} \delta\left( \sqrt{-g} \right) - \delta g^{ac} \nabla_c \xi^b_B \epsilon_{ab}
\\& = \frac{1}{2} \epsilon_{ab} g^{ac} g^{be} \left( \nabla_c \delta g_{de} + \nabla_d \delta g_{ce} - \nabla_e \delta g_{cd} \right) \xi^d_B \\& \; \; \; \; + \frac{1}{2} g^{ac} \epsilon_{ab} g^{de} \delta g_{de} \nabla_c \xi^b_B - \delta g^{ac} \epsilon_{ab} \nabla_c \xi^b_B
\end{split}
\end{equation}
where we have defined $\bar{\epsilon}_{ab} = \epsilon_{ab}/\sqrt{-g}$ and used
\begin{equation}
\bar{\epsilon}_{ab} \delta\left( \sqrt{-g} \right) = \frac{1}{2}\bar{\epsilon}_{ab} \sqrt{-g} g^{cd} \delta g_{cd} = \frac{1}{2} \epsilon_{ab} g^{cd} \delta g_{cd}.
\end{equation}
Also, we have used the expression for the variation of the covariant derivative (\ref{dotGamma})
\begin{equation}
\delta\left( \nabla_c \xi^b_B \right) = \dot{\Gamma}_{cd}^b \xi^d_B = \frac{1}{2} g^{be} \left( \nabla_c \delta g_{de} + \nabla_d \delta g_{ce} - \nabla_e \delta g_{cd} \right)\xi^d_B
\end{equation}
(\ref{vari}) may be reinserted into the expression for $\chi$:
\begin{equation}\label{chiexp}
\begin{split}
\chi_B = - \frac{1}{16 \pi G_N} \bigg[ & \frac{1}{2} \epsilon_{ab} g^{ac} g^{be} \left( \nabla_c \delta g_{de} + \nabla_d \delta g_{ce} - \nabla_e \delta g_{cd} \right) \xi^d_B \\& + \frac{1}{2} g^{ac} \epsilon_{ab} g^{de} \delta g_{de} \nabla_c \xi^b_B - \delta g^{ac} \epsilon_{ab} \nabla_c \xi^b_B  \\& + \xi^b_B \epsilon_{ab} ( \nabla_c \delta g^{ac} - \nabla^a \delta g^{c}_{\, c} ) \bigg]
\end{split}
\end{equation}

We will consider $\chi_B$ restricted to the constant time surface $\Pi$, where all but the time component of $\xi_B$ vanishes. Thus, $\xi_B^t = \frac{\pi}{R} \left[ R^2 - z^2 - (x^i - x^i_0)^2 \right]$ which follows from the expression (\ref{xib}). This entails that $\partial_z \xi^t_B = - \frac{2 \pi z}{R}$ and $\partial_i \xi^t_B = - \delta_{ij} \frac{2 \pi (x^j - x_0^j)}{R}$. One must also use that $\epsilon_{ij}\vert_{\Pi} = \epsilon_{iz}\vert_{\Pi} = \epsilon_{zz}\vert_{\Pi} = 0$ and that $h_tt = \delta^{ij} h_{ij}$. The latter can be verified by requiring that
\begin{equation}
D = g^{ab} g_{ab}
\end{equation}
and 
\begin{equation}
D = (g^{ab} + h^{ab})(g_{ab}+h_{ab}) = g^{ab} g_{ab} + 2 g^{ab} h_{ab} + \order{h^2} \Rightarrow 2 g^{ab} h_{ab} =  - \order{h^2}.
\end{equation}
For small perturbation, the term $\order{h^2}$ vanishes from which it follows that $g^{ab} h_{ab} = 0$. We therefore have 
\begin{equation}
0 = g^{ab} h_{ab} \Rightarrow h_{tt} = \delta^{ij} h_{ij} + h_{zz}
\end{equation}
where $h_{zz} = 0$ assuming radial gauge. Using these, we may evaluate $\chi_B$ on the surface $\Pi$:
\begin{equation}
\begin{split}
\chi_B\vert_{\Pi} = & - \frac{1}{16 \pi G_N} \bigg\lbrace \epsilon_{\; t}^{z} \left[ - \frac{2 \pi z}{R} - \xi^{t}_B \partial_z \right] g^{jk} \delta g_{jk}
\\& \; \; \; + \epsilon_{ \; t}^{i} \left[ - \delta_{il} \frac{2 \pi (x^l - x_0^l)}{R} - \xi^t_B \partial_i \right] g^{jk} \delta g_{jk} - \epsilon_{\; t}^{i} \left[ - \delta_{jl} \frac{2 \pi (x^l - x_0^l)}{R} - \xi^t_B \partial_j \right] g^{jk} \delta g_{ki} \bigg\rbrace.
\end{split} 
\end{equation}

Again, we will consider the case where the metric is approximated by the Fefferman-Graham metric such that $\delta g_{ab} = z^{d-2} h_{ab}$. Using this expression for the metric perturbation and that $g^{ij} = \frac{z^2}{L^2} \delta^{ij}$, we find the final expression for $\chi\vert_{\Pi}$:
\begin{equation}
\begin{split}
\chi_B\vert_{\Pi} & = - \frac{1}{16 \pi G_N} \bigg\lbrace \epsilon_{\; t}^{z} \left[ - \frac{2 \pi z}{R} - \xi^{t}_B \partial_z \right] \delta^{jk} \frac{z^{d}}{L^2} h_{jk}
\\& \; \; \; + \epsilon_{ \; t}^{i} \left[ - \delta_{il} \frac{2 \pi (x^l - x_0^l)}{R} - \xi^t_B \partial_i \right] \delta^{jk} \frac{z^{d}}{L^2} h_{jk} - \epsilon_{\; t}^{i} \left[ - \delta_{jl} \frac{2 \pi (x^l - x_0^l)}{R} - \xi^t_B \partial_j \right] \delta^{jk} \frac{z^{d}}{L^2} h_{ki} \bigg\rbrace
\\& = \frac{z^d}{16 \pi G_N L^2} \bigg\lbrace \epsilon_{\; t}^{z} \left[ \frac{2 \pi z}{R} + \xi^{t}_B \frac{d}{z} + \xi^{t}_B \partial_z \right] \delta^{jk} h_{jk}
\\& \; \; \; + \epsilon_{\; t}^{i} \left[ \delta_{il} \frac{2 \pi (x^l - x_0^l)}{R} + \xi^t_B \partial_i \right] \delta^{jk} h_{jk} - \epsilon_{\; t}^{ i} \left[ \delta_{jl} \frac{2 \pi (x^l - x_0^l)}{R} + \xi^t_B \partial_j \right] \delta^{jk} h_{ki} \bigg\rbrace
\\& = \frac{z^{d+2}}{16 \pi G_N L^4} \bigg\lbrace \epsilon_{zt} \left[ \frac{2 \pi z}{R} + \xi^{t}_B \frac{d}{z} + \xi^{t}_B \partial_z \right] \delta^{jk} h_{jk}
\\& \; \; \; + \epsilon_{it} \left[ \frac{2 \pi (x^i - x_0^i)}{R} + \xi^t_B \partial^i \right] \delta^{jk} h_{jk} - \epsilon_{it} \left[ \frac{2 \pi (x^j + x_0^j)}{R} + \xi^t_B \partial^j \right] \delta^{ik} h_{kj} \bigg\rbrace
\end{split} 
\end{equation}
where we, in the last equality, have ensured that the indices on $\epsilon$ is lowered such that we can use definition (\ref{epsilon}). Note that the index on partial derivatives are still raised and lowered by the Kronecker-Delta, $\partial_i = \delta_{ij} \partial^j$.
\\

With this expression we can now verify that
\begin{align}
\int_{B} \chi_B\vert_{\Pi} = \delta E^{grav}_B
\\ \int_{\tilde{B}} \chi_B\vert_{\Pi} = \delta S^{grav}_B
\\ \chi_B\vert_{\Pi} = -\frac{1}{8 \pi G_N} \xi_B^t \dot{G}^E_{tt} \epsilon^t.
\end{align}
First, consider the integration over $B$. Here, $\epsilon_{it} = 0$ since $dz = 0$ on $B$ which follows since $z=0$ on $B$. Define
\begin{equation}
u \equiv \frac{z^{d+2}}{16 \pi G_N L^4} \bigg( \frac{2 \pi z}{R} + \xi^t_B \frac{d}{z} + \xi^t_B \partial_z \bigg) \delta^{jk} h_{jk}
\end{equation}
and observe that the integral of the $d-1$ form $u \epsilon_{zt}$ is
\begin{equation}\label{intB}
\int_{B} u \epsilon_{zt} = \int_{B} d^{d-1}x \sqrt{\abs{g}} u.
\end{equation}
Inserting this and the expression for $u$ into (\ref{intB}), we find
\begin{equation}
\begin{split}
\int_{B} u \epsilon_{zt} & = \int_{B} d^{d-1}x \frac{L^{d+1}}{z^{d+1}} \frac{z^{d+2}}{16 \pi G_N L^4} \bigg( \frac{2 \pi z}{R} + \xi^t_B \frac{d}{z} + \xi^t_B \partial_z \bigg) \delta^{jk} h_{jk} 
\\& = \frac{L^{d-3}}{16 \pi G_N} \int_{B} d^{d-1}x \; z \left( \frac{2 \pi z}{R} + \xi^t_B \frac{d}{z} + \xi^t_B \partial_z \right) \delta^{jk} h_{jk}
\\& = \frac{L^{d-3} d}{16 \pi G_N} \int_{B} d^{d-1}x \; \xi^t_B \delta^{jk} h_{jk}
\end{split}
\end{equation}
where the last equality follows since $z=0$ on $B$ and $\partial_z \delta^{jk} h_{jk}(x,0) = 0$. Inserting the expression for $\xi^t_B$ at $z=0$ gives
\begin{equation}
\begin{split}
\int_{B} \chi_B\vert_{\Pi} & = \frac{L^{d-3} d}{16 \pi G_N} \int_{B} d^{d-1}x \; \frac{\pi}{R}\left( R^2 - (\vec{x}-\vec{x_0})^2 \right) \delta^{jk} h_{jk} \\& = \frac{L^{d-3}  d}{16 G_N R} \int_{B} d^{d-1}x \; \left( R^2 - (\vec{x}-\vec{x_0})^2 \right) \delta^{jk} h_{jk} = \delta E^{grav}_B.
\end{split}
\end{equation}

Next, consider the integration over $\tilde{B}$. Since $\tilde{B}$ is a bifurcation surface of the killing vector $\xi_B$, it follows that $\xi^t_B = 0$ on $\tilde{B}$. Both $\epsilon_{zt}$ and $\epsilon_{it}$ are non-vanishing, so we have
\begin{equation}
\int_{\tilde{B}} \chi_B\vert_{\Pi} = \int_{\tilde{B}} (u \epsilon_{zt} + v^i \epsilon_{it}).
\end{equation}
Here we define the function $u$ as above but now restricted to the surface $\tilde{B}$:
\begin{equation}
u \equiv \frac{z^{d+2}}{16 \pi G_N L^4} \bigg( \frac{2 \pi z}{R} + \xi^t_B \frac{d}{z} + \xi^t_B \partial_z \bigg) \delta^{jk} h_{jk} = \frac{z^{d+2}}{16 \pi G_N L^4} \frac{2 \pi z}{R} \delta^{jk} h_{jk}
\end{equation}
where the last equality follows since $\xi^t_B = 0$ on $\tilde{B}$. We also define in addition 
\begin{equation}
\begin{split}
v^i & \equiv \frac{z^{d+2}}{16 \pi G_N L^4} \Bigg\lbrace \left[ \frac{2 \pi (x^i - x_0^i)}{R} + \xi^t_B \partial^i \right] \delta^{jk} h_{jk} - \left[ \frac{2 \pi (x^j + x_0^j)}{R} + \xi^t_B \partial^j \right] \delta^{ik} h_{kj} \Bigg\rbrace
\\& = \frac{z^{d+2}}{16 \pi G_N L^4} \left[ \frac{2 \pi (x^i - x^i_0)}{R} \delta^{jk} h_{jk} - \frac{2 \pi (x^j + x^j_0)}{R} \delta^{ik} h_{kj} \right].
\end{split}
\end{equation}
Again, we have
\begin{equation}
\int_{\tilde{B}} u \epsilon_{zt} = \int_{\tilde{B}} d^{d-1}x \sqrt{\abs{g}} u.
\end{equation}
where the integration now is over the surface $\tilde{B}$. For the integration over the $\epsilon_{it}$-part, we have 
\begin{equation}
\begin{split}
\int_{\tilde{B}} v^i \epsilon_{it} & = \int_{\tilde{B}} \sqrt{\abs{g}} v^i dz dx^1 \cdots dx^{i-1} dx^{i+1} \cdots dx^{d-1} 
\\& = - \int_{\tilde{B}} d^{d-1}x \frac{L^{d+1}}{z^{d+1}} v^i \frac{dz}{dx^i} 
\\& = \int_{\tilde{B}} d^{d-1}x \frac{L^{d+1}}{z^{d+1}} v^i \delta_{ji} \frac{(x^j- x_0^j )}{z}
\end{split}
\end{equation}
where we have used the chain rule to replace $dz$ by $dx^i$ and $dz/dx^i = - \delta_{ji} (x^j- x_0^j ) / z$. 

For convenience, we will evaluate the integration over $u \epsilon_{zt}$ and $v^i \epsilon_{it}$ separately:
\begin{equation}
\begin{split}
\int_{\tilde{B}} u \epsilon_{zt} & = \int_{\tilde{B}} \epsilon_{zt} \frac{z^{d+2}}{16 \pi G_N L^4} \frac{2 \pi z}{R} \delta^{jk} h_{jk}
\\& = \int_{\tilde{B}} d^{d-1}x \frac{z^{d+2}}{8 G_N R L^4} \frac{L^{d+1}}{z^{d+1}} z \delta^{jk} h_{jk}
\\& = \frac{L^{d-3}}{8 G_N R} \int_{\tilde{B}} d^{d-1}x z^2 \delta^{jk} h_{jk}.
\end{split}
\end{equation}
and
\begin{equation}
\begin{split}
\int_{\tilde{B}} v^i \epsilon_{it} & = \int_{\tilde{B}} \frac{z^{d+2}}{16 \pi G_N L^4} \left[ \frac{2 \pi (x^i - x^i_0)}{R} \delta^{jk} h_{jk} - \frac{2 \pi (x^j + x^j_0)}{R} \delta^{ik} h_{kj} \right]  \epsilon_{it} 
\\& = - \int_{\tilde{B}} d^{d-1}x \frac{z^{d+2}}{8 G_N R L^4} \frac{L^{d+1}}{z^{d+1}} \frac{dz}{dx^i} \left[ \frac{2 \pi (x^i - x^i_0)}{R} \delta^{jk} h_{jk} - \frac{2 \pi (x^j + x^j_0)}{R} \delta^{ik} h_{kj} \right]
\\& = \frac{L^{d-3}}{8 G_N R} \int_{\tilde{B}} d^{d-1}x \; z \frac{(x^i - x^i_0)}{z} \left[ \frac{2 \pi (x^i - x^i_0)}{R} \delta^{jk} h_{jk} - \frac{2 \pi (x^j + x^j_0)}{R} h_{ij} \right]
\\& = \frac{L^{d-3}}{8 G_N R} \int_{\tilde{B}} d^{d-1}x \left[ (x^i-x_0^i)^2 \delta^{jk} h_{jk} - (x^i-x_0^i) (x^j - x^j_0) h_{ij} \right].
\end{split}
\end{equation}
Putting these elements back together and using $R^2 = z^2 + (\vec{x} - \vec{x_0})^2$, we finally obtain
\begin{equation}
\begin{split}
\int_{\tilde{B}} \chi_B\vert_{\Pi} & = \int_{\tilde{B}} u \epsilon_{zt} + \int_{\tilde{B}} v^i \epsilon_{it}
\\& = \frac{L^{d-3}}{8 G_N R} \int_{\tilde{B}} d^{d-1}x \left[ z^2 \delta^{jk} h_{jk} + (x^i-x_0^i)^2 \delta^{jk} h_{jk} - (x^i-x_0^i) (x^j - x^j_0) h_{ij} \right]
\\& = \frac{L^{d-3}}{8 G_N R} \int_{\tilde{B}} d^{d-1}x \left[ R^2 \delta^{jk} h_{jk} - (x^i-x_0^i) (x^j - x^j_0) h_{ij} \right] = \delta S^{grav}_B.
\end{split}
\end{equation}
\\

Finally, consider the external derivative, $\mathbf{d} \chi\vert_{\Pi}$. Writing $\chi\vert_{\Pi} \equiv u \epsilon_{zt} + v^i \epsilon_{it}$ and using that $\epsilon_{tt} = 0$, we can express the external derivative as
\begin{equation}
\mathbf{d} \chi_B\vert_{\Pi} = \mathbf{d} ( u \epsilon_{zt} + v \epsilon_{it} ) = - \mathbf{d} ( u \sqrt{-g} \bar{\epsilon}_{tz} + v \sqrt{-g} \bar{\epsilon}_{ti} ) = - \left( \frac{\partial (u \sqrt{-g})}{\partial z}+ \frac{\partial(v^i \sqrt{-g})}{\partial x^i} \right) \bar{\epsilon}_t 
\end{equation}
where we have defined $\bar{\epsilon}_{ab} = \epsilon_{ab}/\sqrt{-g}$. The minus sign arises since $t$ must be the first rather than the second index of the Levi-Civita symbol inside $\bar{\epsilon}_t$, and the Levi-Civity is anti-symmetric. Using $\sqrt{-g} =  \frac{L^{d+1}}{z^{d+1}}$ and the expressions for $u$ and $v^i$, we can evaluate the two partial derivatives:
\begin{equation}
\begin{split}
\frac{\partial (u \sqrt{-g})}{\partial z} & = \frac{\partial (u \frac{L^{d+1}}{z^{d+1}})}{\partial z}
\\& = \frac{\partial}{\partial z} \frac{L^{d+1}}{z^{d+1}} \frac{z^{d+2}}{16 \pi G_N L^4} \left( \frac{2 \pi z}{R} + \xi^t_B \frac{d}{z} + \xi^t_B \partial_z \right) \delta^{jk} h_{jk}
\\& = \frac{L^{d-3}}{16 \pi G_N} \bigg( \frac{4 \pi z}{R} + \frac{2 \pi z^2}{R} \partial_z - \frac{2d \pi z}{R} + \xi^t_B d \partial_z + \xi^t_B \partial_z - \frac{2 \pi z^2}{R} \partial_z + z \partial_z^2 \bigg) \delta^{jk} h_{jk}
\\& = \frac{L^{d-3} z}{16 \pi G_N} \bigg( \frac{2 \pi}{R} (2-d) + \frac{\xi^t_B}{z} (d+1) \partial_z + \xi^t_B \partial_z^2 \bigg) \delta^{jk} h_{jk}
\end{split}
\end{equation}
where we have used that $\partial_z \xi^t_B = - 2 \pi z/R$.
The derivative $\frac{\partial(v^i \sqrt{-g})}{\partial x^i}$ becomes
\begin{equation}
\begin{split}
\frac{\partial(v^i \sqrt{-g})}{\partial x^i} & = \frac{\partial(v^i \frac{L^{d+1}}{z^{d+1}})}{\partial x^i} 
\\& = \frac{\partial}{\partial x^i} \frac{L^{d+1}}{z^{d+1}} \frac{z^{d+2}}{16 \pi G_N L^4} \bigg[ \left( \frac{2 \pi (x^i-x_0^i)}{R} + \xi^t_B \partial^i \right) \delta^{jk} h_{jk} 
\\& \; \; \; \; \; \; \; \; \; \; \; \; \; \; \; \; \; \; \; \; \; \; \; \; \; \; \; \; \; \; \; \; \; - \left( \frac{2 \pi (x^j-x_0^j)}{R} + \xi^t_B \partial^{j} \right) \delta^{ik} h_{kj} \bigg]
\\ & = \frac{L^{d-3} z}{16 \pi G_N} \bigg[ \left( \frac{2 \pi}{R} \delta^i_i + \frac{2 \pi (x^i-x_0^i)}{R} \partial_i + (\partial_i \xi^t_B) \partial^i + \xi^t_B \partial_i \partial^i \right) \delta^{jk} h_{jk}
\\& \; \; \; \; - \left( \frac{2 \pi}{R} \delta^i_j + \frac{2 \pi (x^j-x_0^j)}{R} \partial_i + (\partial_i \xi^t_B ) \partial^j + \xi^t_B \partial_i \partial^j \right) \delta^{ik} h_{kj} \bigg]
\\&  = \frac{L^{d-3} z}{16 \pi G_N} \bigg[ \left( \frac{2 \pi}{R} (d-2) + \xi^t_B \partial_i \partial^i \right) \delta^{jk} h_{jk} - \xi^t_B \partial_i \partial^j \delta^{ik} h_{kj} \bigg]
\end{split}
\end{equation}
where we have used that $\partial_i \xi^t_B = - 2 \pi (x^i-x_0^i)/R$ and $\delta^i_i = d-1$.

Collecting the terms, we get
\begin{equation}
\begin{split}
\mathbf{d} \chi_B\vert_{\Pi} & = - \left( \frac{\partial (u \frac{L^{d+1}}{z^{d+1}})}{\partial z}+ \frac{\partial(v \frac{L^{d+1}}{z^{d+1}})}{\partial x^i} \right) \bar{\epsilon}_t
\\& = - \frac{L^{d-3} z}{16 \pi G_N} \bigg[ \bigg( \frac{2 \pi}{R} (2-d) + \frac{\xi^t_B}{z} (d+1) \partial_z + \xi^t_B \partial_z^2 \bigg) \delta^{jk} h_{jk} \\& \; \; \; \; + \left( \frac{2 \pi}{R} (d-2) + \xi^t_B \partial_i \partial^i \right) \delta^{jk} h_{jk} - \xi^t_B \partial_i \partial^j \delta^{ik} h_{kj} \bigg] \bar{\epsilon}_t
\\& = - \frac{L^{d-3} z}{16 \pi G_N} \bigg[ \bigg( \frac{\xi^t_B}{z} (d+1) \partial_z + \xi^t_B \partial_z^2 + \xi^t_B \partial_i \partial^i \bigg) \delta^{jk} h_{jk} \\& \; \; \; \; - \xi^t_B \partial_i \partial^j \delta^{ik} h_{kj} \bigg] \bar{\epsilon}_t.
\end{split}
\end{equation}
To put the exterior derivative on the form $\mathbf{d} \chi_B\vert_{\Pi} = - \frac{1}{8 \pi G_N} \xi^t_B \dot{G}^E_{tt} \epsilon^t$, we have to raise the index of $\bar{\epsilon}_t$ and replace it by $\epsilon^t$
\begin{equation}
\bar{\epsilon}_t = g^{tt} \frac{z^{d+1}}{L^{d+1}} \epsilon^t = - \frac{z^{d-1}}{L^{d-1}} \epsilon^t.
\end{equation}
Inserting this, we get
\begin{equation}
\begin{split}
\mathbf{d} \chi_B\vert_{\Pi} & = - \frac{z^d}{16 \pi G_N L^2}  \xi^t_B \bigg[ - \bigg( \frac{d+1}{z} \partial_z +\partial_z^2 + \partial_i \partial^i \bigg) \delta^{jk} h_{jk} + \partial_i \partial^j \delta^{ik} h_{kj} \bigg] \epsilon^t
\\& = - \frac{1}{8 \pi G_N} \xi^t_B \dot{G}^E_{tt} \epsilon^t
\end{split}
\end{equation}
where 
\begin{equation}
\dot{G}^E_{tt} = \frac{z^d}{2 L^2} \bigg[ - \bigg( \frac{d+1}{z} \partial_z +\partial_z^2 + \partial_i \partial^i \bigg) \delta^{jk} h_{jk} + \partial_i \partial^j \delta^{ik} h_{kj} \bigg]
\end{equation}
in agreement with the result found for the linearized Einstein equations in AdS in section \ref{Perturbing the Poincare Patch}. Thus, we see that the linearized Einstein equations in vacuum follows from the holographic interpretation of the first law of entanglement entropy.

\chapter{Beyond Linearized Gravity}\label{Beyond linearized gravity}
So far, we have on the AdS side considered the limit $G_N \rightarrow 0$ of general relativity. In this limit, the source term appearing in the Einstein equations, $8 \pi G_N T_{ab}$, is suppressed since it is of order $G_N$. Thus, to go beyond vacuum solutions we must include also the first subleading order in $G_N$ in the expansion of the Einstein equations. The limit $G_N \rightarrow 0$ is -- following the identification (\ref{gsGN}) -- equivalent to the limit $g_s \rightarrow 0$. According to the AdS/CFT dictionary, $g_s \propto \frac{1}{N}$ (\ref{gsl/N}) and therefore the first subleading order in $G_N$ corresponds to an expansion on the CFT side that includes the first subleading order in $\frac{1}{N}$. In the previous chapter, the CFT side relation
\begin{equation}
\frac{d}{d\zeta} S_B = \frac{d}{d\zeta} E^{Hyp}_B
\end{equation}
was derived in full generality, i.e. to any order in $\zeta$. This relation therefore holds also to subleading orders in $\frac{1}{N}$.

Our interest here will be to see what this first law of entanglement entropy corresponds to on the AdS side, if we include the first subleading order in $\frac{1}{N}$. We will find that this gives rise to a source term in linearized Einstein gravity exactly corresponding to the source term, $8 \pi G_N T_{ab}$, in the Einstein Equations. This result was first obtained in \citep{swingle_universality_2014} which the present chapter is based on.

\section{$\frac{1}{N}$ corrections to the Ryu-Takayanagi formula}
As already indicated, once we include subleading orders in $\frac{1}{N}$ on the CFT side, we will have to consider matter fields on the AdS side. To first subleading order in $G_N$, we can approximate the AdS side as a semi-classical gravity, i.e. by a classical spacetime $M$ and a quantum matter field living on that spacetime, $\ket{\Phi}_{Bulk}$ such that the dual of some quantum state $\ket{\Psi}$ is the pair $(M, \ket{\Phi}_{Bulk})$. The approximation remains semi-classical since we assume that the spacetime $M$ is a classical spacetime, i.e. it cannot be a superposition of spacetimes, which supposedly would be allowed in a full theory of quantum gravity.

Going to subleading order in $\frac{1}{N}$ or equivalently $G_N$ induces corrections to the Ryu-Takayanagi formula. More precisely, it is conjectured in \citep{faulkner_quantum_2013} that subleading order corrections to the Ryu-Takayanagi formula is due to bulk entanglement over the surface $\tilde{B}$, that is the entanglement between the degrees of freedom in the region $\Pi$ -- the region bounded by $B \cup \tilde{B}$ -- and the rest of the system. The correction takes the form of the entanglement entropy of the quantum field on $\Pi$
\begin{equation}
S_B(\ket{\Psi}) = \frac{A(\tilde{B})}{4 G_N} + S_\Pi (\ket{\Phi}_{Bulk})
\end{equation}
where we have signified that the CFT entanglement entropy, $S_B$, is a function of the CFT state $\ket{\Psi}$ and that he bulk entanglement entropy, $S_\Pi$, is a function of the state of the bulk field, $\ket{\Phi}_{Bulk}$. Just as we motivated the Ryu-Takayangi formula as a generalization of the Bekenstein-Hawking formula, we can motivate this correction to the Ryu-Takayanagi formula by continuing the analogy to the case of black holes where corrections are also supposed to arise due to entanglement over the black hole horizon. It is essentially this result that has been generalized in \citep{faulkner_quantum_2013} to arbitrary surfaces that minimizes the area functional. 

Subsequently, we will simply assume that this corrected Ryu-Takayanagi formula is sound for quantum states with a semi-classical spacetime dual. With this formula, we can -- in analogy to the previous chapter -- obtain holographic interpretations of $S_B$ and $E^{Hyp}_B$. Again, a holographic interpretation of $S_B$ is immediately provided now by the corrected Ryu-Takayanagi formula. The variation $\delta S_B$ therefore consists of the already derived variation of the area of the surface $\tilde{B}$ (\ref{dSB}) and the variation of the bulk entanglement entropy over the surface $\tilde{B}$. The variation of the latter, $\delta S_\Pi (\ket{\Phi}_{Bulk})$, in principle originates in two sources: variation of the surface $\tilde{B}$ and variation of the bulk field $\delta \ket{\Phi}_{Bulk})$. The former, however, vanishes to first order. This follows since entanglement entropy -- just like the area of the minimal surface $\tilde{B}$ -- is extremized by the embedding function $X_{ext}^{a}(x_i)$ and the first order variation always vanishes for such an extremum. What remains, therefore, is to find an expression for the variation of $S_\Pi (\ket{\Phi}_{Bulk})$ due to the variation of the bulk field $\delta \ket{\Phi}_{Bulk})$.

\section{A holographic interpretation of $\delta S_B$ and $\delta E^{Hyp}_B$}
In analogy to the entanglement entropy of a quantum subsystem on the CFT side, the bulk entanglement entropy, $S_\Pi (\ket{\Phi}_{Bulk})$, is given in terms of the reduced density matrix, $\rho^{Bulk}_\Pi$, for the region $\Pi$
\begin{equation}
S_\Pi (\ket{\Phi}_{Bulk}) = - \tr( \rho^{Bulk}_\Pi \log( \rho^{Bulk}_\Pi ) ).
\end{equation}
Depending only on the variation, $\delta \ket{\Phi}_{Bulk}$, we may express the variation of the bulk entanglement entropy in terms of the variation of the $\delta S_\Pi (\ket{\Phi}_{Bulk})$ by an argument parallel to that of (\ref{dSB=dtr})
\begin{equation}
\delta S_\Pi (\ket{\Phi}_{Bulk}) = - \tr( \delta \rho^{Bulk}_\Pi \log( \rho^{Bulk}_\Pi ) )
\end{equation}
where we have again used that the trace of the density matrix must be unity. Defining the bulk modular Hamiltonian
\begin{equation}
H^{Bulk}_{\Pi} \equiv - \log( \rho^{Bulk}_\Pi )
\end{equation}
we find
\begin{equation}
\begin{split}
\delta S_\Pi (\ket{\Phi}_{Bulk}) & = \tr( \delta \rho^{Bulk}_\Pi H^{Bulk}_{\Pi} )
\\& = \delta \tr( \rho^{Bulk}_\Pi H^{Bulk}_{\Pi} )
\\& = \delta \expval{H^{Bulk}_{\Pi}}
\end{split}
\end{equation}
where the last equality follows since $H^{Bulk}_{\Pi}$ is independent of the variation and from the definition of the expectation value. 

Similarly to the modular Hamiltonian on the CFT side, the bulk modular Hamiltonian cannot in general be expressed in terms of local operators. However, for the particular case where the full spacetime is AdS spacetime and we are interested in the modular Hamiltonian of the region $\Pi$, the modular Hamiltonian can be expressed as an integral over the bulk energy momentum tensor
\begin{equation}
H^{Bulk}_{\Pi} = \int_{\Pi} \xi^a_B T^{Bulk}_{ab} \epsilon^b
\end{equation}
where $\xi^b_B$ and $\epsilon^a$ are given as in (\ref{xib}) and (\ref{epsilon}). The derivation of this expression is analogous to the derivation in section \ref{Entropy and the Modular Hamiltonian}. Further details can be found in \citep{swingle_universality_2014}. With this expression for $H^{Bulk}_{\Pi}$ we find that
\begin{equation}
\delta S_\Pi (\ket{\Phi}_{Bulk}) = \delta \expval{H^{Bulk}_{\Pi}} = \int_{\Pi} \xi^a_B \delta \expval{T^{Bulk}_{ab}} \epsilon^b.
\end{equation}

%Given this expression for $\delta S_\Pi (\ket{\Phi}_{Bulk})$, we then find the following holographic interpretation for the variation of $S_B$ to the first subleading order in $\frac{1}{N}$
%\begin{equation}
%\delta S_B = \frac{\delta A(\tilde{B})}{4 G_N} + \int_{\Pi} \epsilon^a \xi^b_B \expval{T^{Bulk}_{ab}} 
%\end{equation}
%where $\frac{\delta A(\tilde{B})}{4 G_N}$ is given by (\ref{dSB}).

For the hyperbolic energy, we found in section \ref{Holographic interpretation of dE} the first order holographic interpretation 
\begin{equation}
\delta E^{Hyp}_B = 2 \pi \int_B d^{d-1} x \frac{R^2-r^2}{2R} \delta \expval{T^{tt}(x,z=0)}.
\end{equation} 
This was obtained by considering an infinitesimal ball on the boundary; subleading corrections to $\delta E^{Hyp}_B$ should therefore be related to the energy-momentum tensor of the bulk region associated with this infinitesimal boundary ball. Thus, if we assume that the expectation value of bulk energy-momentum tensor, $\expval{T^{Bulk}_{ab}}$, vanishes on the boundary, it follows that the correction to $\delta E^{Hyp}_B$ vanishes to first subleading order in $\frac{1}{N}$.
\section{Linearized gravity coupled to matter}
In the previous chapter, we found (\ref{0=intX})
\begin{equation}
\delta S^{Grav}_B - \delta E_B^{Grav} = \int_{\Pi} \mathbf{d}\chi_B(h)
\end{equation}
where $\text{d}\chi_B = -\frac{1}{8 \pi G_N} \xi_B^a \dot{G}^{E}_{ab} \epsilon^b$. Including the correction to $S^{Grav}_B$, we have the following relation to first subleading order in $G_N$ in the bulk
\begin{equation}
\begin{split}
\delta S^{Grav}_B - \delta E_B^{Grav} & = - \int_{\Pi} \frac{1}{8 \pi G_N} \xi_B^a \dot{G}^{E}_{ab} \epsilon^b + \int_{\Pi} \xi^a_B \delta \expval{T^{Bulk}_{ab}} \epsilon^b 
\\& = - \frac{1}{8 \pi G_N} \int_{\Pi} \xi_B^a \left( \dot{G}^{E}_{ab} - 8 \pi G_N \delta \expval{T^{Bulk}_{ab}} \right) \epsilon^b.
\end{split}
\end{equation}
Now, imposing as a constraint the holographic interpretation of the first law of entanglement entropy $\delta S^{Grav}_B = \delta E_B^{Grav}$, we find
\begin{equation}
0 = - \frac{1}{8 \pi G_N} \int_{\Pi} \xi_B^a \left( \dot{G}^{E}_{ab} - 8 \pi G_N \delta \expval{T^{Bulk}_{ab}} \right) \epsilon^b.
\end{equation}
An argument analogous to that in section (\ref{Combining the holographic results}) implies that this constraint is satisfied only if $\dot{G}^{E}_{ab} - 8 \pi G_N \delta \expval{T^{Bulk}_{ab}} = 0$, i.e. if the linearized Einstein equations with source term $8 \pi G_N \delta \expval{T^{Bulk}_{ab}}$ are satisfied in the bulk. 

This, therefore, demonstrates that imposing the constraint $\delta S_B = \delta E_B^{Hyp}$ to subleading order in $\frac{1}{N}$ on a quantum state with a semi-classical asymptotically AdS spacetime dual is equivalent to imposing the Einstein equations on that spacetime to subleading order in $G_N$ in the bulk, i.e. the spacetime satisfies the equations 
\begin{equation}\label{lEe}
\dot{G}^{E}_{ab} = 8 \pi G_N \delta \expval{T^{Bulk}_{ab}}.
\end{equation}
\\

The constraint, $d S_B = d E_B^{Hyp}$, we argued holds to any order in $\frac{1}{N}$. This may trigger the consideration whether it is equivalent to the Einstein equations to all orders. In order for $\delta \expval{T^{Bulk}_{ab}}$ to occur in (\ref{lEe}), it must be the case that $\expval{T^{Bulk}_{ab}}$ is present in the general spacetime constraint equivalent the general CFT constraint $d S_B = d E_B^{Hyp}$. \textit{Prima facie}, one could think that there were other operators, $O^{Bulk}_{ab}$, in this general spacetime constraint. The idea would be that the source term could be of the form $8 \pi G_N \expval{T^{Bulk}_{ab} + O^{Bulk}_{ab}}$. As argued in \citep{swingle_universality_2014}, a rather simple observation renders such a construction unlikely. From this general source term, we must obtain $8 \pi G_N \delta \expval{T^{Bulk}_{ab}}$ to subleading order in $G_N$, i.e. $\delta \expval{O^{Bulk}_{ab}}$ must vanish to this order. This would obtain if the operator $O^{Bulk}_{ab}$ was such that it annihilated the bulk vacuum state. However, any expansion in terms of creation and annihilation operators of a local quantum field theory operator will include terms without annihilation operators -- in other words, only with creation operators -- and therefore, the operator $O^{Bulk}_{ab}$ cannot be a local quantum field theory operator since $\delta \expval{O^{Bulk}_{ab}} = 0$. There may be more subtle ways to make this construction, however, the argument indicates that the full semi-classical Einstein equations, 
\begin{equation}\label{G=expT}
G^{E}_{ab} = 8 \pi G_N \expval{T^{Bulk}_{ab}}
\end{equation}
are the general constraints on the spacetime equivalent to CFT constraint $d S_B = d E_B^{Hyp}$. 

A more indirect argument to this effect is the observation that entanglement may occur between all types of degrees of freedom. All degrees of freedom in a subsystem contribute to the entanglement entropy and are therefore taken into account when the entanglement entropy for the subsystem is evaluated. In this way, entanglement is universal. As is well known, the Einstein equations signifies that all stress energy are sources of gravity. Regardless of the origin of the stress energy it sources the gravitational field. As they write in \citep{swingle_universality_2014}: ``we can say that \textit{the universality of the gravitational interaction comes directly from
the universality of entanglement}". This, we may argue, is why the full semi-classical Einstein equations should be equivalent to some constraint on entanglement in the dual quantum state. 

It is worth remarking that even if we can obtain the full semi-classical Einstein equations from $d S_B = d E_B^{Hyp}$, there still is an important restriction to the domain of application of this result essentially due to the fact that the spacetime is still classical. This is explicitly seen from the fact that the Einstein tensor (with cosmological constant) $G^{E}_{ab}$, on the LHS of (\ref{G=expT}), is not a quantum operator. Thus, (\ref{G=expT}) as a whole is not an operator relation; which can be regarded as the reason why we have to take the expectation value of $T^{Bulk}_{ab}$. A more subtle point leading to the same conclusion is that the Ryu-Takayanagi formula is ill-defined if the geometry under consideration is a superposition of spacetimes. Since the whole machinery above is founded on the Ryu-Takayanagi formula, it follows that we can at most obtain semi-classical gravity using it as our holographic dictionary. Thus, unless further correction are made to the Ryu-Takayanago formula, we cannot go beyond this domain of validity.

\chapter{Conclusion}
As demonstrated in chapter \ref{Entropy and Field Equations} and \ref{Beyond linearized gravity}, the linearized Einstein equations both with and without matter in a $d+1$-dimensional AdS background can be derived from the first law of entanglement entropy in a $d$-dimensional CFT. Furthermore, it was speculated that the full semi-classical Einstein gravity follows from imposing to first law of entanglement to all order in $\frac{1}{N}$.

The result is quite remarkable. Gravity in $d+1$ dimensions may be encoded in an intrinsically quantum mechanical structure -- entanglement structure -- in a $d$-dimensional CFT. The dynamics described by the Einstein's field equations can be reinterpreted as a first law like dynamical constraint on entanglement. But however remarkable this may be, the import of the result is initially less significant. The holographic relation between gravity and entanglement does not in the form developed here offer new insights the overarching problem of contemporary theoretical physics: Quantum gravity. By offering the alternative interpretation of semi-classical gravity as entanglement structure, we are offered an alternative description of a theory that we already know well and which arguably is less than full theory of quantum gravity.\footnote{It should be noted that there are those who argue for the possibility of semi-classical gravity on the fundamental level. See for instance \citep{tilloy_sourcing_2016} for a recent discussion.} For this and other reasons, there are continued research on the holographic relation between entanglement and gravity. It is beyond the scope of the present project to account for this research in detail, however, we will here just mention a few of the recent developments.

In \citep{faulkner_nonlinear_2017}, the holographic relation between entanglement and gravity is rigorously extended to non-linear gravity by showing that the second order perturbation of Einstein's field equations follows from the first law of entanglement entropy. Again, the Ryu-Takayanagi formula is employed as the only ingredient of the AdS/CFT dictionary. This may prove relevant, since a holographic relation between entanglement entropy and an emergent spacetime structure similar to the Ryu-Takayanagi formula has recently been recovered in the context of loop quantum gravity \citep{han_loop_2017}. This may suggest an independence from the AdS/CFT correspondence (and therefore from string theory) for a holographic prescription like the Ryu-Takayanagi formula. Obtaining the Ryu-Takayanagi in loop quantum gravity is driven by the relation between loop quantum gravity and tensor networks. Interesting, recent research has also explored the links from the relation between entanglement and gravity studied here to (random) tensor networks \citep{swingle_entanglement_2012,hayden_holographic_2016}.

In the chapters above, the Ryu-Takayanagi formula played a crucial role in the derivation of the holographic relation between gravity and entanglement. In \citep{jacobson_entanglement_2016} it is argued that one can obtain the same result using only the Bekenstein-Hawking formula under the assumption that the entanglement entropy is maximized for small fixed volume balls.

A research program relevant for classical general relativity is the conjectured relation between the Fisher information defined in terms of relative entropy for CFT subsystems and canonical energy in a dual AdS spacetime.  It has been shown in \citep{lashkari_canonical_2016} that the Fisher information is positive definite which entails that the corresponding canonical energy is positive definite as well. This in turn suggests a new positive energy theorem in general relativity.

There are also work that explores other constraints on entanglement in CFTs and how these translate holographically as constraints in the bulk. For instance, mutual entanglement information is considered in \citep{hayden_holographic_2013} and strong subadditivity (as well as mutual information) is considered in \citep{bao_holographic_2015}.
\subsection*{\small Acknowledgements}
\footnotesize This project is the result of many insightful discussion with more people than I can list here. However, I would in particular like thank my supervisor Troels Harmark for his valuable comments on earlier drafts of this work. Honourable mention also goes to my most frequent discussion partners Bjarke Nielsen and Niels Linnemann who have helped me sort out the numerous questions that arose during the genesis of this project.
\normalsize
\bibliography{physics,philosophy}
\end{document}